\RequirePackage{fix-cm}
\documentclass[smallextended]{svjour3}       % onecolumn (second format)
\smartqed  % flush right qed marks, e.g. at end of proof
\usepackage{graphicx}
\usepackage{subfigure}
\usepackage{float}
\usepackage{bm}
\usepackage{epsfig}
\usepackage{epstopdf}
\usepackage[usenames]{color}
\usepackage{amsmath,bm}
\usepackage{amsmath}
\usepackage{tikz}
\usepackage{lineno}
\usepackage{todonotes}
\usepackage{xcolor}
\usepackage{ulem}
\usepackage[hyperfootnotes=false,pdftex]{hyperref}
\usepackage{amsmath, amsthm, amsfonts}
\usepackage{cite}
\usepackage{caption}
\usepackage{bbm}

\theoremstyle{remark}
\numberwithin{figure}{section}
\numberwithin{table}{section}
\numberwithin{equation}{section}

\makeindex %para crear el indice de palabras
\hypersetup{
 pdfmenubar=true,
 pdftoolbar=true,
 pdfnewwindow=true,
 colorlinks=true,
 citecolor=blue,
 linkcolor=red,
 urlcolor=blue,
 bookmarks=true,}
 
\begin{document}
\normalem
\title{Mathematical modelling  of the first  HIV/ZIKV co-infection cases in Colombia and Brazil}

%\subtitle{Mathematical modeling  of   HIV/ZIKV}


\author{Jhoana P. Romero-Leiton \and Idriss Sekkak \and Julien Arino \and Bouchra Nasri 
}


\institute{J. P. Romero-Leiton \& Julien Arino \at
   Department of Mathematics, University of Manitoba,\\
Winnipeg MB R3T 1E9 Canada.\\
\email{jhoana.romero@umanitoba.ca \&  julien.arino@umanitoba.ca}
   \and
        I. Sekkak \& B. Nasri \at
              Département de Médecine Sociale et Préventive, 
              \\École de Santé Publique de l’Université de Montréal, Montreal QC H3N 1X9 Canada.\\
              \email{idriss.sekkak@umontreal.ca \& bouchra.nasri@umontreal.ca}           %  \\
        }

\date{Received: date / Accepted: date}

\maketitle
\begin{abstract}
This paper presents a mathematical model to investigate the co-infection between Human Immunodeficiency virus (HIV) and Zika virus (ZIKV) in Colombia and Brazil, where the first cases were reported during 2015-2016. The model considers the sexual transmission dynamics of both viruses and vector-host interactions. We begin by exploring the qualitative behaviour of each model separately. Then, we analyze the dynamics of the co-infection model using the thresholds and results defined separately for each model. The model also considers the impact of intervention strategies, such as personal protection, antiretroviral therapy (ART), and sexual protection (condom use). Using available and assumed parameter values for Colombia and Brazil, the model is calibrated to predict the potential effect of implementing combinations of those intervention strategies on the co-infection spread. According to these findings, transmission through sexual contact is a determining factor in the long-term behaviour of these two diseases. Furthermore, it is important to note that co-infection between HIV and ZIKV may result in higher rates of HIV transmission and an increased risk of severe congenital disabilities in children linked to ZIKV infection. As a result, control interventions have been implemented to limit the number of infected individuals and mosquitoes, with the aim of interrupting disease transmission. This study provides novel insights into the dynamics of HIV/ZIKV co-infection. It highlights the importance of integrated intervention strategies in controlling the spread of these viruses, which may impact positively public health. 

\end{abstract}

\section{Introduction}\label{Section_introduction}
%%%%%

Human immunodeficiency virus (HIV) and Zika virus (ZIKV) are two major public health concerns worldwide, particularly in Latin America and Caribbean countries \cite{machado2019scientific,garcia2014changing}.  While HIV is a chronic infection that attacks the immune system, ZIKV is transmitted by mosquitoes and can even cause congenital malformations in children, such as Guillain-Barré syndrome \cite{cao2016,oehler2014,smith2016} and microcephaly \cite{mlakar2016,calvet2016}. If HIV is not promptly treated, it can cause Acquired Immunodeficiency Syndrome (AIDS) \cite{kapila2016review}. This virus can be transmitted through sexual contact, syringe misuse, and vertically (from mother to child) \cite{kabapy2020attributes}. HIV/AIDS still has no cure, so treatments seek to lower or reduce the level of virus replication within the host, which consists of several medications, commonly called antiretroviral  therapy (ART) \cite{tilton2010entry}. ZIKV, unlike other arboviruses, also presents transmission through sexual contact. Some studies have demonstrated its detection and transmission through semen, urine, and saliva \cite{atkinson2016,gourinat2015,khurshid2019human}. 
\par
Although  HIV  is not a zoonotic disease, it has similar specific transmission mechanisms with ZIKV. Both viruses can be transmitted through sexual contact and vertically from mother to fetus. In endemic zones, the sexual transmission route can substantially worsen the vulnerability of both mother and fetus to other sexually transmitted infections, particularly HIV \cite{rothan2018}. 
Until now, there have been few cases of ZIKV infection in HIV-infected individuals worldwide. 
The first documented case of HIV/ZIKV co-infection was confirmed in a 38-year-old patient in a Rio de Janeiro (Brazil) laboratory in 2015 \cite{calvet2016}. In the same region, a Zika case was reported in an HIV-infected pregnant woman \cite{brasil2016,joao2018}. The fetus displayed significant abnormalities, consistent with findings from previous studies conducted on pregnant women who contracted the Zika virus in Brazil. This particular case concluded with the fetus's death \cite{brasil2016}.
In 2018, five individuals from the departments of Risaralda and Sucre were reported with HIV/ZIKV co-infection in Colombia \cite{villamil2018}, who demonstrated adequate immune and virological control for ZIKV, as compared to those who were only infected with ZIKV \cite{villamil2018}. 
\par
Therefore, additional research is necessary to understand better the interactions between HIV and ZIKV and the impact of this co-infection on the immune response, disease severity, further complications, and control \cite{villamil2018,rothan2018}. 
It is not yet well understood how HIV infection increases the risk of ZIKV infection and vice-versa, and how this the co-infection affects pregnant women and fetuses. Nevertheless, laboratory investigations have demonstrated that placental tissues are susceptible to ZIKV infection \cite{rothan2018}. ZIKV not only causes immunity response problems but also, in the presence of co-infections, causes placental dysfunction \cite{rothan2018}.  In pregnant women especially, ZIKV mainly targets CD14+ monocytes, which leads to an inflammatory responses and immune tolerance \cite{rothan2018}. However, there are several ways that ZIKV can play a role in HIV infection, especially via cytokines and the activation of CD4+T cells or linking to HIV  \cite{koblischke2018structural}. Unlike HIV, the relationship between transmission of ZIKV by mother to child and fetal disease infection has not yet been established \cite{aschengrau2021international}. However, even in the presence of ART, severe viral infection is likely to exacerbate the disorder of the immune system for pregnant women with HIV and increase the risk of transmission from mother to child of HIV and ZIKV \cite{aschengrau2021international}. Therefore, the potential interaction between HIV and ZIKV has recently garnered significant attention \cite{mittal2017zika}. These interactions can modify infections' epidemiology, pathogenesis, immune response, and therapy. For instance, co-infection can expedite HIV pathogenesis and enhance transmission by boosting viral replication efficiency. Furthermore, ZIKV transmission and infections can potentially cause serious symptoms and conditions for immunocompromised individuals when co-infections occur \cite{rothan2018}.
\par 
Given the potential impact of HIV/ZIKV co-infection on public health, it is crucial to understand the transmission dynamics of these viruses and evaluate the effectiveness of intervention strategies. Mathematical models are useful for understanding and providing insights into public health policy decisions. To our knowledge, there is no evidence of mathematical models studying this phenomenon in the literature. Therefore, this study aimed to formulate and analyze an HIV/ZIKV co-infection model, assuming that both viruses are sexually transmitted and ZIKV is also mosquito-transmitted. From the analysis of this model is expected to identify important transmission outcomes that would help to design and evaluate different control and prevention strategies to minimize their impact on public health.  
\par
The organization of this study is as follows: In Section \ref{sec-model-formulation}, the co-infection model is introduced, offering an in-depth overview of its structure and assumptions. Subsequently, Sections \ref{sec-hiv-only} and \ref{sec-zikv-only} present the individual dynamics of the HIV-only and ZIKV-only models, respectively, highlighting the unique characteristics of each infection. Section \ref{sec-coinfection} focuses on the analysis of the co-infection model, examining the interactions and joint effects of HIV and ZIKV. The optimal control problem is addressed analytically in Section \ref{sec-optimal-control}, emphasizing the exploration of combined strategies to mitigate the spread and impact of the co-infection. Furthermore, in Section \ref{sec:case-of-study}, a case study centred in Colombia and Brazil is presented, whereby the uncontrolled and controlled models are numerically analyzed using data derived from available literature and convenient assumptions. Finally, in Section \ref{sec-discussion},  discussion and concluding remarks are provided, covering the modelling approach and its outcomes, as well as limitations, opportunities for further study and open-ended questions.

%%%%%%%%%%%%%sec2
\section{The HIV/ZIKV mathematical model formulation }\label{sec-model-formulation}
%%%%%
This model examines two distinct groups: the human (host) and mosquito (vector) populations. We assumed that an individual susceptible to either disease could not be infected with both diseases simultaneously at the same time. Additionally, interactions between susceptible, co-infected, and infected individuals with a single pathogen can result in numerous complex scenarios. Determining outcomes is challenging because of the different paths of infection and possible combinations of outcome. For example, a susceptible person who has sexual contact with a co-infected individual may be infected with Zika, HIV, or both simultaneously. Given this complexity, the mathematical model assumes that co-infected individuals transmit the virus through vectorial (non-sexual) contact, as reflected in the equations for $S_m$ and $I_m$.
The total number $N(t)$ of people in the human population at a given time is divided into six categories: those who are susceptible to both viruses, denoted as $S(t)$; those who are infected with only ZIKV but still susceptible to HIV, denoted as $I_z(t)$; those who are infected with only HIV but still susceptible to ZIKV, denoted as $I_h(t)$; those who are infected with both HIV and ZIKV simultaneously, denoted as $I_{hz}(t)$; those who are infected with AIDS, denoted as $A(t)$; and those who have recovered from ZIKV, denoted as $R(t)$. Thus, the total human population can be represented as the sum of all these compartments ($N(t)=S(t)+I_z(t)+I_h(t)+I_{hz}(t)+A(t)+R(t)$), as shown in Tables \ref{table01}-\ref{table02}.
\par 
The total mosquito population $N_m(t)$ at time $t$ is classified into two compartments:
the susceptible mosquito population $S_m(t)$ and the ZIKV-carrying mosquitoes $I_m(t)$. Thus,
$N_m(t) = S_m(t) + I_m(t)$ (see Tables \ref{table01} and \ref{table02}).

%%%%%
\begin{table}[H]
\centering
{\footnotesize
\begin{tabular}{ll}
  \hline
  Variable  & Description   \\ \hline
  $N(t)$ & The total human population at time $t$ \\
  $S(t)$ & Susceptible human population at time $t$ \\
  $I_z(t)$ &Infected human population with only ZIKV at time $t$  \\
  $I_h(t)$ &Infected human population with only HIV at time $t$  \\
  $I_{hz} (t)$&Infected human population with ZIK/HIV at time $t$  \\
  $A (t)$&Infected human population with AIDS  at time $t$  \\
   $R(t)$ & Recovered human population of ZIKV   at time $t$  \\
   $N_m(t)$ & The total mosquito population at time $t$ \\
   $S_m(t)$ &Susceptible mosquito population at time $t$ \\
   $I_m(t)$ &ZIKV-carrying mosquito population at time $t$ \\
  \hline 
\end{tabular}}
\caption{Description of the state variables involved  in Model (\ref{model1}). }
\label{table01}
\end{table}

%%%%%%table parameters
\begin{table}[H]
\centering
\resizebox{14cm}{!}{
{\footnotesize
\begin{tabular}{lll}
  \hline
  Parameter & Description & Dimension  \\ \hline
   $\Lambda$ & Recruitment of humans  &$pop\times time^{-1}$  \\  
   $\beta_m$ &Infection transmission of  humans by contact with  infected mosquitoes with ZIKV &$(pop \times time)^{-1}$ \\
   $\beta_z$ & Infection transmission of  humans by contact with  humans infected with ZIKV through sexual contact   & $(pop\times time)^{-1}$ \\
   $\beta_h$ &Infection transmission of  humans by contact with  humans infected with HIV through sexual contact  & $(pop \times time)^{-1}$\\
  %deleted  $\gamma_h$ & Infection rate by HIV through other factors (blood transfusion or infected syringes )  & time$^{-1}$\\
   $1/\sigma_1$ &Mean duration of the immunodeficiency period & time \\
      $1/\sigma_2$ & Mean duration of the immunodeficiency period in co-infected individuals  & time \\
   $1/\mu$ & Human mean lifespan  & time\\
   $\mu_z$ & Mortality rate by Zika  & time$^{-1}$\\
    $\mu_h$ & Mortality rate by AIDS  & time$^{-1}$\\
     $\mu_{hz}$ & Mortality rate by ZIKV/AIDS  & time$^{-1}$\\
   $1/\delta_z$ & Mean duration of the Zika infection & time\\
   $\omega_1$ &Transition probability from HIV to HIV/ZIKV co-infection& Dimensionless  \\
    $\omega_2$ & Transition probability from ZIKV to HIV/ZIKV co-infection& Dimensionless  \\
   $\epsilon$ & Zika recovery rate scaling factor& Dimensionless   \\
   $\Lambda_m$ & Recruitment of mosquitoes  & $pop\times time^{-1}$ \\
   $\alpha_m $ &Infection transmission of mosquitoes by contact with infected humans with ZIKV & $(pop \times time)^{-1}$\\
   $1/\mu_m$ & Mosquito mean lifespan  & time\\ 
  \hline 
\end{tabular}}}
\caption{Description and dimension of the parameters involved in Model (\ref{model1}). }
\label{table02}
\end{table}

\begin{equation}\label{model1}
\left\{
\begin{array}{ll}
%%%%1
 \dfrac{dS}{dt}&=\Lambda-\left(\beta_mI_m+\beta_z I_z+\beta_h I_h\right)S-\mu S \\ \\
 %%%%2
 \dfrac{dI_z}{dt}&=\left(\beta_mI_{m}+\beta_z I_z\right)S-\omega_2\beta_hI_z I_h-(\mu_z+\delta_z+\mu)I_z \\ \\
 %%%%3
\dfrac{dI_h}{dt}&=\beta_hI_hS-\omega_1(\beta_mI_m+\beta_zI_z)I_h-(\sigma_1+\mu )I_h \\ \\
%%%%%%%%%%%%4
\dfrac{dI_{hz}}{dt}&=\omega_1(\beta_mI_m+\beta_zI_z)I_h+\omega_2\beta_hI_h I_z-\epsilon\delta_zI_{hz}-(\sigma_2+\mu_{hz}+\mu)I_{hz} \\ \\
%%%%%5
\dfrac{dA}{dt}&=\sigma_1 I_h+\sigma_2 I_{hz}-(\mu_h+\mu)A\\ \\
%%%%6
\dfrac{dR}{dt}&=\delta_zI_z+\epsilon\delta_zI_{hz}-\mu R \\ \\
%%%%%7
\dfrac{dS_m}{dt}&=\Lambda_m-\alpha_m (I_z+I_{hz})S_m-\mu_mS_m \\ \\
%%%%%8
\dfrac{dI_m}{dt}&=\alpha_m (I_z+I_{hz}) S_m-\mu_mI_m \\ \\
\end{array}
\right.
\end{equation}

\begin{figure}[H]
\centering
\includegraphics[width=16cm, height=10cm]{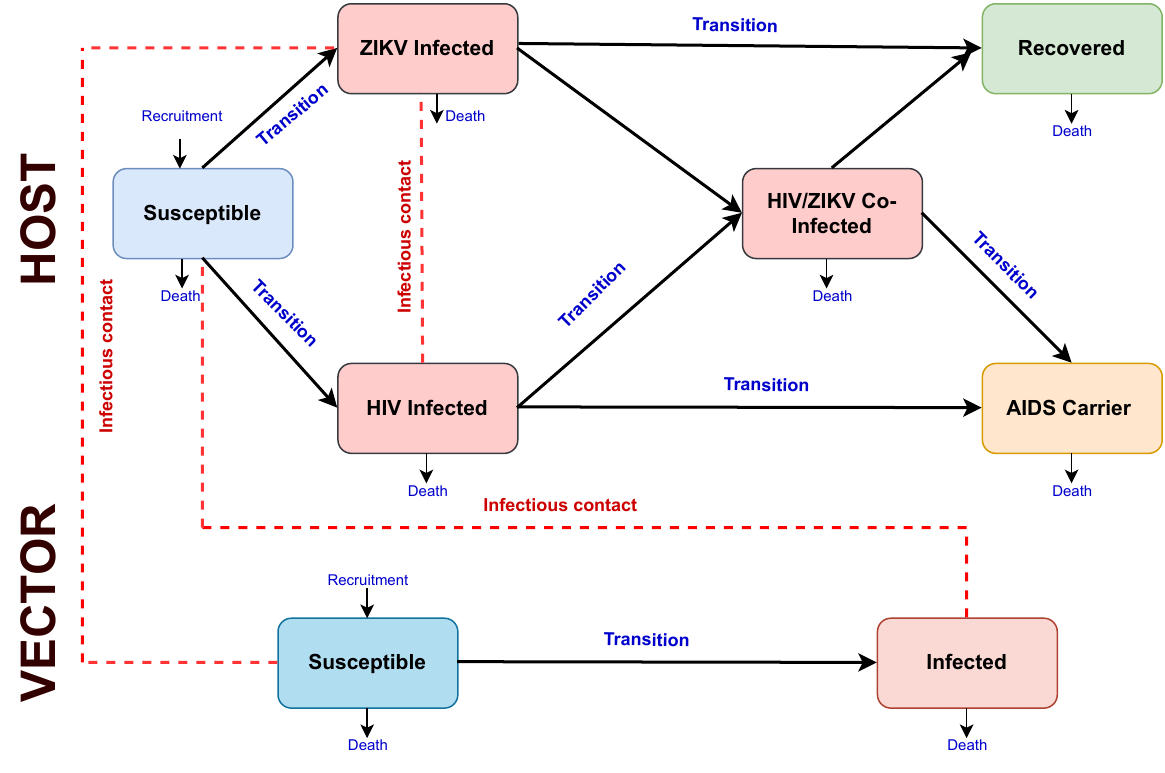}
\caption{HIV/ZIKV co-infection model represented in  Model (\ref{model1}).}
\label{fig01}
\end{figure}

Figure \ref{fig01} illustrates the dynamics depicted by the equations involved in Model \eqref{model1}.

In the following two sections, we qualitatively analyze the properties of System \eqref{model1}. We will start by analyzing the dynamics of the two-component models: the HIV/AIDS model and the ZIKV model. 

%%%%%%%%% HIV only model 
\subsection{Qualitative behaviour of the HIV/AIDS  model  }\label{sec-hiv-only}
The HIV/AIDS model is obtained by setting $I_z=I_{hz}=R=S_m=I_m=0$ in System \eqref{model1}. Thus, the ODEs described in \eqref{model1} can be rewritten as: 
\begin{equation}\label{model-hiv}
\left\{
\begin{array}{ll}
%%%%1
 \dfrac{dS}{dt}&=\Lambda-\beta_h I_h S-\mu S \\ \\
 
 %%%%3
\dfrac{dI_h}{dt}&=\beta_hI_hS-(\sigma_1+\mu )I_h \\ \\

%%%%%5
\dfrac{dA}{dt}&=\sigma_1 I_h-(\mu_h+\mu)A,\\ \\

\end{array}
\right.
\end{equation}
where the total human population is  $N_h(t)=S(t)+I_h(t)+A(t)$. For this model, the region of biological interest is

\begin{equation}\label{omega_h}
\Omega_h=\left\{ (S, I_h, A)\in \mathbb{R}_+^{3}: 0\leq N_h\leq \frac{\Lambda}{\mu} \right\}.
\end{equation}

It can be proved that $\Omega_h $ is positively-invariant under the flow of \eqref{model-hiv} (see e.g., \cite{romero2019optimal}), that is, all solutions of System \eqref{model-hiv} starting in $\Omega_h$ remain in $\Omega_h$ for all $t\geq 0$. 
Therefore, it is enough to consider the dynamics of \eqref{model-hiv} in $\Omega_h$. 
\par 
The disease-free equilibrium (DFE) of Model \eqref{model-hiv} is given by 
\begin{equation}\label{DFE-hiv}
\textbf{E}_{h_0}=\left(\frac{\Lambda}{\mu}, 0,0\right),
\end{equation}
which is obtained when $I_h=A=0$. The stability of this equilibrium point can be analyzed in terms of the basic reproduction number for the HIV/AIDS model ($\mathcal{R}_h$), which can be computed using the next-generation operator \cite{van2002reproduction}.  Using the notation of \cite{romero2019optimal} in Model \eqref{model-hiv} the matrices $\textbf{F}$ and $\textbf{V}$ are given by
\begin{equation*}
\textbf{F}=\begin{bmatrix}
\beta_h\frac{\Lambda}{\mu} &0 \\
0 &0
\end{bmatrix}, \quad \text{and} \quad \textbf{V}=\begin{bmatrix}
\sigma_1+\mu & 0 \\
-\sigma_1 & \mu_h+\mu
\end{bmatrix}.
\end{equation*}
Therefore, it follows that $\mathcal{R}_h$ associated to Model \eqref{model-hiv} is given by

\begin{equation}\label{R0-hiv}
\mathcal{R}_h=\rho(\textbf{FV}^{-1})=\dfrac{\beta_h}{\sigma_1+\mu}\dfrac{\Lambda}{\mu}, 
\end{equation}
where $\rho$ represents the spectral radius of the matrix $\textbf{FV}^{-1}$. 
\par 
To determine the endemic equilibrium points of Model\eqref{model-hiv}, we must solve  the system of algebraic equations 
\begin{equation}\label{equilibrium_hiv}
\left\{
\begin{array}{ll}
%%%%1
& 0=\Lambda-\beta_h I_h S-\mu S\\ \\
 
 %%%%3
&0=\beta_hI_hS-(\sigma_1+\mu )I_h \\ \\

%%%%%5
&0=\sigma_1 I_h-(\mu_h+\mu)A.
\end{array}
\right.
\end{equation}
After some algebraic manipulations  for $I_h, A\neq 0$, we find that the solutions of System \eqref{equilibrium_hiv} are 

\begin{equation*}
S^*= \frac{\sigma_1+\mu}{\beta_h}, \quad I_h^*=\frac{\mu}{\beta_h}(\mathcal{R}_h-1)\quad \text{and}\quad A^*=\frac{\mu\sigma_1}{\beta_h(\mu_h+\mu)}(\mathcal{R}_h-1).
\end{equation*}

Thus, as long as  $\mathcal{R}_h>1$,  System \eqref{model-hiv} has an endemic equilibrium point given by

\begin{equation}\label{endemic-hiv}
\textbf{E}^*_{h}=\left(\frac{\sigma_1+\mu}{\beta_h}, \frac{\mu}{\beta_h}(\mathcal{R}_h-1), \frac{\mu\sigma_1}{\beta_h(\mu_h+\mu)}(\mathcal{R}_h-1)\right).
\end{equation}

The local stability of $\textbf{E}_{h_0}$ defined on \eqref{DFE-hiv} and $\textbf{E}^*_h$ defined on \eqref{endemic-hiv} is determined by the sign of the eigenvalues  of the  linearisation matrix (Jacobian matrix) of System \eqref{model-hiv} around them. The Jacobian matrix of System \eqref{model-hiv} at an arbitrary point $\textbf{E}=(S, I_h, A)$ is 
\begin{equation}\label{jacobian-hiv}
\textbf{J}(\textbf{E})=\begin{bmatrix}
-(\beta_h I_h+\mu) &-\beta_h S &0 \\
\beta_h I_h & \beta_h S-(\sigma_1+\mu) &0 \\
0 & \sigma_1 &-(\mu_h+\mu )
\end{bmatrix}
=\begin{bmatrix}
    \mathbf{J}_{11}(\mathbf{E}) & \mathbf{0} \\
    \mathbf{\star} & -(\mu_h+\mu )
\end{bmatrix},
\end{equation}
so eigenvalues of $\mathbf{J}(\mathbf{E})$ are $-(\mu_h+\mu)<0$ and those of $\mathbf{J}_{11}(\mathbf{E})$. 
At $\mathbf{E}_{h_0}$,
\[
\mathbf{J}_{11}(\mathbf{E}_{h_0})=\begin{bmatrix}
    -\mu &-\beta_h \frac{\Lambda}{\mu} \\
    0 & (\sigma_1+\mu)(\mathcal{R}_h-1)
\end{bmatrix}
\]
is upper triangular with eigenvalues $-\mu<0$ and $(\sigma_1+\mu)(\mathcal{R}_h-1)$. At $\textbf{E}^*_h$,
\[
\mathbf{J}_{11}(\mathbf{E}_h^*)
=\begin{bmatrix}
-\mu\mathcal{R}_h &-(\sigma_1+\mu) \\
\mu\mathcal{R}_h & 0
\end{bmatrix}.
\]
Thus $\mathbf{J}_{11}(\mathbf{E}_h^*)$ has positive determinant and negative trace, implying eigenvalues with negative real parts when $\mathbf{E}_h^*$ is biologically relevant, i.e., when $\mathcal{R}_h>1$.

These results are summarized in the following proposition.

%%%%
%%%%%%%

\begin{proposition}\label{lemma_hiv}
System \eqref{model-hiv} always has a DFE $\textbf{E}_{h_0}$ given in \eqref{DFE-hiv}, and for $\mathcal{R}_h>1$, an endemic equilibrium point $\textbf{E}^*_h$ given in \eqref{endemic-hiv} exists. Additionally, the following stability results hold:
\begin{itemize}
\item [(i)] If $\mathcal{R}_h<1$ the DFE is both locally asymptotically stable (\textit{LAS}) and globally asymptotically stable (GAS) in $\Omega_h$ defined in \eqref{omega_h}, whereas $\textbf{E}^*_h$ is unestable.
\item [(ii)] If $\mathcal{R}_h>1$, the endemic equilibrium point $\textbf{E}^*_h$ is LAS-GAS and in $\Omega_h$, whereas the DFE $\textbf{E}{h_0}$ becomes an unstable hyperbolic point.
\end{itemize}
\end{proposition}

Appendix \ref{appendix:A1} establishes the global asymptotic stability of $\mathbf{E}_{h_0}$ under the condition $\mathcal{R}h<1$, while Appendix \ref{appendix:A2} demonstrates the global asymptotic stability of $\mathbf{E}{h}^*$ for the case where $\mathcal{R}_h>1$.
\par 
As it is well known, the basic reproduction number of an infection $\mathcal{R}_h$  is the average number of new cases generated by a given case throughout an infectious period. Thus, Proposition \ref{lemma_hiv} tells us that HIV infection can be controlled in the community if the initial values of the subpopulation of the model are in the region of attraction of $\textbf{E}_{h_0}$. Additionally, to ensure that the control of the virus is independent of initial conditions, in the same proposition we also stated the global asymptotic stability (\textit{GAS}) of the equilibrium points.

%%%%%%%ZIKA ONLY MODEL SECTION
 \subsection{Qualitative behaviour of the ZIKV model }\label{sec-zikv-only}
   
By   setting $I_h=I_{hz}=A=0$ in System \eqref{model1}, we obtain the ZIKV model as follows

\begin{equation}\label{model-zikv}
\left\{
\begin{array}{ll}
%%%%1
 \dfrac{dS}{dt}&=\Lambda-\left(\beta_mI_m+\beta_z I_z\right)S-\mu S \\ \\
 %%%%2
 \dfrac{dI_z}{dt}&=\left(\beta_mI_{m}+\beta_z I_z\right)S-(\mu_z+\delta_z+\mu)I_z \\ \\
%%%%6
\dfrac{dR}{dt}&=\delta_zI_z-\mu R \\ \\
%%%%%7
\dfrac{dS_m}{dt}&=\Lambda_m-\alpha_m I_zS_m-\mu_mS_m \\ \\
%%%%%8
\dfrac{dI_m}{dt}&=\alpha_m I_z S_m-\mu_mI_m 
\end{array}
\right.
\end{equation}  
For this model, the total human population is  $N_{z}(t)=S(t)+I_{z}(t)+R(t)$, and the total mosquito population is $N_{m}(t)=S_{m}(t)+I_{m}(t)$.  We are interested in analyzing the solutions within the biological region of interest
\begin{equation} \label{omega_z}
    \Omega_{z}=\left\{ (S,I_z,R,S_m,I_m) \in  \mathbb{R}_+^{5}: 0\leq N_h\leq \frac{\Lambda}{\mu} ; 0\leq N_{m} \leq \frac{\Lambda_{m}}{\mu_{m}}  \right\}. 
\end{equation}
It can also be shown that  $\Omega_{z}$ is positively invariant under the flow of \eqref{model-zikv}.
\par
The DFE for Model   \eqref{model-zikv} is given by 
\begin{equation}\label{DFE-zikv}
\textbf{E}_{z_{0}}=\left(\frac{\Lambda}{\mu}, 0,0,\frac{\Lambda_{m}}{\mu_{m}},0\right).
\end{equation}
Now, we will determine the basic reproduction number $\mathcal{R}_{z}$ for Model \eqref{model-zikv}.  Similarly as in Section \ref{sec-hiv-only}, the matrices  $\textbf{F}$  and  $\textbf{V}$ are given by 

\begin{equation*}
    \textbf{F}=\begin{bmatrix}
        \beta_z\frac{\Lambda}{\mu} & \beta_m\frac{\Lambda}{\mu} \\
        \alpha_m\frac{\Lambda_m}{\mu_m} & 0 \\ 
    \end{bmatrix},
    \quad \text{and} \quad \textbf{V}=\begin{bmatrix}
        \delta_z+\mu+\mu_z & 0 \\ 
        0 & \mu_m
    \end{bmatrix}.    
\end{equation*}

Thus, the basic reproduction number for Model \eqref{model-zikv} is given by 

\begin{equation}\label{R_{z}}
        \mathcal{R}_{z}=\rho(\textbf{FV}^-1)=\mathcal{R}_{z_{1}}+\sqrt{\left( \mathcal{R}_{z_1}  \right)^2+\bar{\mathcal{R}}_{z_2} }=:\mathcal{R}_{z_1}+\mathcal{R}_{z_2},
\end{equation}
where 
\begin{equation}\label{R_{z12}}
\mathcal{R}_{z_1}=\dfrac{\beta_z\Lambda}{2\mu(\mu_z+\delta_z+\mu)}, \quad 
\bar{\mathcal{R}}_{z_2}=\dfrac{\beta_m\alpha_m\Lambda_m\Lambda}{\mu\mu_m^2(\mu_z+\delta_z+\mu)}  \quad \text{and} \quad \mathcal{R}_{z_2}=
\sqrt{\left( \mathcal{R}_{z_1}  \right)^2+\bar{\mathcal{R}}_{z_2} }.
\end{equation}

\begin{remark}
Note that if the transmission of ZIKV by sexual contact is not considered  ($\beta_z=0$),  $\mathcal{R}_{z_1}=0$ and $ \mathcal{R}_{z}$ reduces to 
\begin{equation*}\label{R-z}
        \left.\mathcal{R}_{z}\right|_{(\beta_z=0)}=\sqrt{\bar{\mathcal{R}}_{z_2}}=\sqrt{\dfrac{\beta_m\alpha_m\Lambda_m\Lambda}{\mu\mu_m^2(\mu_z+\delta_z+\mu)}}, 
\end{equation*}
indicating that sexual contact transmission of ZIKV has an impact on $\mathcal{R}_{z}$.  
\end{remark}

The following lemma makes it easier to determine the sign of $\mathcal{R}_z$.

\begin{lemma}\label{lemmaR*}
Let us define 
\begin{equation}\label{R*-zikv}
\begin{array}{ll}
\mathcal{R}_z^*&=2\mathcal{R}_{z_1}+\bar{\mathcal{R}}_{z_2}\\ \\
 &=:\mathcal{R}_z^2+2\mathcal{R}_{z_1}(1-\mathcal{R}_z).
\end{array}
\end{equation}

\begin{enumerate}
\item [i.] If $\mathcal{R}_z^*<1$, then $2\mathcal{R}_{z_1}<1$ and $\mathcal{R}_z<1$.
\item [ii.] If $\mathcal{R}_z^*>1$ and $2\mathcal{R}_{z_1}<1$, then $\mathcal{R}_z>1$.
\item [iii.] If $\mathcal{R}_z^*>1$ and $2\mathcal{R}_{z_1}>1$, then $\mathcal{R}_z<1$.
\item [iv.] If $\mathcal{R}_z^*=1$, then $\mathcal{R}_z=1$.
\end{enumerate}
\end{lemma}

The proof can be found in Appendix \ref{appendix:A3}.
\par
From the above lemma we can see that the sign of $\mathcal{R}_z$ is  determined by the signs of $\mathcal{R}^*_z$ and $2\mathcal{R}_{z_1}$.  
\par 

%%%%%%%%%%%%%%%%%LAS of DFE
We can now determine the local asymptotic stability of the DFE    $\textbf{E}_{z_0}$.  To this end, we   order equations and variables as $S,S_m,I_z,I_m,R$ and compute the Jacobian matrix of System \eqref{model-zikv} at an arbitrary point $\textbf{E}=(S,S_m,I_z,I_m,R)$, which is given by 

\begin{equation}\label{jacobian-zikv}
\textbf{J}(\textbf{E})=\begin{bmatrix}
    \textbf{J}_{11}(\textbf{E}) & \mathbf{0} \\
    \mathbf{\star} & -\mu, 
\end{bmatrix}.
\end{equation}

Thus, the eigenvalues of $\textbf{J(E)}$ are $-\mu < 0 $ and those of $\textbf{J}_{11}(\textbf{E})$.  At $\textbf{E}_{z_0}$ 

\begin{equation*}
\textbf{J}_{11}(\textbf{E}_{z_0})=\begin{bmatrix}
%%f1
-\mu &0&-\beta_z\Lambda/\mu &-\beta_m\Lambda/\mu \\
%%f2
0 &-\mu_m &-\alpha_m\Lambda_m/\mu_m&0\\ 
%%%f3
0&0&\beta_z\Lambda/\mu-(\mu_z+\delta_z+\mu)&\beta_m\Lambda_m/\mu_m\\
%%%%%%f4
0 & 0&\alpha_m\Lambda_m\mu_m &-\mu_m
\end{bmatrix}=\begin{bmatrix}
\mathbf{D}_{11} & \mathbf{\star} \\
0 & \mathbf{D}_{22}
\end{bmatrix}.
\end{equation*}

Again, the above matrix is $2\times 2$ block upper triangular with (1,1) block being a diagonal matrix with diagonal entries (eigenvalues) $-\mu<0$ and $-\mu_m<0$. So all that remains to determine the eigenvalues of the $2\times 2$ matrix

\begin{equation*}
\mathbf{D}_{22}=\begin{bmatrix}
\beta_z\dfrac{\Lambda}{\mu}-(\mu_z+\delta_z+\mu)& \beta_m\dfrac{\Lambda_m}{\mu_m} \\ \\
%%%%%%f4
\alpha_m\dfrac{\Lambda_m}{\mu_m} &-\mu_m
\end{bmatrix}.
\end{equation*}

After some algebraic manipulations, we  found that the above matrix has associated  the following characteristic polynomial. 

\begin{equation}\label{quadratic_Ez0}
q(x)= x^2+a_1x+a_2, \quad \text{where}
\end{equation}

\begin{equation*}
\begin{array}{ll}
&a_1= \mu_m-(\mu_z+\delta_z+\mu)(2\mathcal{R}_{z_1}-1) \\ \\
&a_2=\mu_m(\mu_z+\delta_z+\mu)(1-R_z^*).
\end{array}
\end{equation*}

Note that all roots of the characteristic polynomial $q(x)$ have a negative real part if  $\mathcal{R}^*_z$ defined in \eqref{R*-zikv} satisfies $\mathcal{R}^*_z<1$ (see Lemma \ref{lemmaR*}).

%%%%%%%%%%%%%%%%Endemic equilibrium 
\par
Now, in order to determine the endemic equilibrium points of System \eqref{model-zikv}, we have to solve  the following system of algebraic equations
\begin{equation}\label{equilibrium_zikv}
\left\{
\begin{array}{ll}
%%%%1
0=\Lambda-\left(\beta_mI_m+\beta_z I_z\right)S-\mu S\\ \\
 
 %%%%2
 0 =\left(\beta_mI_{m}+\beta_z I_z\right)S-(\mu_z+\delta_z+\mu)I_z \\ \\
%%%%6
0=\delta_zI_z-\mu R \\ \\
%%%%%7
0 =\Lambda_m-\alpha_m I_zS_m-\mu_mS_m \\ \\
%%%%%8
0 =\alpha_m I_z S_m-\mu_mI_m.
\end{array}
\right.
\end{equation}
To this end, we first define the following threshold 

\begin{equation}\label{Rmax-zikv}
\begin{array}{ll}
&R_{max}=\dfrac{\Lambda}{\mu}\dfrac{\delta_z}{\mu_z+\delta_z+\mu}.
\end{array}
\end{equation}

Therefore, after some algebraic manipulations in \eqref{equilibrium_zikv}, we can write the variables $S$, $I_z$, $S_m$ and $I_m $ in terms of the variable $R$ and the thresholds defined in \eqref{Rmax-zikv} as follows. 

\begin{equation}\label{coordinates-zikv}
\begin{array}{llll}
& S=\dfrac{\Lambda}{\mu}\left( 1- \dfrac{R}{R_{max}}\right), 
&I_z= \dfrac{\mu}{\delta_z}R, \\ \\
&S_m= \dfrac{\Lambda_m}{\mu_m}\left( 1-\dfrac{\alpha_m \mu R}{\mu_m\delta_z+\alpha_m \mu R}\right), 
& I_m= \dfrac{\alpha_m\Lambda_m\mu}{\mu_m\delta_z+\alpha_m\mu R}R.
\end{array}
\end{equation}

From the above expressions, it can be observed  that $S_m$ is always positive (because $\mu_m\delta_z+\alpha_m \mu R>\alpha_m\mu R$ and,  if $R<R_{max}$ then $S>0$). Replacing the above values into the second equation of \eqref{equilibrium_zikv}, we get the following quadratic equations in the variable $R$:

\begin{equation}\label{R-zikv}
aR^2+bR+c=0, \quad \text{where}
\end{equation}

\begin{equation}\label{coeficients-R-zikv}
\begin{array}{ll}
a&=\dfrac{\beta_z\Lambda\alpha_m\mu}{\delta_z R_{max}} \\ \\
b&=\dfrac{\mu(\mu_z+\delta_z+\mu)}{\delta_z}\left[ \dfrac{\beta_m\alpha_m\Lambda_m}{\mu_m}+\beta_z\mu_m+\alpha_m\mu(1-2\mathcal{R}_{z_1})\right] \\ \\
c&=(\mu_z+\delta_z+\mu)\mu\mu_m\left( 1-\mathcal{R}_z^*  \right).
\end{array}
\end{equation}

Note that $a>0$, and the signs of $b$ and $c$ depends on the sign of $\mathcal{R}^*_z$ and $2\mathcal{R}_{z_1}$.  We have the following possibilities. 

\begin{itemize}
\item[P1)] If $\mathcal{R}^*_z<1$, then $2R_{z_1}<1$ (Lemma \ref{R*-zikv} item $i$).  Therefore $b>0$ y $c>0$ and thus, the quadratic equation \eqref{R-zikv}  has not any positive root. 
%%%%%%%%%%
\item[P2)] If $\mathcal{R}^*_z>1$, then $c<0$ and regardless of the sign of $b$,  the quadratic equation \eqref{R-zikv}  has only one positive root given by
\begin{equation}\label{R*}
  R^*=\frac{-b + \sqrt{b^2-4ac}}{2a}.   
\end{equation}
\item[P3)] If $\mathcal{R}^*_z=1$ and $2R_{z_1}<1$, then $c=0$ and $b>0$. Therefore, the  quadratic equation \eqref{R-zikv}  has as solution $R=-b/a$.  Thus, there are no positive roots.
\item[P4)] If $\mathcal{R}^*_z=1$ and $b<0$,   the  quadratic equation \eqref{R-zikv}  has a positive root given by  $R=-b/a$. 
\end{itemize}

\begin{remark}
The case $\mathcal{R}^*_z<1$ and  $2R_{z_1}>1$ ($c>0$ and $b<0$), which gives the possibility of the existence of two positive roots for the quadratic equation  \eqref{R-zikv} whenever $b^2-4ac>0$,  is not considered since by Lemma \ref{R*-zikv} item $i$,  when $\mathcal{R}^*_z<1$ leads to $2R_{z_1}<1$.  
\end{remark}

Based on the information presented earlier, the quadratic equation \eqref{R-zikv} has only one positive root defined in  \eqref{R*} whenever $\mathcal{R}^*_z>1$.   The proof that the condition  $R^*<R_{max}$ is satisfied can be found in Appendix \ref{appendix:A4}.

Finally, we determine the sign of the eigenvalues of the matrix defined in \eqref{jacobian-zikv} evaluated in $\textbf{E}^*_z$.  In order to simplify algebraic manipulations we make the following change of variables

\begin{equation*}
\begin{array}{lll}
&\kappa=\mu_z+\delta_z+\mu,  &t_1= \beta_mI_m^*+\beta_zI_z^*, \\ \\
& t_2=\beta_zS^*, & t_3= \beta_m S^*,\\ \\
 & t_4= \alpha_m S_m^*,  &t_5= \alpha_m I_z^*, 
\end{array}
\end{equation*}
 thus, the matrix $\textbf{J}(\textbf{E}^*_z)$ can be rewritten as
 
\begin{equation*}
\textbf{J}(\textbf{E}^*_z)=\begin{bmatrix}
%%f1
-(t_1+\mu) &-t_2&0 &0&-t_3\\ 
%%f2 
t_1&t_2-\kappa &0&0 &t_3  \\ 
%%%f3
0 & \delta_z &-\mu &0 &0 \\ 
%%%%%%f4
0 & -t_4 &0  &-(t_5+\mu_m)  &0 \\ 
%%%f5
0 & t_4  &0   & t_5   &-\mu_m
\end{bmatrix}.
\end{equation*}

After some algebraic manipulations, we find that the characteristic polynomial associated with the matrix $\textbf{J}(\textbf{E}^*_z)$ is 

\begin{equation}\label{polynimial-quadratic-endemic-zikv}
r(y)= (y+\mu)(y+\mu_m)(y^3+b_1y^2+b_2y+b_3), \quad \text{where}
\end{equation}

\begin{equation*}
\begin{array}{ll}
&b_1=t_6+\kappa(1-2\mathcal{R}_{z_1}) \\ \\
&b_2=t_7+\kappa(\mathcal{R}_z^*-1) \\ \\
&b_3=t_8(\mathcal{R}_z^*-1),
\end{array}
\end{equation*}

and  $t_6$, $t_7$ and $t_8$ defined as a positive linear combination of $t_i$, $i=1,...,5$. The characteristic polynomial $r(y)$ gives five roots, two of which are $y=-\mu$ and $y=-\mu_m$, whereas 
the Routh-Hurtwiz criteria assures that the other three roots have negative real part if $b_i>0$  for $i = 1, 2, 3$  and $b_1b_2-b_3>0$.  Clearly the coefficients  $b_i>0$,  $i = 1, 2, 3$ are all positive if $\mathcal{R}_z^*>1$ and $1-2\mathcal{R}_{z_1}>0$.  Thus, it follows that the endemic equilibrium $\textbf{E}^*_z$ of System \eqref{model-zikv}  is \textit{LAS} if $\mathcal{R}_z^*>1$.  To conclude the qualitative analysis of Model \eqref{model-zikv},  we prove the global stability of the DFE  ($\textbf{E}_{z_0}$) using similar techniques to those used in Section \ref{sec-hiv-only}. All results of this section are summarized in the following proposition.  

\begin{proposition}\label{lemma_zikv}
System \eqref{model-zikv} always has a DFE $\textbf{E}_{z_0}$ defined in \eqref{DFE-zikv}. If  $\mathcal{R}_z^*>1$ the system has an endemic equilibrium point given by 
\begin{equation}\label{endemic-zikv}
\textbf{E}^*_z=\left(\dfrac{\Lambda}{\mu}\left( 1- \dfrac{R^*}{R_{max}}\right), \dfrac{\mu}{\delta_z}R^*, R^*, \dfrac{\Lambda_m}{\mu_m}\left( 1-\dfrac{\alpha_m \mu R^*}{\mu_m\delta_z+\alpha_m \mu R^*}\right), \dfrac{\alpha_m\Lambda_m\mu}{\mu_m\delta_z+\alpha_m\mu R^*}R^*  \right),
\end{equation}
with $R_{max}$ given in \eqref{Rmax-zikv} and $R^*$ in \eqref{R*}.
Additionally, the following stability results hold:
\begin{itemize}
\item [(i)] If $\mathcal{R}^*_z<1$, then $\textbf{E}_{z_0}$ is \textit{LAS-GAS} in $\Omega_z$ defined in \eqref{omega_z}.
\item [(ii)] If $\mathcal{R}^*_z>1$, the endemic equilibrium point $\textbf{E}^*_{z}$ is LAS in  $\Omega_z$ defined in \eqref{omega_z}. 
\end{itemize}
\end{proposition}

The proof of the global stability of $\textbf{E}_{z_0}$ when  $\mathcal{R}_{z}^*<1$ can be found in Appendix \ref{appendix:A5}.

%%%%%%%%%%%%%%%%%%%%%%%%%%%%%%%%%%%%%%%
%%%%%%%%HIV/ZIKV model
\section{Qualitative behaviour of the HIV/ZIKV model}\label{sec-coinfection}
In this section, we discuss the qualitative properties of the HIV/ZIKV co-infection model \eqref{model1}.  To achieve this purpose, we use the existence and stability results as well as the definition of the basic reproduction number for the HIV model $\mathcal{R}_h$ in \eqref{R0-hiv} and $\mathcal{R}_z$ in \eqref{R_{z}} obtained in Sections \ref{sec-hiv-only} and \ref{sec-zikv-only}.  
\par 
In Model \eqref{model1}, the total human population is denoted by $N(t)= S(t)+I_z(t)+I_h(t)+I_{hz}(t)+A(t)+R(t)$, and the total mosquito population is $N_{m}(t)=S_{m}(t)+I_{m}(t)$. Additionally,  to simplify algebraic calculations we rename parameters:
\begin{equation}\label{kappas}
\begin{array}{ll}
\kappa_1=\mu_z+\delta_z+\mu, &\kappa_2=\sigma_1+\mu,  \\ \\ \kappa_3=\sigma_2+\mu_{hz}+\mu, &\kappa_4= \mu_h+\mu. 
\end{array}
\end{equation}  

The interest region set is given by 
\begin{equation} \label{omega}
    \Omega=\left\{ (S,I_z, I_h, I_{hz}, A,R,S_m,I_m) \in  \mathbb{R}_+^{8}: 0\leq N\leq \frac{\Lambda}{\mu} ; 0\leq N_{m} \leq \frac{\Lambda_{m}}{\mu_{m}}  \right\}. 
\end{equation}
As in the previous sections, it can be proved that $\Omega$ is positively invariant under the flow of \eqref{model1}.

\subsection{Computation of the basic reproduction number}\label{sec:R0calculation}

The DFE for Model   \eqref{model1} is given by 
\begin{equation}\label{DFE}
\textbf{E}_{0}=\left(\frac{\Lambda}{\mu}, 0,0,0,0,0,\frac{\Lambda_{m}}{\mu_{m}},0\right).
\end{equation}

Similarly to Sections  \ref{sec-hiv-only} and \ref{sec-zikv-only}, the basic reproduction number associated to Model \eqref{model1} can be determined through  the matrices  $\textbf{F}$  and  $\textbf{V}$ given by 

\begin{equation*}
    \textbf{F}=\begin{bmatrix}
    %%f1
        \beta_z\frac{\Lambda}{\mu} & 0&0&0& \beta_m\frac{\Lambda}{\mu} \\
        %%%%f2
      0&    \beta_h\frac{\Lambda}{\mu}&0&0&0 \\
      0&0&0&0&0 \\
      0&0&0&0&0 \\
      \alpha_m\frac{\Lambda_m}{\mu_m} &0& \alpha_m\frac{\Lambda_m}{\mu_m} &0&0
    \end{bmatrix}, 
    \quad \text{and} \quad \textbf{V}=\begin{bmatrix}
        \kappa_1 &0&0&0&0\\
        0 &\kappa_2 &0&0&0 \\
        0&0&\epsilon\delta_z+\kappa_3 &0&0 \\
        0&0&0&\kappa_4&0 \\
        0&0&0&0&\mu_m
    \end{bmatrix}.   
\end{equation*}

Here,  the matrix $\textbf{FV}^{-1}$ is given by 

\begin{equation*}
   \textbf{FV}^-1=\begin{bmatrix}
    %%f1
        \beta_z\frac{\Lambda}{\mu\kappa_1} & 0&0&0& \beta_m\frac{\Lambda}{\mu\mu_m} \\
        %%%%f2
      0&    \beta_h\frac{\Lambda}{\mu\kappa_2}&0&0&0 \\
      0&0&0&0&0 \\
      0&0&0&0&0 \\
      \alpha_m\frac{\Lambda_m}{\mu_m\kappa_1} &0& \alpha_m\frac{\Lambda_m}{\mu_m(\epsilon\delta_z+\kappa_3)} &0&0
    \end{bmatrix}.
\end{equation*}

The above matrix has as eigenvalues $\lambda_{1,2}=0$ (twice)  and $\lambda_3=\dfrac{\beta_h\Lambda}{\mu\kappa_2}=\mathcal{R}_h$, whereas the other two eigenvalues are 
 
\begin{equation*}
\lambda_{4,5}=\dfrac{\beta_z\Lambda}{2\mu\kappa_1}\pm \sqrt{\left(\dfrac{\beta_z\Lambda}{2\mu\kappa_1}\right)^2+\dfrac{\alpha_m\beta_m\Lambda_m\Lambda}{\mu\mu_m^2\kappa_1}}=\mathcal{R}_{z_1}\pm \mathcal{R}_{z_2},
\end{equation*}
with the positive eigenvalues being 
\begin{equation*}
\lambda_{4}=\dfrac{\beta_z\Lambda}{2\mu\kappa_1}+ \sqrt{\left(\dfrac{\beta_z\Lambda}{2\mu\kappa_1}\right)^2+\dfrac{\alpha_m\beta_m\Lambda_m\Lambda}{\mu\mu_m^2\kappa_1}}=\mathcal{R}_{z_1}+ \mathcal{R}_{z_2}=\mathcal{R}_z.
\end{equation*}

Thus, the  basic reproduction number for Model \eqref{model1} is given by 

\begin{equation}\label{R0}
        \mathcal{R}_0=\rho(\textbf{FV}^-1)=\max\{\mathcal{R}_h, \mathcal{R}_z\},
\end{equation}
where $\mathcal{R}_h$ is the basic reproduction number for the HIV model defined in the equation \eqref{R0-hiv} and $\mathcal{R}_z=\mathcal{R}_{z_1}+\mathcal{R}_{z_2}$ is the basic reproduction number of the ZIKV model, which is defined in the equation \eqref{R_{z}}.

%%%%%%%%%%%%%%%%%%%
%%%%%%%local sensitivity section 
\subsection{Local sensitivity analysis of  the parameters}\label{sec:R0indices}

The local sensitivity analysis of $\mathcal{R}_0$ with respect to the model parameters allows quantifying parameter variations' effect on the value of $\mathcal{R}_0$. The sign of the sensitivity index denotes the direction of the change, where a positive index for a particular parameter indicates that increasing that parameter will increase $\mathcal{R}_0$ and vice-versa. In addition, the sensitivity index's magnitude provides insight into each parameter's impact on the predictions \cite{saltelli2008global}.
\par
The normalized sensitivity index of a variable concerning a parameter is a measure of how much the variable relatively changes to the change in the parameter\cite{chitnis2008determining} and is defined as:

\begin{equation}\label{sensitivity-index}
\Gamma_p^{X}=\dfrac{\partial X}{\partial p}\dfrac{p}{X}.  
\end{equation}

Because $\mathcal{R}_0$ is defined as $\max\{\mathcal{R}_h, \mathcal{R}_z\}$,  the sensitivity indices of $\mathcal{R}_0$ with respect to the eleven parameters $\{ \beta_h, \Lambda, \mu, \sigma_1, \beta_z,  \mu_z, \delta_z, \beta_m, \alpha_m, \Lambda_m, \mu_m\}$ in
the expression of $\mathcal{R}_0$ in \eqref{R0}, can be determined for the sensitivity indices of $\mathcal{R}_h$ and $\mathcal{R}_z$, respectively.  A calculation example of the sensitivity index of $\mathcal{R}_z$  with respect to the parameter $\beta_z$ can be found in Appendix \ref{appendix:A6}.

%%%%%%%%%%%new section 

\subsection{Stability analysis}
Similarly to in Sections \ref{sec-hiv-only} and \ref{sec-zikv-only}, the local stability of the DFE $\textbf{E}_{0}$ is determined by the linearization of \eqref{model1} around an arbitrary equilibrium. By ordering the equations and variables as $S,S_m,I_z, I_m, I_h, I_{hz},A,R$ and computing the Jacobian matrix of System \eqref{model-zikv} at an arbitrary equilibrium point $\textbf{E}=(S,S_m,I_z, I_m, I_h, I_{hz},A,R)$, we obtain

\begin{equation}\label{jacobian}
\textbf{J}(\textbf{E} ) =
\begin{bmatrix}
\textbf{J}_{11}(\textbf{E} ) &\textbf{0}&\textbf{0}\\
 \textbf{0}&-\kappa_4 &\textbf{0} \\
 \textbf{0} &\textbf{0} &-\mu
\end{bmatrix}.
\end{equation}
The matrix shown above has eigenvalues $-\mu < 0 $, $-\kappa_4=-(\mu_h+\mu)<0$ and the eigenvalues of $\textbf{J}_{11}(\textbf{E})$.  By evaluating  the matrix $\textbf{J}_{11}$ at the DFE $\textbf{E}_0$ we obtain

\begin{equation*}
\textbf{J}_{11}(\textbf{E}_0 ) =
\begin{bmatrix}
-\mu &\textbf{0} &\mathbf{\star} &\mathbf{\star} \\
\textbf{0} &-\mu_m & \mathbf{\star} &\mathbf{\star} \\
\textbf{0} & \textbf{0} & \textbf{M}_{33} & \mathbf{\star} \\
\textbf{0} & \textbf{0} & \textbf{0}& \textbf{D}_{44}
\end{bmatrix}, \, \text{where} \, \textbf{M}_{33}= 
\begin{bmatrix}
-\kappa_1+\dfrac{\Lambda \beta_z}{\mu} &\dfrac{\Lambda \beta_m}{\mu} \\ \\
\dfrac{\Lambda_m\alpha_m}{\mu_m} & -\mu_m
\end{bmatrix}
\end{equation*}

and $\textbf{D}_{44}$ is a diagonal matrix with entries $-(\epsilon\delta_z+\kappa_3)<0$ and $\kappa_2-\frac{\beta_h\Lambda}{\mu}=\sigma_1+\mu -\frac{\beta_h\Lambda}{\mu}=(\sigma_1+\mu)(1-\mathcal{R}_h)<0$ if and only if $\mathcal{R}_h<1$. Therefore, the remaining eigenvalues of $\textbf{J}_{11}(\textbf{E}_0 )$ are those of the matrix $\textbf{M}_{33}$, whose characteristic equation is 

\begin{equation}\label{characteristic}
\lambda^2+c_1\lambda+c_2=0, 
\end{equation}

where
\begin{equation*}
\begin{array}{ll}
&c_1= \kappa_1+\mu_m-\frac{\beta_z\Lambda}{\mu}=\mu_m+\kappa_1(1-2\mathcal{R}_{z_1}) \\ \\
&c_2=\kappa_1\mu_m-\frac{\alpha_m\beta_m\Lambda_m\Lambda}{\mu\mu_m}-\frac{\beta_z\Lambda\mu_m}{\mu}=k_1\mu_m(1-\mathcal{R}_z^*).
\end{array}
\end{equation*}

Thus, the roots of the equation \eqref{characteristic} have negative real part if and only if $\mathcal{R}_z^*<1$ and $2\mathcal{R}_{z_1}<1$, where $\mathcal{R}_z^*$ and  $\mathcal{R}_{z_1}$ are defined in \eqref{R*-zikv} and \eqref{R_{z12}}, respectively. We have the following result:
\begin{proposition}\label{prop-las-dfe}
If $\mathcal{R}_0=\max\{\mathcal{R}_h, \mathcal{R}_z\}<1$,  then $\textbf{E}_{0}$ defined in \eqref{DFE} is \textit{LAS} in $\Omega$ defined in \eqref{omega}.
\end{proposition}

%%%%%%%%%%%%%%%%%%%%
Techniques similar to those used in Sections \ref{sec-hiv-only} and \ref{sec-zikv-only}  can be applied to confirm the presence of endemic solutions and assess their local and global stability. 

%%%%%%%%%%%%%%%%%
\section{The control problem analysis}\label{sec-optimal-control}
Here, we present an optimal control problem (OCP) by including three different methods of intervening to manage how HIV/ZIKV co-infection spreads within our Model \eqref{model1}. The proposed approach mitigates HIV and Zika infections through the implementation of personal protection measures (such as the use of repellents) using control $\eta_1$, the use of ART  with control $\eta_2$ and preventive sexual contact (such as condom use) with control   $\eta_3$. The control functions $\eta_1$, $\eta_2$ and $\eta_3$ are defined in the interval $[0, T $], where $T$ denotes the final time of the controls, $0 \leq \eta_i(t) \leq  1$ and $t \in [0,T]$ for $ i = 1, 2, 3$. 
Besides, the choice of the cost function is contingent upon the specific goals of the optimization problem. Different types of cost functions may be appropriate for different scenarios, such as an exponential cost function \cite{avusuglo2023leveraging, grandits2019optimal}, where it can be used if the decision-maker under uncertainty, and its concave nature of exponential cost describes a risk-averse attitude, or a quadratic cost function, which its convexity ensures that any local minimum is also a global minimum. It offers an extensive description of the trade-off between disease control and intervention costs applied in an infectious disease model and can be additive.
Based on the considerations mentioned earlier, the following  OCP is formulated by a hybrid cost function combining linear and quadratic terms, where the controls are shown in red for emphasis.

\begin{equation}\label{model_ocp}
	\left\{
	\begin{array}{ll}
	%%%%%%funcional
	&\min  \mathcal{J} (\eta)= \int_0^T \left(c_1 I_z +c_2 I_h +c_3 I_{hz} + c_4 I_m + c_5 A +d_1 \dfrac{\eta_1^2}{2}+d_2 \dfrac{\eta_2^2}{2}+d_3 \dfrac{\eta_3^2}{2} \right)dt \\\\
	%%%%%%1
	& \dfrac{dS}{dt}=\Lambda-\left[\textcolor{red}{(1-\eta_1)}\beta_mI_m+\textcolor{red}{(1-\eta_3)}(\beta_z I_z+\beta_h I_h)\right]S-\mu S \\ \\
 %%%%2
 &\dfrac{dI_z}{dt}=\left[\textcolor{red}{(1-\eta_1)}\beta_mI_{m}+\textcolor{red}{(1-\eta_3)}\beta_z I_z\right]S-\omega_2\textcolor{red}{(1-\eta_3)}\beta_hI_z I_h-(\mu_z+\delta_z+\mu)I_z \\ \\
 %%%%3
&\dfrac{dI_h}{dt}=\textcolor{red}{(1-\eta_3)}\beta_hI_h S-\omega_1[\textcolor{red}{(1-\eta_1)}\beta_mI_m+\textcolor{red}{(1-\eta_3)}\beta_zI_z] I_h-[\textcolor{red}{(1-\eta_2)}\sigma_1+\mu]I_h \\ \\
%%%%%%%%%%%%4
&\dfrac{dI_{hz}}{dt}=\omega_1[\textcolor{red}{(1-\eta_1)}\beta_mI_m+\textcolor{red}{(1-\eta_3)}\beta_zI_z]I_h+\omega_2\textcolor{red}{(1-\eta_3)}\beta_hI_h I_z-\epsilon\delta_zI_{hz}-[\textcolor{red}{(1-\eta_2)}\sigma_2+\mu_{hz}+\mu]I_{hz} \\ \\
%%%%%5
&\dfrac{dA}{dt}=\textcolor{red}{(1-\eta_2)}\sigma_1 I_h+\textcolor{red}{(1-\eta_2)}\sigma_2 I_{hz}-(\mu_h+\mu)A\\ \\
%%%%6
&\dfrac{dR}{dt}=\delta_zI_z+\epsilon\delta_zI_{hz}-\mu R \\ \\
%%%%%7
&\dfrac{dS_m}{dt}=\Lambda_m-\textcolor{red}{(1-\eta_1)}\alpha_m (I_z+I_{hz}) S_m-\mu_mS_m \\ \\
%%%%%8
&\dfrac{dI_m}{dt}=\textcolor{red}{(1-\eta_1)}\alpha_m (I_z+I_{hz}) S_m-\mu_mI_m \\ \\
	&{\bf X}(0)=(S_0, I_{z0}, I_{h0}, I_{hz0}, A_0, R_0, S_{m0}, I_{m0})={\bf X}_0 \\ \\
	&{\bf X}(T)=(S_f, I_{zf}, I_{hf}, I_{hzf}, A_f, R_f, S_{mf}, I_{mf})={\bf X}_f.
	\end{array}
	\right.
	\end{equation}
In the above formulation  $\eta= (\eta_1(t),\eta_2(t),\eta_3(t))$, and $c_1,c_2,c_3,c_4,c_5,d_1,d_2,$ and $d_3$ are positive weights. Therefore, we seek an optimal control $\eta^*(t)$ determined as
\begin{equation}\label{Jcontrol}
    \mathcal{J} (\eta^*(t))= \min \left\lbrace  \mathcal{J}  (\eta(t) | \eta \in \mathcal{A}) \right\rbrace,
\end{equation}
with a set $\mathcal{A}$ of controls defined as
\begin{equation*}
    \mathcal{A}= \left\lbrace \eta(t)= (\eta_1(t),\eta_2(t),\eta_3(t)) | 0 \leq \eta_1(t) \leq \eta_1^{\max}, 0 \leq \eta_2(t) \leq \eta_2^{\max}, 0 \leq \eta_3(t) \leq \eta_3^{\max} \right\rbrace,
\end{equation*}
where $\eta_i^{\max} \leq 1$, $i=\{ 1,2,3\}$ and $\eta$ is Lebesgue measurable.
In order to define the formulation of our OCP using Pontryagin's Maximum Principle (PMP) \cite{pontryagin2018ls}, we have the Lagrangian as
\begin{equation}\label{Lagrangian}
    L= c_1 I_z(t) +c_2 I_h(t) +c_3 I_{hz}(t) + c_4 I_m(t)+ c_5 A(t) +d_1 \dfrac{\eta_1^2(t)}{2}+d_2 \dfrac{\eta_2^2(t)}{2}+d_3 \dfrac{\eta_3^2(t)}{2},
\end{equation}
and we determine the Hamiltonian function as
\begin{equation}\label{Hamiltonian}
    H= L(I_z, I_h, I_{hz}, I_m,A, \eta ) +p_1 \dfrac{dS}{dt}+p_2 \dfrac{dI_z}{dt}+p_3 \dfrac{dI_h}{dt}+p_4 \dfrac{dI_{hz}}{dt}+p_5 \dfrac{dA}{dt}+p_6 \dfrac{dR}{dt}+p_7 \dfrac{dS_m}{dt}+p_8 \dfrac{dI_m}{dt}.
\end{equation}

In the remainder, we investigate the minimum value of Lagrangian \eqref{Lagrangian}. 
Firstly, we must prove the existence of the optimal control $\eta^*$ according to the controlled system \eqref{model_ocp}.
%%%%
\begin{proposition}\label{Prop_existence_OCP}
    There exists an optimal control $\eta^*$ such that
    \begin{equation*}
        \mathcal{J} \left( \eta^* (t) \right) = \min_{\eta \in \mathcal{A} } (\mathcal{J}( \eta(t))),
    \end{equation*}
    subject to Control System \eqref{model_ocp} with initial conditions as $\bf{X}_0$.
\end{proposition}

%%%%%%%%
The proof of the above proposition can be found in Appendix \ref{appendix:A7}.
\par 
In the following, we apply PMP \cite{pontryagin2018ls} to provide a characterization of an optimal control solution to the Hamiltonian \eqref{Hamiltonian} subject to the OCP \eqref{model_ocp}. If $(X^*, \eta^*)$ is an optimal solution for the controlled system \eqref{model_ocp}, then there exists a non trivial vector function $p=(p_1,p_2,p_3,p_4,p_5,p_6,p_7,p_8)$, such that
\begin{equation}
    \dfrac{\partial H}{\partial \eta_i}=0, \, i=1,2,3 \quad \text{ and } \dot p_i=\dfrac{dp_i}{dt}=-\dfrac{\partial H}{\partial X_i}, \, i=1,\ldots,8.
\end{equation}

\begin{proposition}\label{Prop-OCP-2}
    Let $(S^*, I_{z}^*, I_{h}^*, I_{hz}^*, A^*, R^*, S_{m}^*, I_{m}^*)$ be the optimal state variables solution associated to the optimal control variable $\eta^*$ subject to the control problem \eqref{Jcontrol}. Then, there exists an adjoint vector $p$ that satisfies the controlled system \eqref{model_ocp}, with transversality conditions $p_i(T)=0$, for $i=1,\ldots,8 $, where the optimal controls are
    \begin{equation}
    \begin{array}{ll}
        \eta_1^* =  & \dfrac{\left( (p_2 - p_1) S +  (p_4 - p_3) I_h \right)  \beta_m I_m + (p_8 - p_7) \alpha_m (I_z + I_{hz}) S_m}{d_1}\\ \\
        \eta_2^*  = & \dfrac{\left( p_5 - p_3  \right)  \sigma_1 I_h + \left( p_5 - p_4  \right)  \sigma_2 I_{hz}}{d_2} \\ \\
        \eta_3^*  = & \dfrac{(p_2 - p_1) \beta_z I_z S   + (p_3 - p_1) \beta_h I_h S + (p_4 - p_3) \omega_1 \beta_z I_z I_h  + (p_4 - p_2) \omega_2 \beta_h I_z I_h}{d_3}. 
        \end{array}
    \end{equation}\label{asser1-3}
\end{proposition}

The proof is in Appendix \ref{appendix:A8}. 

%%%%%%%%%%%%%%%%%%

\section{HIV/ZIKV co-infection in Brazil and Colombia}\label{sec:case-of-study}
As mentioned previously, between 2015 and 2016, there was a concerning event in Brazil and Colombia associated with the co-infection of two major viral diseases: Zika and HIV/AIDS  \cite{brasil2016,calvet2016, villamil2018}. This problem became a complicated public health issue that posed a challenge to the healthcare systems in these countries \cite{brasil2016,calvet2016, villamil2018 }. This co-infection not only required costly medical resources but also highlighted the need for further research, extensive surveillance, and implementation of prevention and prompt reaction strategies to mitigate the impact of these two infections \cite{rothan2018}. 
\par
Due to the absence of temporal records on HIV/ZIKV co-infection to date, it was not possible to estimate the parameters of Model \eqref{model1} in this section. However,  specific parameter values were derived from available demographic information, previous research on Zika and HIV/AIDS in Colombia and Brazil, and epidemiological assumptions. When there was insufficient data available, these values were either estimated based on specific assumptions or adapted from research conducted on different regions or diseases. The following outlines the main assumptions for extracting these parameter values.
\par 
Firstly, the precise relationship between humans and population densities of \textit{Aedes aegypti} remains uncertain. These mosquitoes' entomological and biological characteristics, including their population size and geographical location or distribution, are impacted by meteorological factors including but not limited to temperature and precipitation \cite{heinisch2019seasonal}. Therefore, to facilitate our numerical experiments and to not lose generality, we assume that during 2015-2016, there existed an approximate ratio of one human to three female \textit{Aedes aegypti} mosquitoes (1:3) in Colombia and Brazil. Secondly, the modification parameter related to the Zika recovery rate of humans  $\epsilon$ was assumed to be much smaller than one (1) because, according to research in this area, an individual with HIV/ZIKV co-infection recovers more slowly than an individual with only Zika \cite{rothan2018, bidokhti2018siv, aschengrau2021international}, and can lead to serious neurological issues in children \cite{rothan2018}. Thirdly, since only a few cases were reported between 2015 and 2016 in Colombia and Brazil \cite{calvet2016,villamil2018}, the parameters associated with the co-infection probability  $w_1$ and $w_2$ were assumed to be small enough. However,  a hypothetical scenario in which these parameters are escalated is depicted to demonstrate the correlation between the rising probabilities of co-infection and the subsequent increase in the number of co-infected individuals. Finally, some parameter values regarding HIV/AIDS were conveniently adjusted at a population level, utilizing estimates obtained from Luxembourg, the Czech Republic, Japan, Croatia, the United Kingdom, and Mexico \cite{prieto2021current}.
\par 
We use the same methodological strategy employed in the reference \cite{gao2016} for Brazil, Colombia, and El Salvador, supposing that Colombia and Brazil share some common parameter values (see Table \ref{values-parameters1}), excluding those related to the magnitude of the nation's population and its initial conditions  (see Table \ref{values-parameters2}).  

Due to the dispersion of the units of measurement of the parameters in the different sources consulted, all parameter values were adapted to {\it day} as the standard unit of measurement.

%%%%shared parameters
\begin{table}[H]
\centering
\begin{tabular}{llll}
  \hline
Parameter &  Symbol&  Range& Reference   \\ \hline
Recovery rate of Zika&$\delta_z$ & [1.11e-2, 7.14e-2]     & \cite{gao2016}\\
Immunity recovery rate in HIV-only infected people&$\sigma_1$ & [2.1e-4, 5.46e-4]      & \cite{prieto2021current}\\
Mortality rate by AIDS&$\mu_h$  & 5.46e-5  &\cite{prieto2021current} \\  
Mortality rate of mosquitoes &$\mu_m$ & [5e-2, 1.2e-1]            &\cite{gao2016}\\  
Mortality rate by Zika&$\mu_z$ &   [3.42e-5, 5.46e-5]      &Assumed \\
Mortality rate by ZIKV/AIDS &$\mu_{hz}$ &   0      &\cite{calvet2016,villamil2018}\\
Immunity recovery rate in co-infected people&$\sigma_2$ &  4.55e-5    &\cite{calvet2016,villamil2018}\\ 
Transition probability from HIV to HIV/ZIKV co-infection&$\omega_1$ &  [1e-3, 1e-2]      &Assumed\\  
Transition probability from ZIKV to HIV/ZIKV co-infection&$\omega_2$&   [1e-3, 1e-2]         &Assumed\\ 
Zika recovery rate scaling factor&$\epsilon$ &    1e-10      &Assumed\\   
\hline 
\end{tabular}
\caption{Shared (Colombia-Brazil) range of parameter values (minimum and maximum)  involved in Model (\ref{model1}). Time in days.}
\label{values-parameters1}
\end{table}

%%%%No shared parameters
\begin{table}[H]
\centering
\begin{tabular}{lllll}
  \hline
  Parameter&Symbol &Colombia   & Brazil &Reference   \\ \hline
Infection transmission of humans by mosquitoes&   $\beta_m$ &  [6.25e-10, 1.25e-9]  &[2.42e-10, 2.9e-10]  & \cite{gao2016}  \\
Infection transmission of Zika by sexual contact&   $\beta_z$ &  [2.08e-12, 1.04e-11]  &[4-83e-13, 2.42e-12]  & \cite{gao2016}  \\
Infection transmission of HIV by sexual contact &  $\beta_h$ &  [6.25e-12, 1.46e-11]  & [1.45e-12, 3.38e-12]  &\cite{prieto2021current} \\
Infection transmission of mosquitoes by humans  &    $\alpha_m $&  [2.08e-10, 4.17e-10]  & [4.83e-11, 9.66e-11] &\cite{gao2016}\\
Recruitment of humans&  $\Lambda$ & [1.73e+03,	1.73e+03]   & [7.44e+03, 7.44e+03]&Computed \\ 
Mortality rate of humans&     $\mu$ &   [3.59e-5, 3.74e-5]  &[3.50e-5, 384e-5] & Computed\\
Recruitment of mosquitoes&       $\Lambda_m$  &  [1.03e+7, 1.03e+7] & [4.43e+7, 4.43e+7]                &Computed\\ \hline 
\end{tabular}
\caption{Range of parameter values (minimum and maximum)  involved in Model (\ref{model1}) that differ in Colombia and Brazil. Time in days. }
\label{values-parameters2}
\end{table}

%%%%%initial conditions
\begin{table}[H]
\centering
\begin{tabular}{llll}
  \hline
  Parameter &Colombia   & Brazil  &Reference  \\ \hline
   $S(0)$ &28,800,000 &	124,200,000&\cite{datacolombia,databrazil} \\
  $I_z(0)$ &9,600,000&	41,400,000 &Computed\\
  $I_h(0)$ &4,320,000 &	18,630,000 & Computed\\
  $I_{hz}(0)$&4.8	&20.7 &\cite{calvet2016,villamil2018}\\
  $A(0)$&2,832,000&	12,213,000&Computed\\
  $R(0)$&2,448,000	&10,557,000&Computed \\
  $S_m(0)$ &33,600,000	&144,900,000&Computed \\
  $I_m(0)$ &14,400,000&	62,100,000 &Computed \\ \hline
  \end{tabular}
\caption{Initial conditions involved  in Model (\ref{model1}), which correspond to the population sizes in Colombia \cite{datacolombia} and Brazil \cite{databrazil}  in 2015-2016. }
\label{values-parameters3}
\end{table}

Therefore, our case study is organized into three separate stages. 
In the initial stage, we seek to determine numerically  the basic reproduction number related to the co-infection epidemical model \ref{model1}. 
Furthermore, numerical values of the sensitivity indices determined analytically in Section \ref{sec:R0indices} are also presented. 
In the second stage, the uncontrolled model defined by \eqref{model1} is simulated numerically. 
Finally, the third stage focuses on numerically simulating the control problem described in \eqref{model_ocp}.  
This sequential framework allows for a thorough exploration of the various aspects under investigation, providing a complete analysis of the case study.

%%%%%%%%stahe 1
\subsection{Value of the basic reproduction number and its sensitivity indices}

\begin{figure}[H]
\centering
\subfigure[Colombia]{\includegraphics[width=8cm, height=4.5cm]{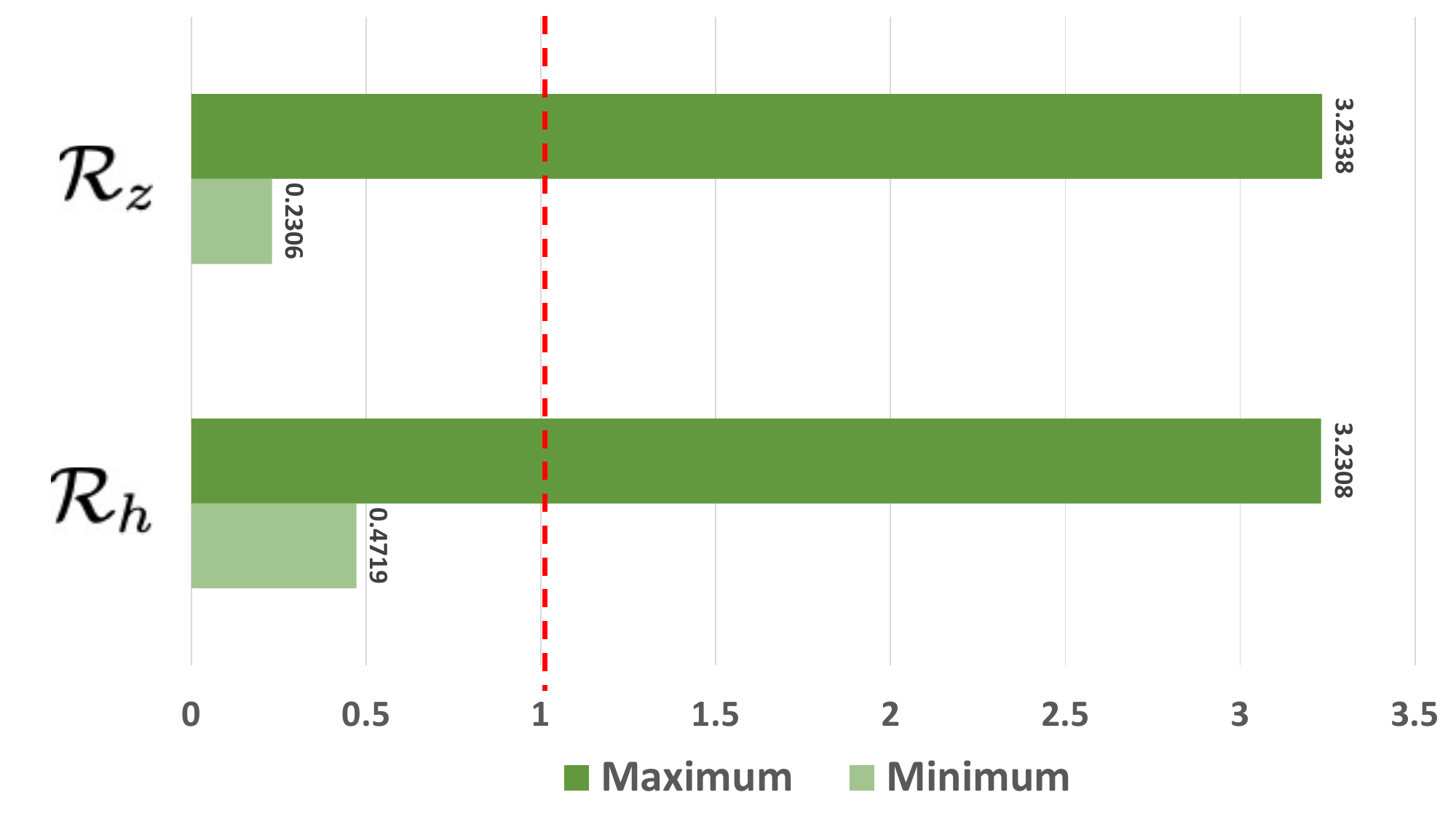}}
\subfigure[Brazil]{\includegraphics[width=8cm, height=4.5cm]{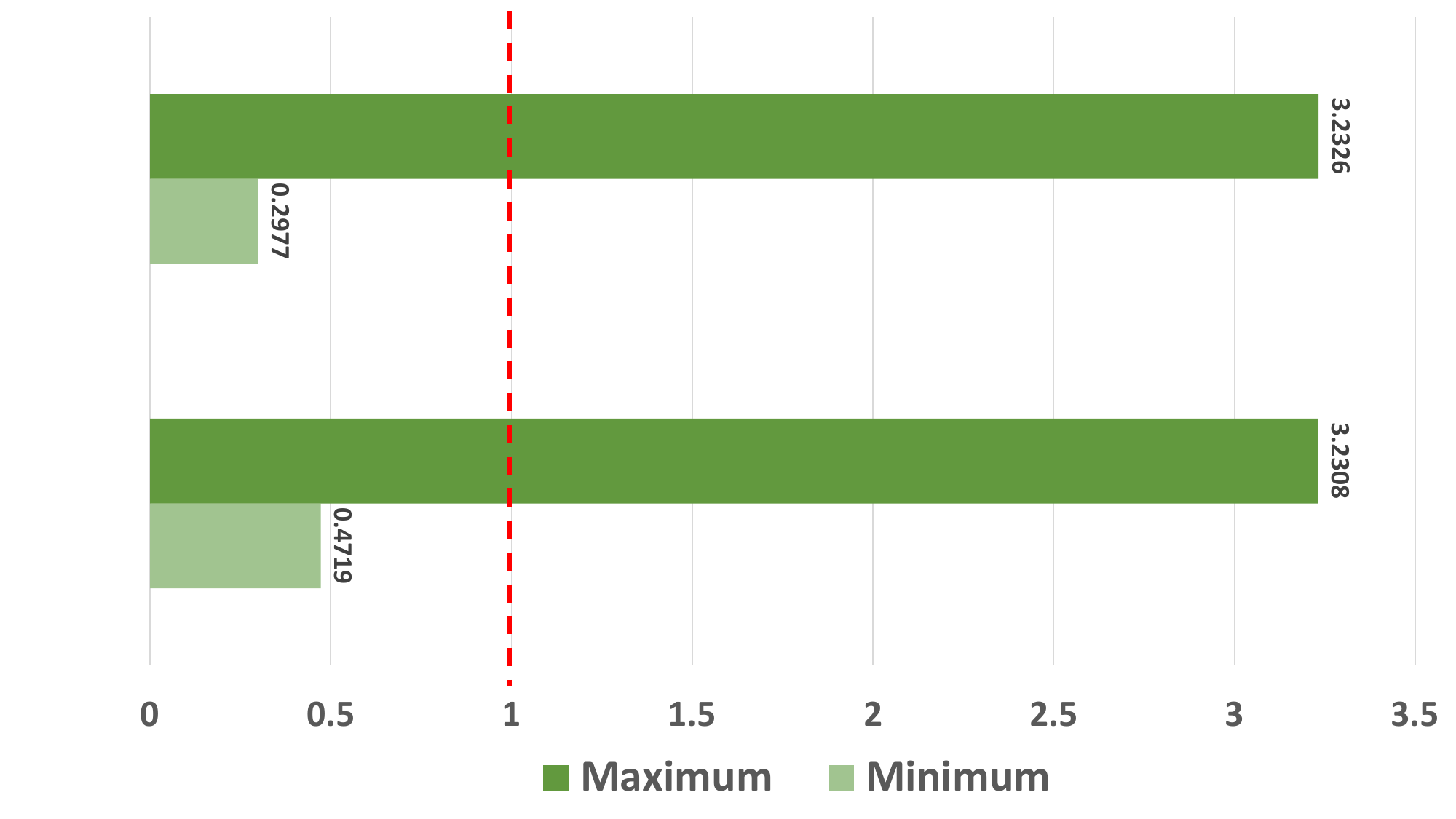}}
\caption{Possible values for the thresholds using the extreme parameter values given in Tables \ref{values-parameters1} and  \ref{values-parameters2} for Colombia and Brazil, that result in the minimum or maximum value of $\mathcal{R}_0$, respectively. In each case, $\mathcal{R}_0=\max\{ \mathcal{R}_h, \mathcal{R}_z \}$. The vertical red line represents $\mathcal{R}_{h,z} = 1$.}
\label{fig04}
\end{figure}

In Section \ref{sec:R0calculation}, we calculated the basic reproduction number \eqref{R0} for Model \eqref{model1}. Figure \ref{fig04} shows  the 
possible values for the basic reproduction number using the extreme parameter values given in Tables \ref{values-parameters1} and  \ref{values-parameters2} for Colombia and Brazil, that result in the minimum or maximum value of $\mathcal{R}_0$, respectively.

%%%%%

\begin{figure}[H]
\centering
\includegraphics[width=14cm, height=7cm]{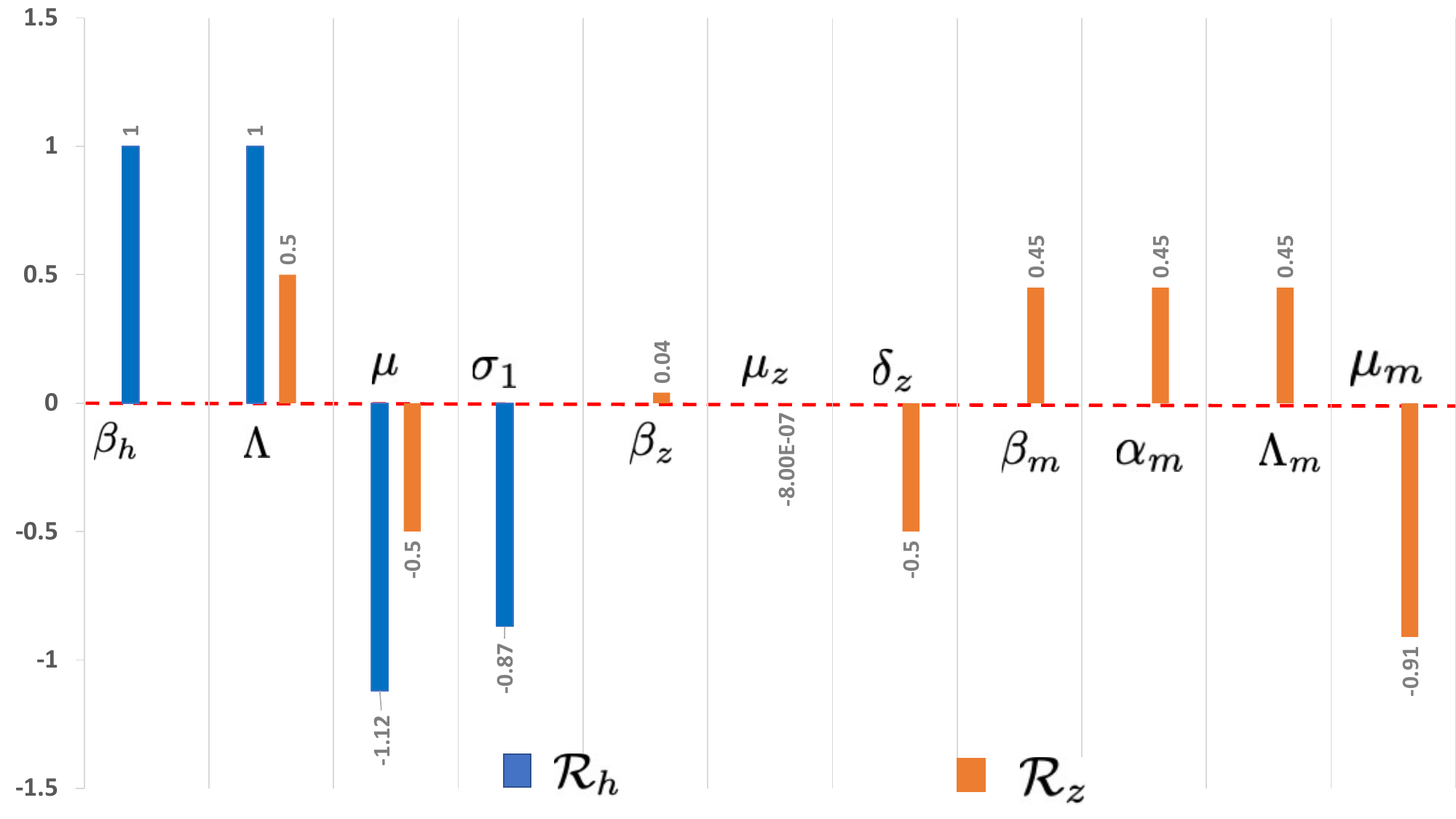}
\caption{Normalized sensitivity index of $\mathcal{R}_h$ to the parameters $\{ \beta_h, \Lambda, \mu, \sigma_1\}$ and  $\mathcal{R}_z$ to the parameters $\{ \Lambda, \mu, \beta_z,  \mu_z, \delta_z, \beta_m, \alpha_m, \Lambda_m, \mu_m\}$. Thus, the red dot line represents a sensitivity of zero.   A calculation example using Equation \eqref{sensitivity-index} can be found in Appendix \ref{appendix:A6}.}.
\label{fig03}
\end{figure}

The normalized sensitivity indices of $\mathcal{R}_h$ and $\mathcal{R}_z$  summarized in Figure \ref{fig03}, are obtained using the values of the parameters in Tables \ref{values-parameters1} and \ref{values-parameters2}.

%%%%%%%%
Regarding $\mathcal{R}_h$, the primary driver is the human death rate $\mu$, followed by the infection transmission $\beta_h$ of humans through sexual contact with HIV-infected individuals and the human recruitment parameter $\Lambda$. This implies that a reduction in the human death rate $\mu$ would result in the most significant increase in the likelihood of HIV infection among humans. 
Regarding $\mathcal{R}_z$, the outcomes reveal that the parameter with the highest impact is the mosquitoes death rate $\mu_m$: longer surviving mosquitoes lead to an increase in $\mathcal{R}_z$.
Additionally, the human recruitment parameter $\Lambda$ and death rate $\mu$ are also essential factors influencing $\mathcal{R}_z$.

\subsection{Evaluation of uncontrolled population behaviour over time}
\begin{figure}[H]
\centering
\subfigure[Colombia]{
\includegraphics[width=8.cm, height=5cm]{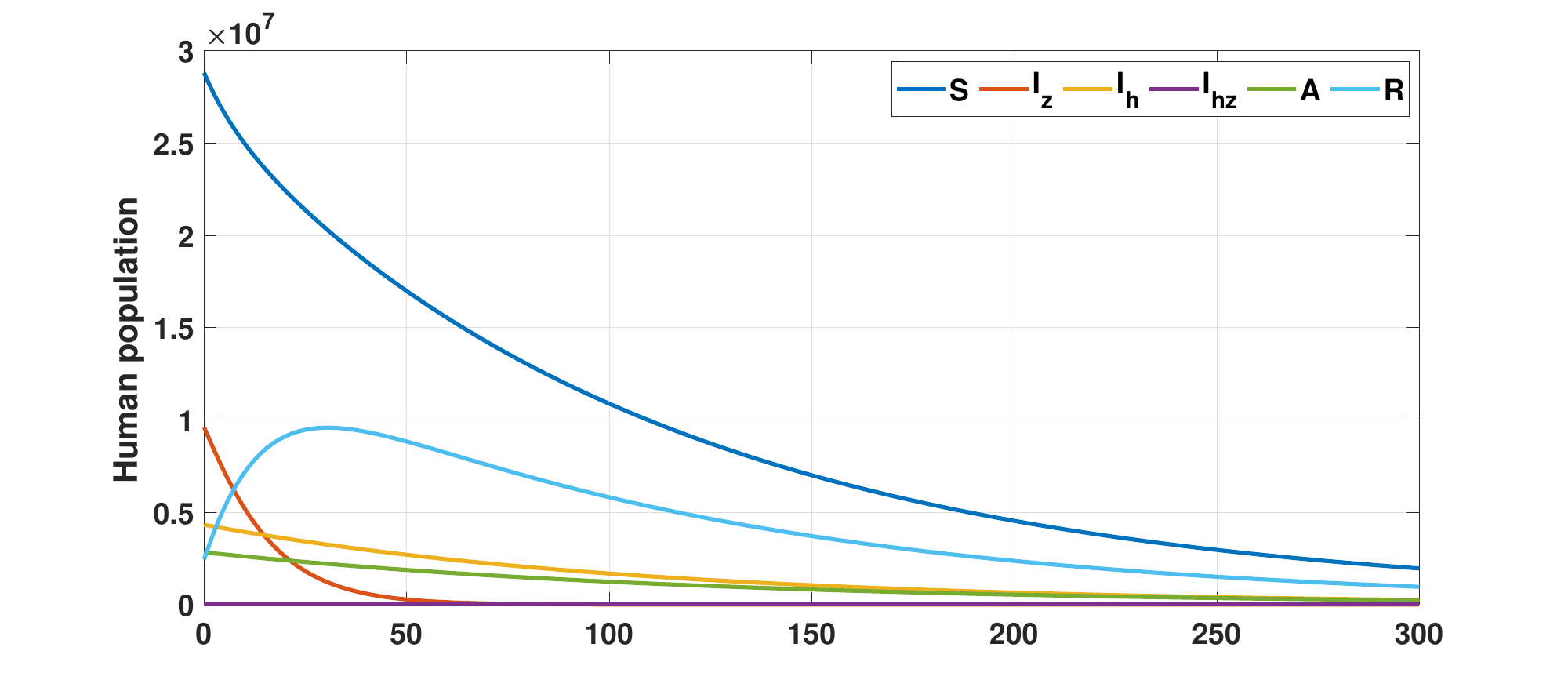}}
\subfigure[Colombia]{
\includegraphics[width=8.cm, height=5cm]{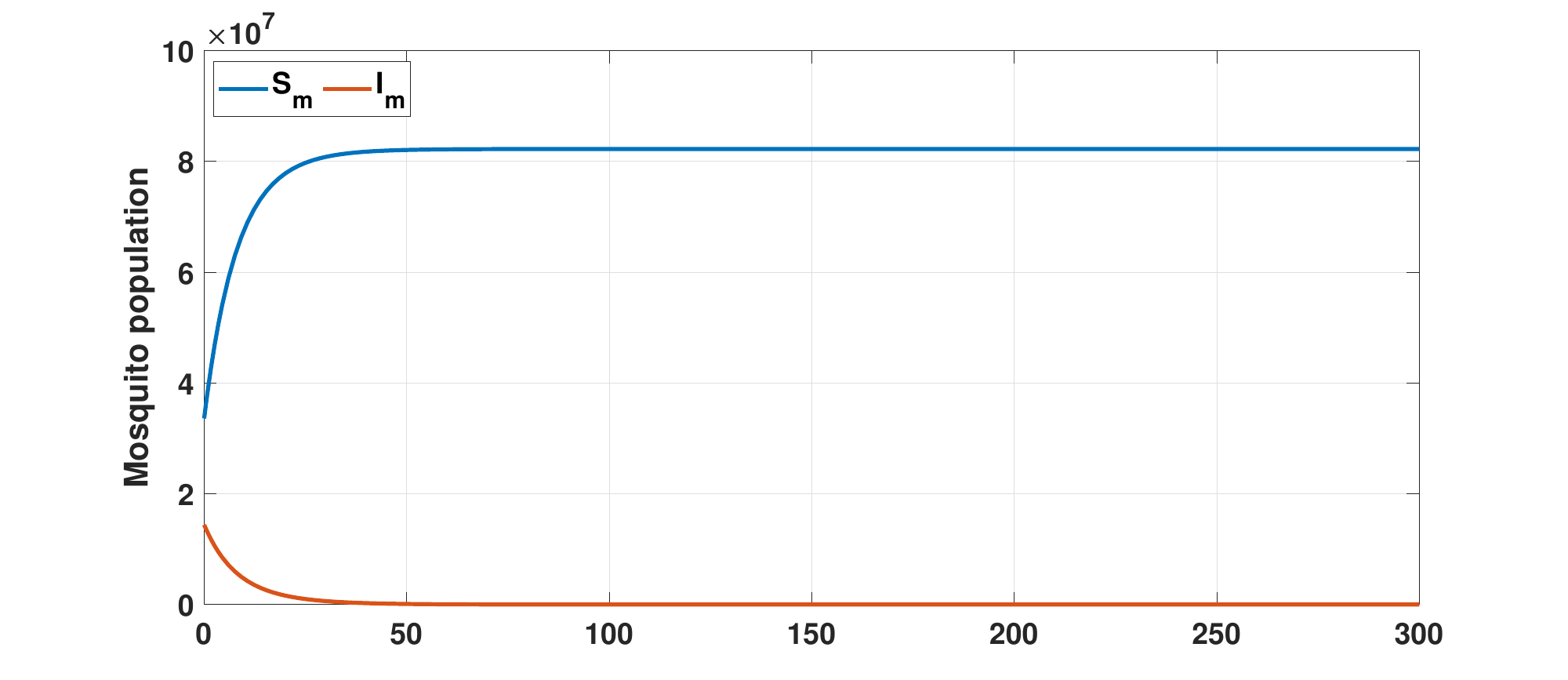}}
\subfigure[Brazil]{
\includegraphics[width=8.cm, height=5cm]{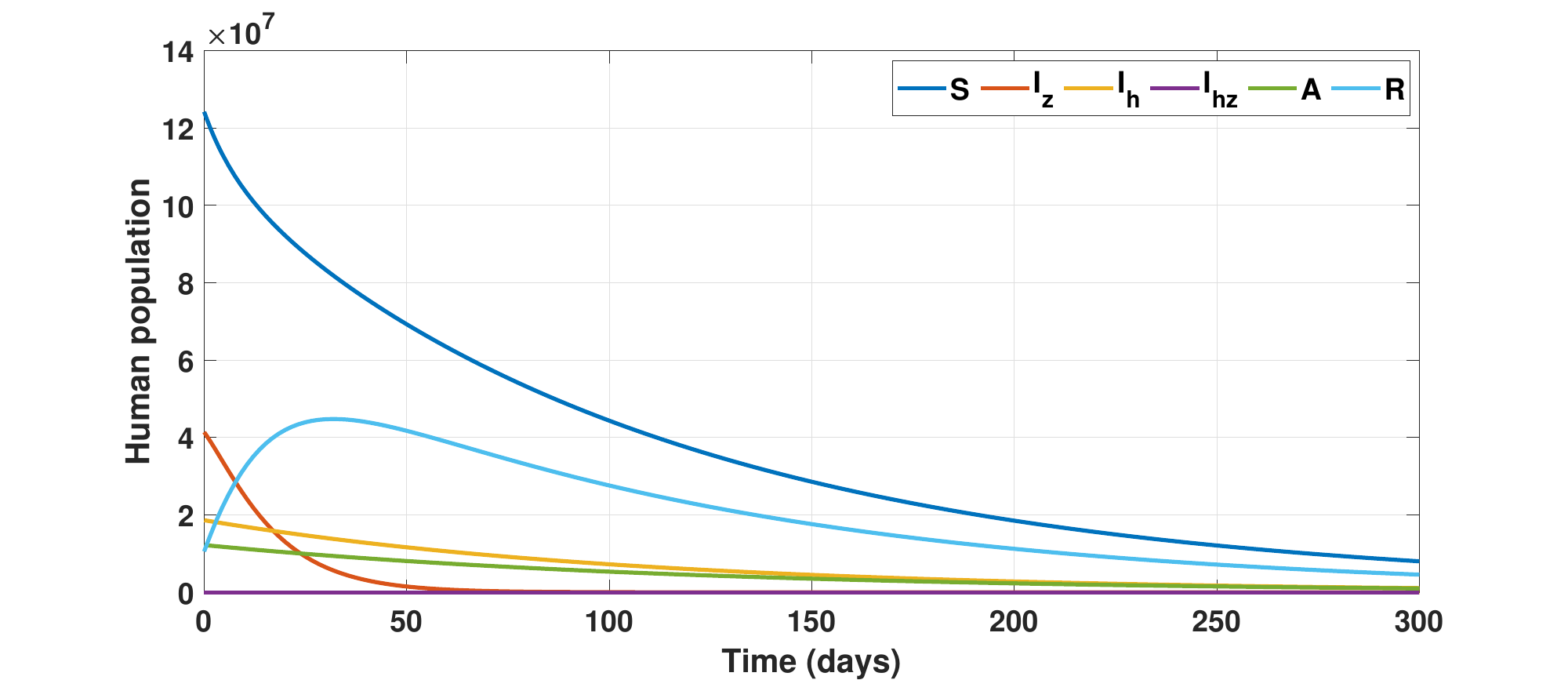}}
\subfigure[Brazil ]{
\includegraphics[width=8.cm, height=5cm]{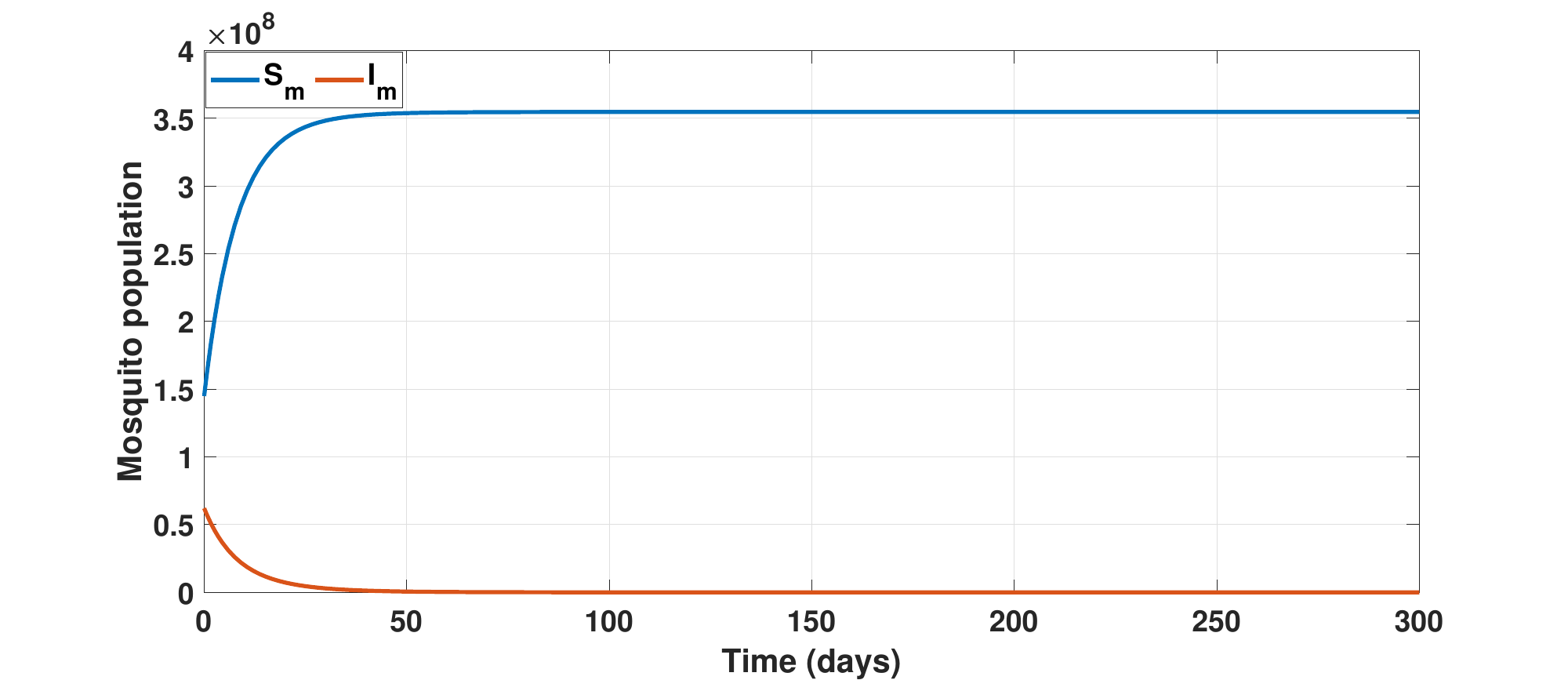}}
\caption{Simulations of the DFE for Colombia and Brazil.  For Colombia, $\mathcal{R}_h$=  0.4719, $\mathcal{R}_z$= 0.2306. For Brazil,  $\mathcal{R}_h$=  0.4719, $\mathcal{R}_z$=  0.2977. In both countries, $\mathcal{R}_0$= 0.4719.  The solutions tend to the DFE.}
\label{DFEboth}
\end{figure}

\begin{figure}[H]
\centering
\subfigure[Colombia]{
\includegraphics[width=8.cm, height=5cm]{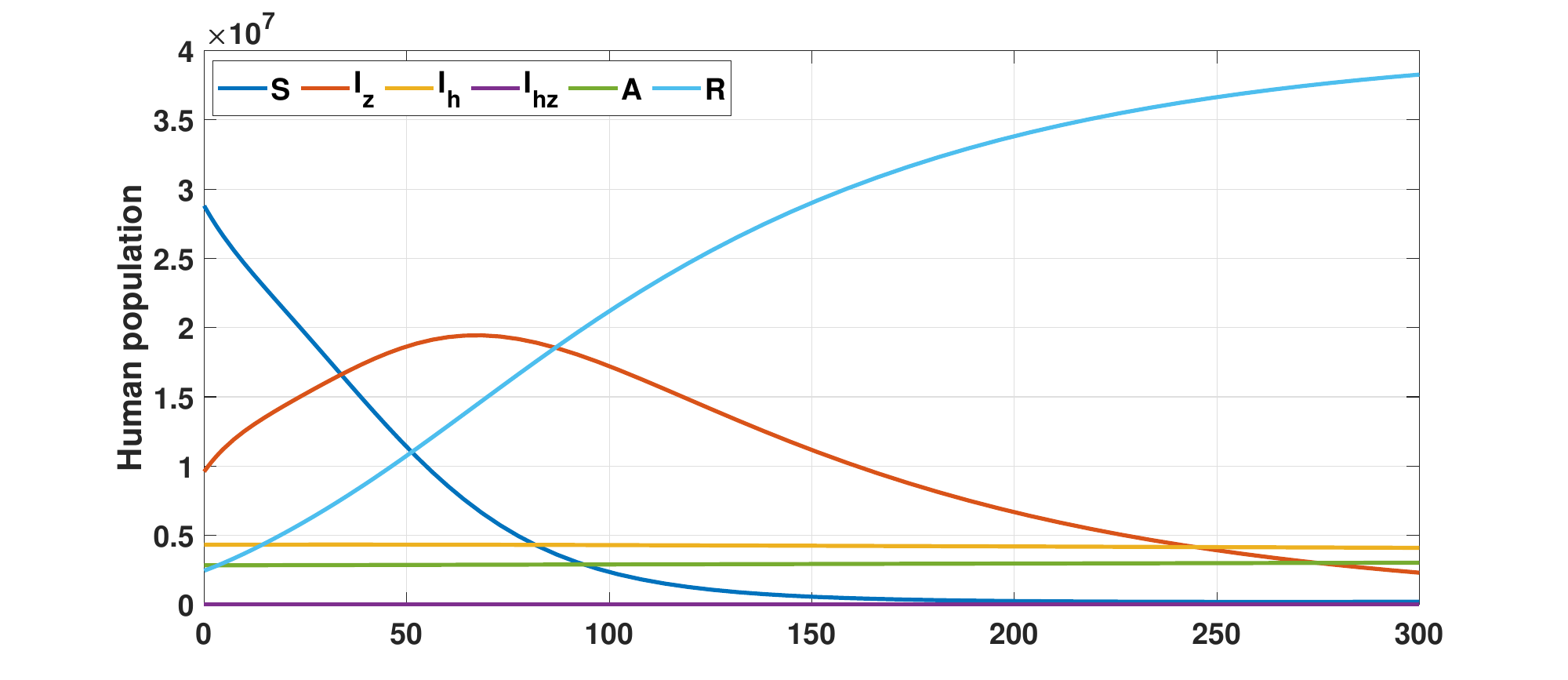}}
\subfigure[Colombia]{
\includegraphics[width=8.cm, height=5cm]{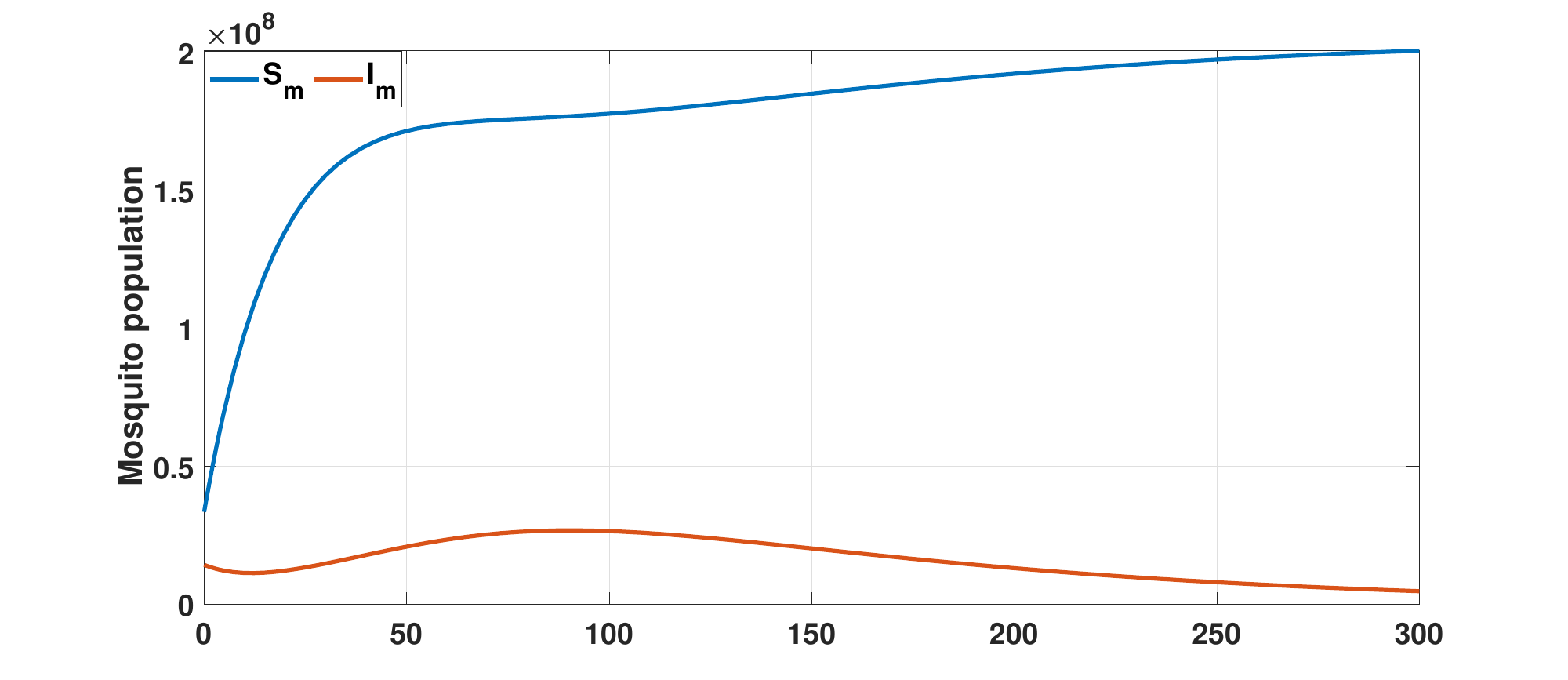}}
\subfigure[Brazil]{
\includegraphics[width=8.cm, height=5cm]{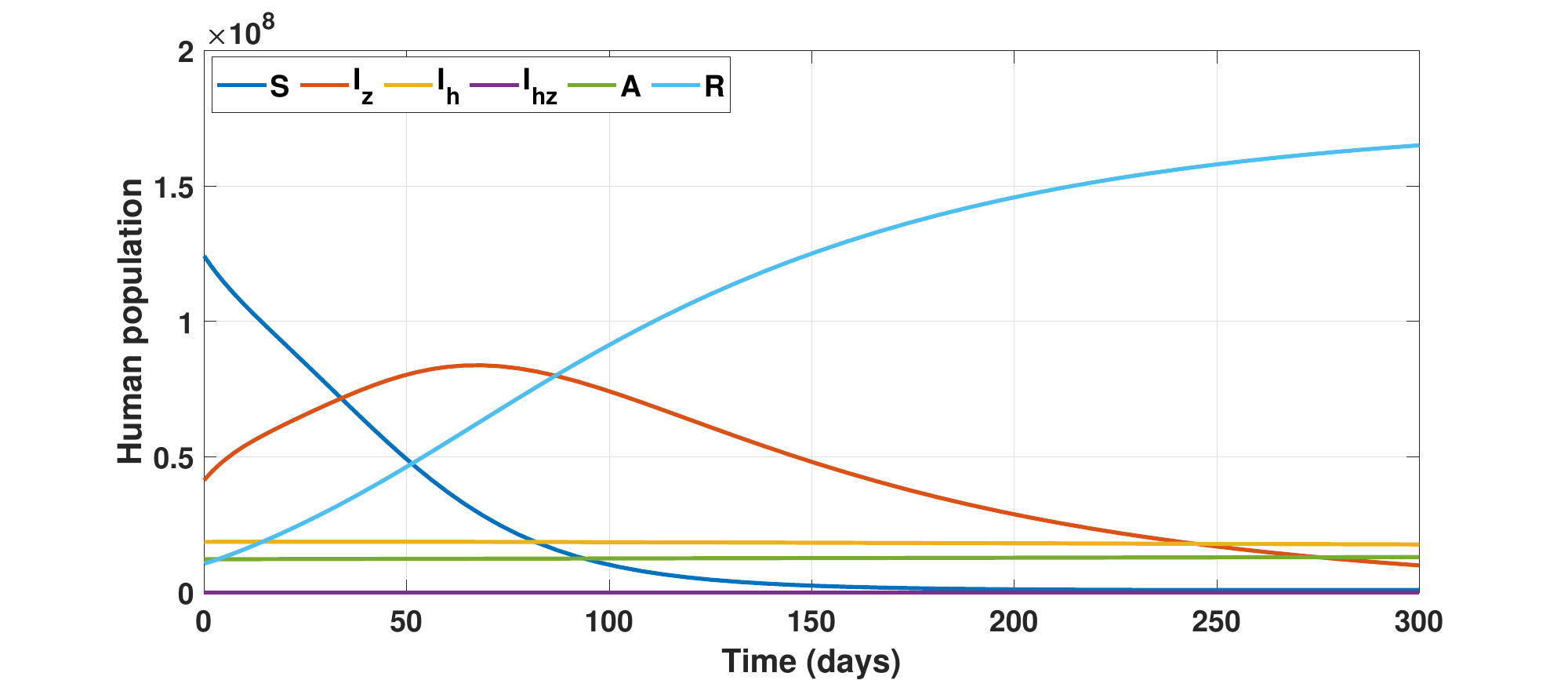}}
\subfigure[Brazil]{
\includegraphics[width=8.cm, height=5cm]{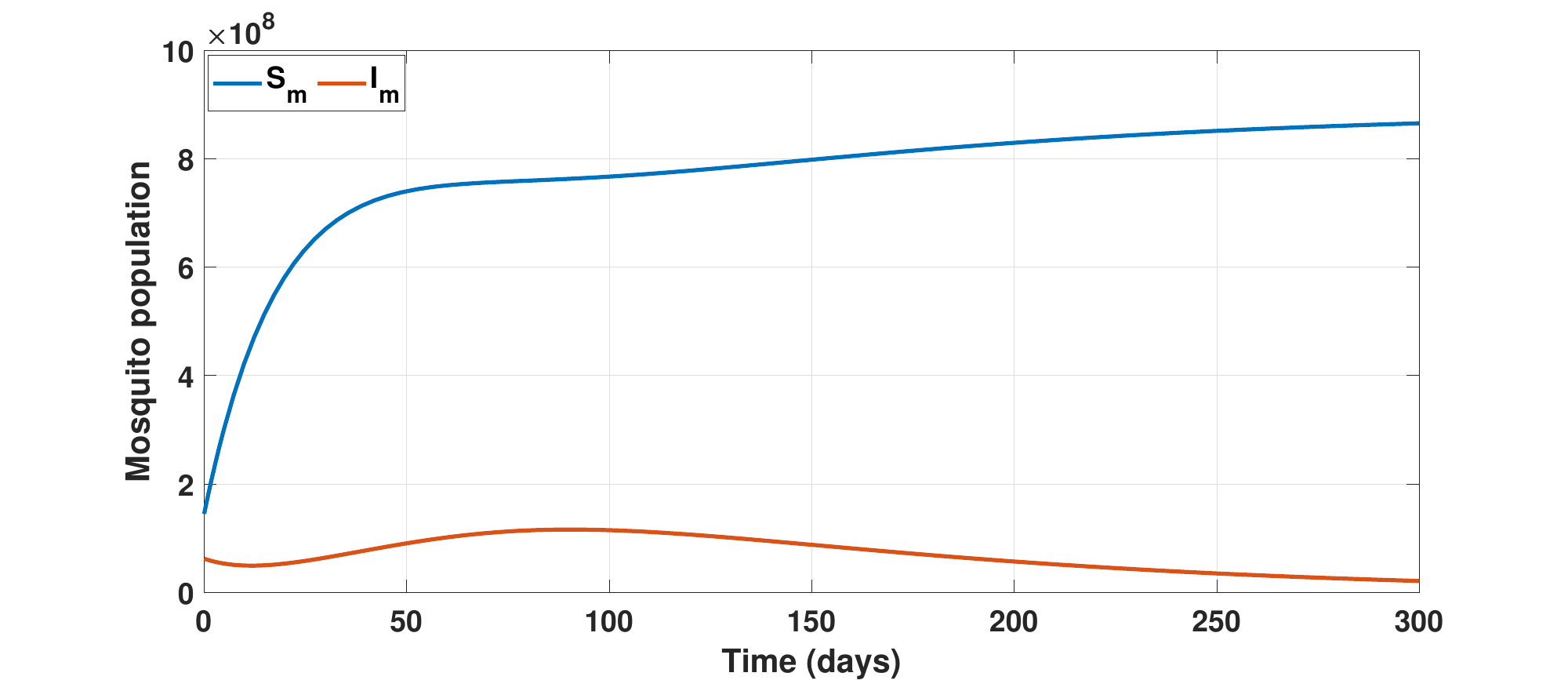}}
\caption{Simulations of the endemic equilibrium. For Colombia, $\mathcal{R}_h$= 3.2308, $\mathcal{R}_z$= 3.2338.  For Brazil, $\mathcal{R}_h$= 3.2308, $\mathcal{R}_z$= 3.2326.  The solutions tend  to the endemic equilibrium.}
\label{endemicboth}
\end{figure}

We then performed numerical simulations of the uncontrolled Model \eqref{model1}. 
We begin by representing the DFE for Colombia and Brazil.  
The DFE is obtained when the value of $\mathcal{R}_0$ is less than one.  As shown in Figure \ref{DFEboth}, for Colombia and Brazil, $\mathcal{R}_0$=0.47192. Figure \ref{DFEboth} shows the behaviour of the solutions for both human and mosquito populations. In both cases, on day 50 after the first observation, infected humans (with either disease) tended to zero, whereas the susceptible human population gradually decreased and appeared to stabilize after day 250. A similar scenario was observed in the mosquito population. On day 25 of the first observation, infected mosquitoes decreased to zero, and susceptible mosquitoes stabilized.
In this scenario, HIV/AIDS cases outnumbered ZIKV cases. No outbreaks of infected individuals (humans or mosquitoes) were observed under the initial conditions used.  
Similarly, the endemic equilibrium is obtained when the value of $\mathcal{R}_0$ is greater than one (see Figure \ref{fig04}). This scenario is simulated and depicted in Figure \ref{endemicboth}. For  Colombia   $\mathcal{R}_0$=3.2338 and for Brazil $\mathcal{R}_0$=3.2326. We can observe a significant increase in the number of humans infected with ZIKV during the first 70 days of the first observation, with the number of HIV/AIDS-infected humans being lower. Humans infected with ZIKV reached maximum values on day 70 of the first observation. In contrast, humans infected with HIV/AIDS did not generate peaks and appear to have constant behaviour throughout the observed period (approximately a year). During the first 250 days from the first observation, ZIKV-infected humans outnumber HIV/AIDS-infected humans.  However, after day 250, HIV/AIDS-infected humans constantly outnumber ZIKV-infected humans. 
\par 
For the mosquito population, there is always a higher population size of susceptible than infected mosquitoes, but there is stable coexistence between them. Both countries differ mainly in \textcolor{red}{population size} (Brazil has a larger population than Colombia). The simulations indicated that the behaviour of both mosquito populations in Colombia and Brazil was highly similar, with a notable distinction in terms of population sizes. This suggests that the dynamics and patterns of infection transmission were comparable between the two countries, with variations primarily related to the size of the populations being studied.
An interesting observation from the simulations was that the behaviour of humans infected with ZIKV displayed more significant variability than those infected with HIV/AIDS. This implies that the progression and manifestations of the ZIKV infection were more diverse and fluctuating, potentially influenced by various factors such as environmental conditions, viral load, or other epidemiological characteristics. In contrast, the behaviour of HIV/AIDS-infected individuals appeared to be more consistent and stable over time.
These findings highlight the distinctive characteristics of ZIKV and HIV/AIDS infections, particularly in terms of the dynamic nature of Zika  infection compared to the relatively more constant nature of HIV/AIDS infection.

\begin{figure}[H]
\centering
\subfigure[Colombia]{\includegraphics[width=8.cm, height=4.5cm]{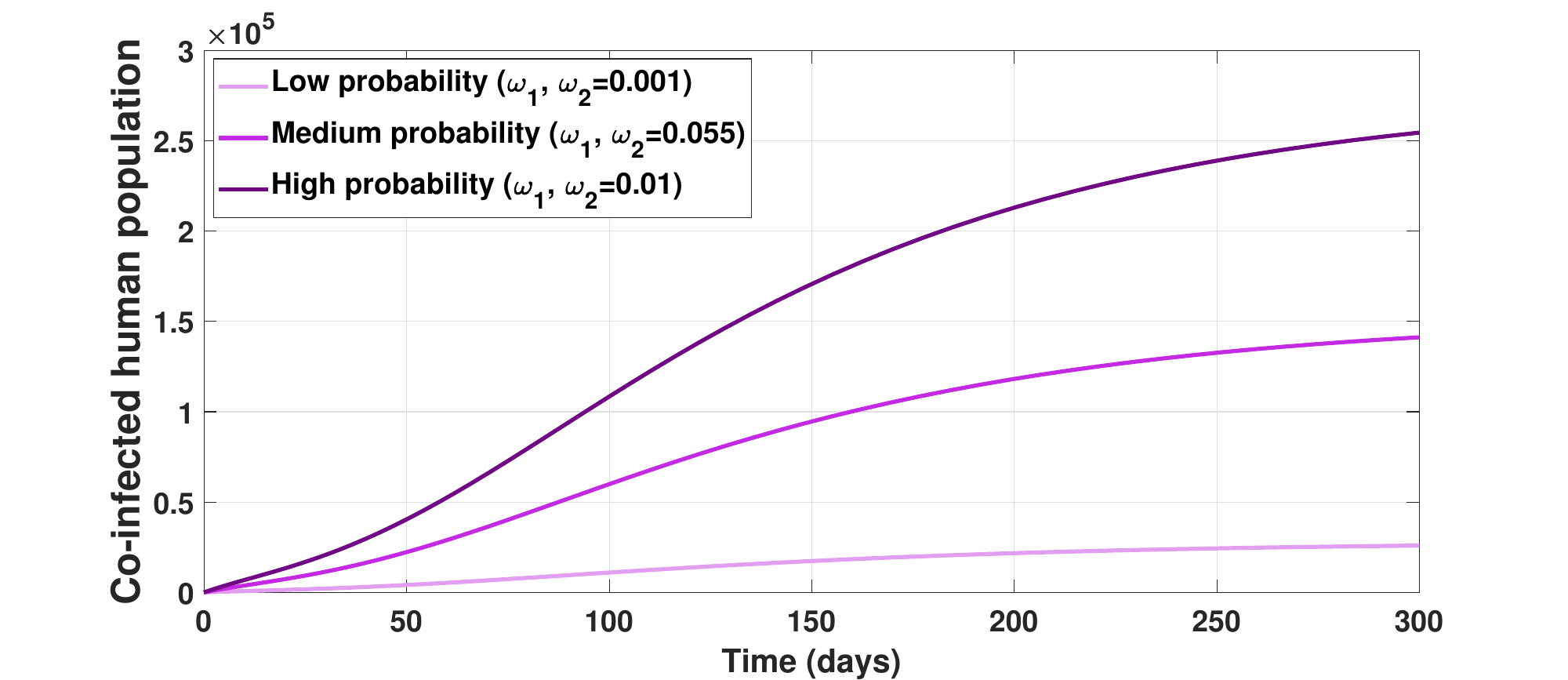}}
\subfigure[Brazil]{\includegraphics[width=8.cm, height=4.5cm]{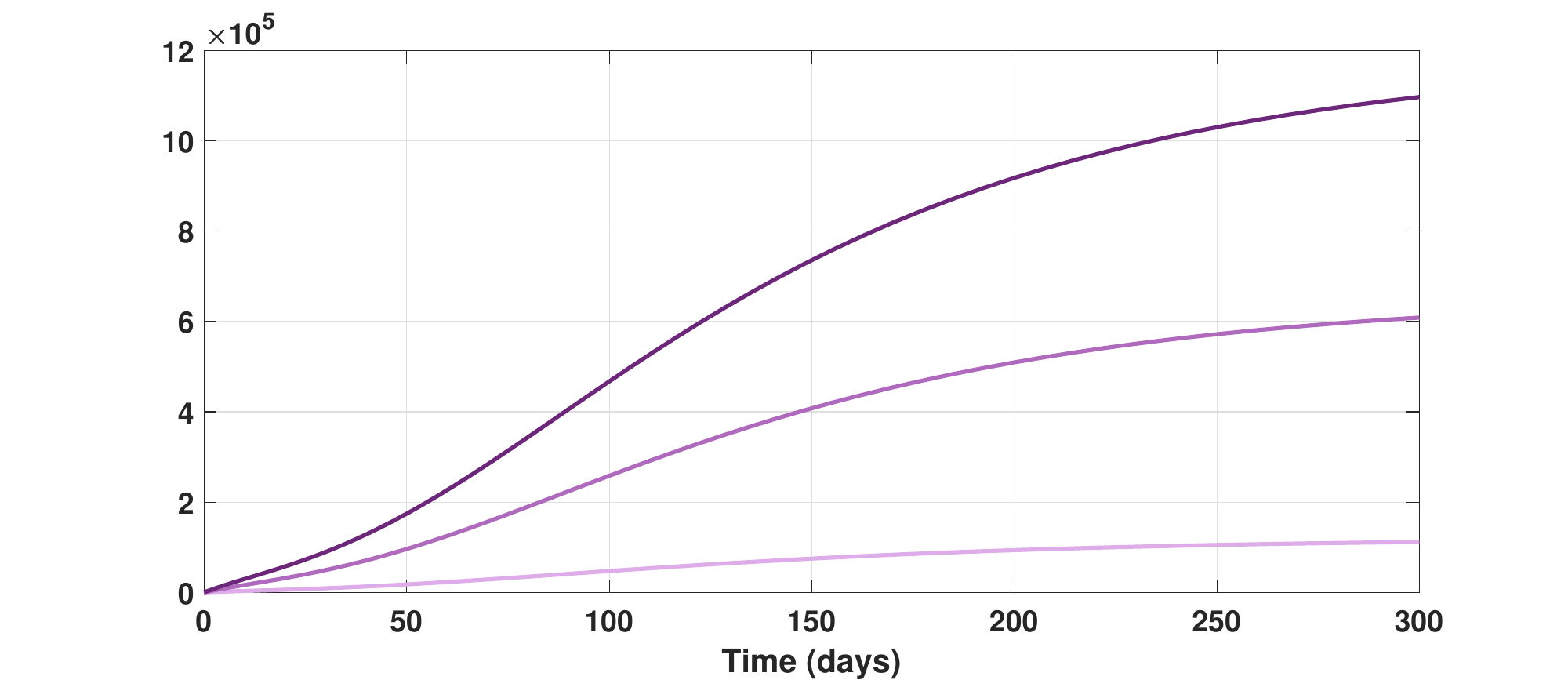}}
\caption{Hypothetical scenario of ZIKV/HIV co-infection for different values of the co-infection probabilities $\omega_1$  and $\omega_2$ (high, medium, and low). The illustration shows that probabilities exceeding 0.01 are associated with a significant increase in co-infected individuals. This emphasizes the crucial role of those parameters in impacting the transmission intensity of HIV/ZIKV co-infection. }
\label{hypboth}
\end{figure}

In reality, the number of co-infections is small and maintained in the previous figures. Therefore, we discuss a drastically hypothetical scenario for  HIV/ZIKV co-infected individuals as shown in Figure \ref{hypboth} to investigate some assumptions that can lead to an increase of the co-infected population size.  We contrasted three different possibilities for the probability of co-infection: $\omega_{1,2}$=0.001, 0.0055, 0.01. The first value (0.001) was named low probability, the second value (0.0055) medium probability, and the last value (0.01)  high probability.  It can be seen that for higher values of this pair of parameters, a higher population size of individuals co-infected with ZIKV and HIV follows. Evidently, an increase in these two probabilities increased the number of co-infected individuals.

\par

 %%%%%%%%
These numerical experiments showed that for values of $\mathcal{R}_0<1$, the HIV/ZIKV co-infection epidemics tend to fade away after some time, with infected individuals tending to zero and the susceptible population reaching a stable level. The same trend was observed for the mosquito population, with infected mosquitoes decreasing to zero and susceptible mosquitoes stabilizing. A noticeable increase in the number of individuals infected with ZIKV, with a lower number of individuals infected with HIV/AIDS, was observed during the first 70 days of the first observation for the values of $\mathcal{R}_0>1$. Infected individuals with ZIKV peaked on day 70, whereas HIV-infected individuals did not generate any peaks during the 300 simulated days. Regarding the mosquito population, susceptible mosquitoes always outnumber infected mosquitoes; however, both coexist.

\par 
In addition, Figure \ref{hypboth} provides insights into the hypothetical scenario of ZIKV/HIV co-infection. Comparing three different scenarios for co-infection probabilities (high, medium and low), an increase in these probabilities above 0.01, leads to an increase in the number of co-infected individuals. These findings suggest that the values of these parameters can significantly impact the spread and severity of HIV/ZIKV co-infection, highlighting the importance of considering multiple factors when designing effective strategies to prevent and control the spread of these viruses, particularly in pregnant women to avoid congenital malformations in children. 

\par 
These findings suggest that the behaviour of populations in Colombia and Brazil was similar but different in population densities. This suggests that the patterns of infection transmission were comparable, primarily influenced by population size. Notably, individuals infected with ZIKV exhibited more significant behaviour variability than those infected with HIV/AIDS. This indicates that ZIKV infection showed diverse and fluctuating progression and manifestations, potentially influenced by various factors. In contrast, HIV/AIDS-infected individuals displayed more consistent and stable behaviour over time.

%%%%%%
\subsection{Evaluation of controlled population behaviour over time}
After establishing some numerical simulations to discuss the dynamics of the epidemic model \eqref{model1} in Columbia and Brazil, and building hypothetical scenarios to investigate the impact of a change in transition probabilities from HIV and ZIKV to the HIV/ZIKV co-infection compartment. Hence, we focus at this stage to investigate numerically the control problem described by \eqref{model_ocp}. To this end,  we investigate the effects of mixing strategies to control HIV/ZIKV co-infection in Colombia and Brazil.  Table \ref{values-parameters4} shows the balancing and weighting constants values associated with the OPC \eqref{model_ocp}.

\begin{table}[H]
\centering
\begin{tabular}{lccccccc}
  \hline \\
 Parameter &$c_1$ &$c_1$&$c_3$ &$c_4$ &$d_1$&$d_2$& $d_3$\\\hline \\
Value & 0.5 &0.5 &0.5&0.5 &$10^5$  &$10^7$  &$7 \times 10^6$   \\
  \hline \\
\end{tabular}
\caption{Balancing and weighting constants values associated to the OCP. \eqref{model_ocp}. }
\label{values-parameters4}
\end{table}

To curb the spread of ZIKV and HIV in Colombia, targeted intervention strategies are needed to address the particular issues that the population faces. To combat ZIKV, efforts must be directed toward preventing mosquito bites using insect repellents, protective clothing, and mosquito control measures. To achieve this goal, the Zika Strategic Response Plan reported in 2016 \cite{world2016zika} was a priority for the Colombian government and focused on improving VBDs public health surveillance systems, invest in health campaigns, distributing funds for necessary tools like insecticides, larvicides, and mosquito nets, deal with climate change-related emergencies, and support maternal and child health \cite{forero2020zika}.   Secure sexual behaviours need to be prioritized in public health initiatives targeting disadvantaged populations in order to stop the propagation of STDs \cite{prachniak2016hiv}. Therefore, ART and condoms are essential for controlling the spread of HIV \cite{prachniak2016hiv}. To ensure that people living with HIV in Colombia receive the care they require to be healthy and to lower their risk of spreading the virus to others, it is mandatory to increase access to HIV laboratory testing and treatment \cite{galindo2014hiv}. Condom use and other methods for reducing infections must also be promoted in public health campaigns, especially among high-risk groups \cite{donoghoe2006injecting}. By addressing these challenges through an inclusive approach, including education, prevention, and access to health systems, Colombia could make significant progress in controlling the spread of ZIKV and HIV. 
\par 
Conversely, Brazil has also implemented a nationwide strategy to manage {\it Aedes} mosquito populations and to stop the spread of ZIKV.  This endeavour has included activities, such as the use of insecticides, the removal of standing water, and public health campaigns to motivate people to take action to reduce the number of mosquito breeding sites \cite{bancroft2022vector}, which are incorporated as $\eta_1$ in the epidemic Model \eqref{model_ocp}. Additionally, given that it has been noted that Zika can be transmitted sexually; hence, condom use has been advised as a preventative measure. Brazil has also implemented a complete strategy for HIV prevention and care \cite{benzaken2019antiretroviral}. This has involved administering ART to all HIV-positive individuals since 2013. Brazil started a campaign to distribute condoms, encouraging condom use among vulnerable populations. As a result, the number of deaths from AIDS in Brazil has significantly dropped \cite{bastos2009aids,pereira2019decline}.

\begin{figure}[H]
\centering

\subfigure[Colombia]{
\includegraphics[width=7.cm, height=4cm]{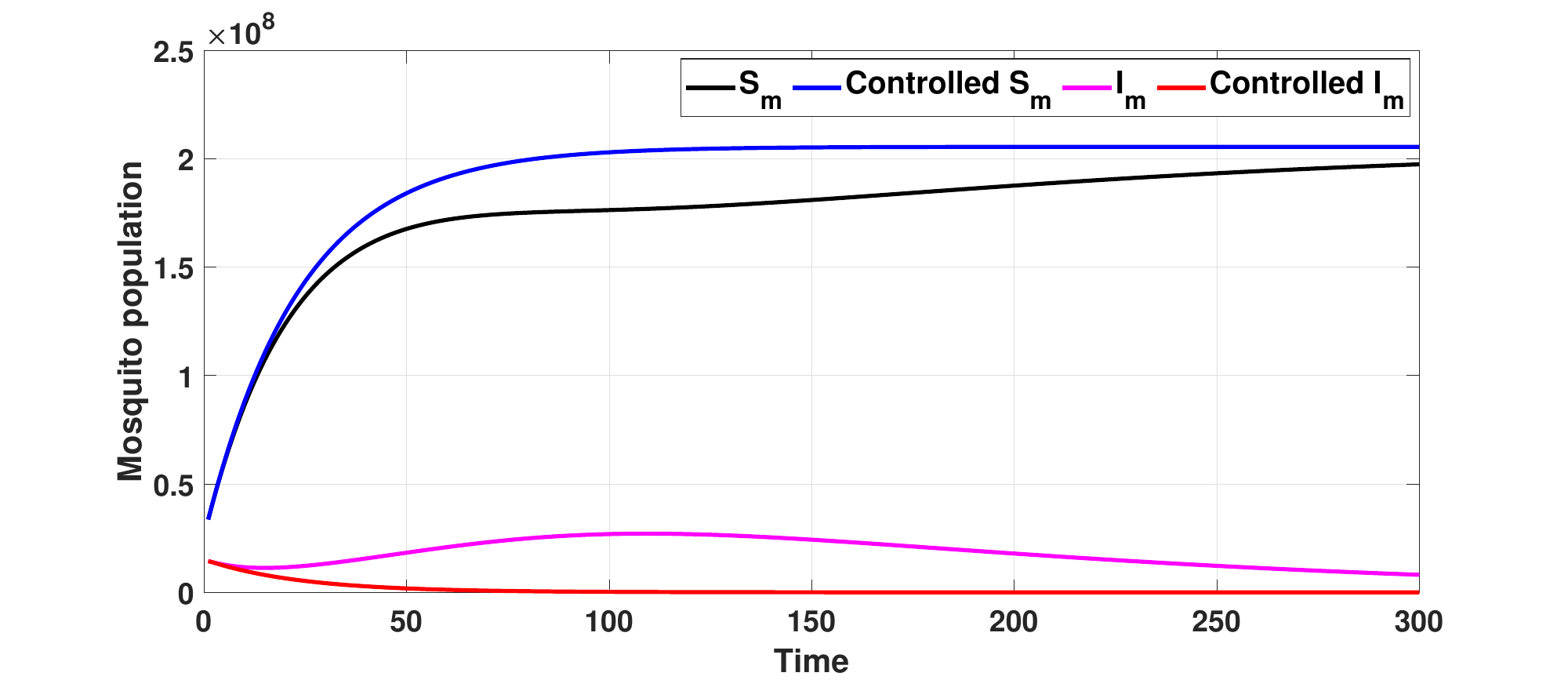}}
\subfigure[Brazil ]{
\includegraphics[width=7.cm, height=4cm]{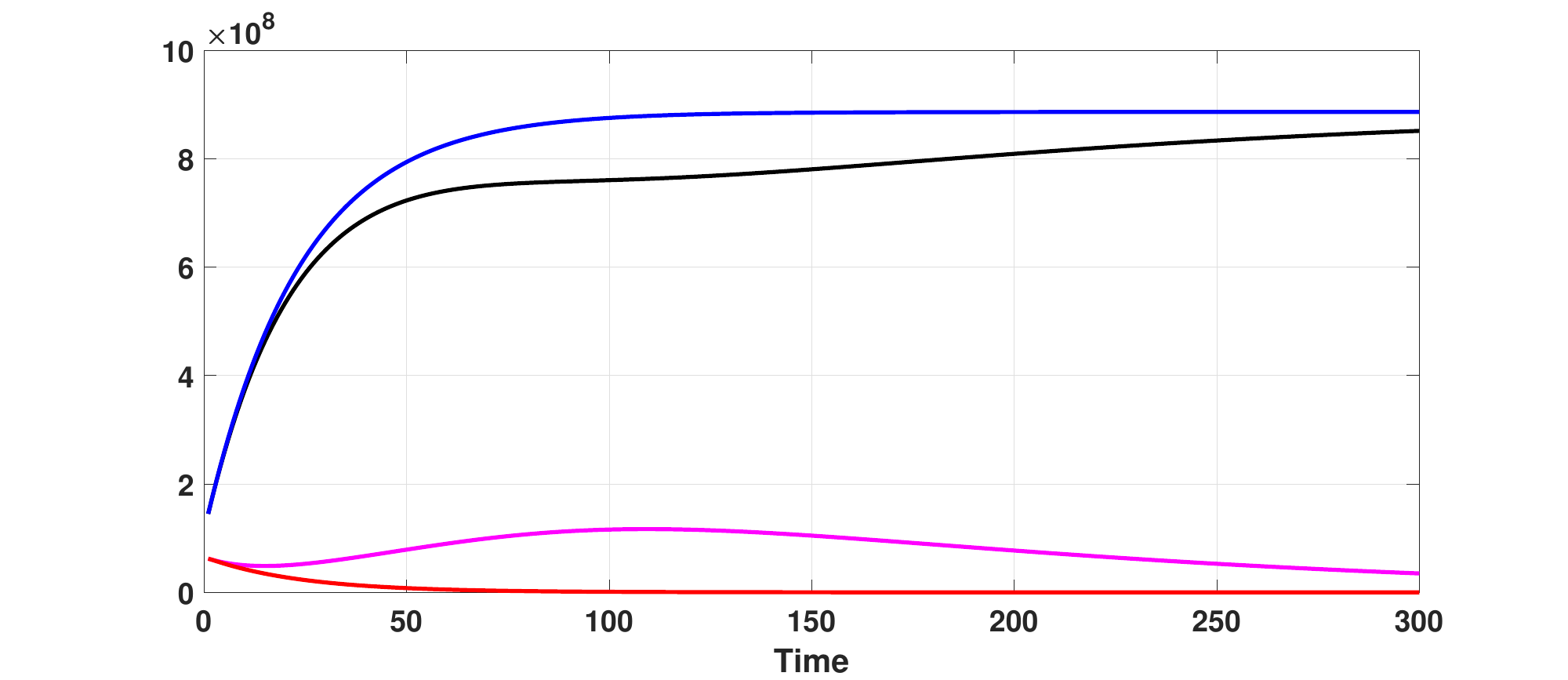}}
\caption{Simulations of the mosquito population in Colombia and Brazil with the activation of different controls.}
\label{controledboth-mosquitos}
\end{figure}

\begin{figure}[H]
\centering
\subfigure[Colombia]{
\includegraphics[width=7.cm, height=4cm]{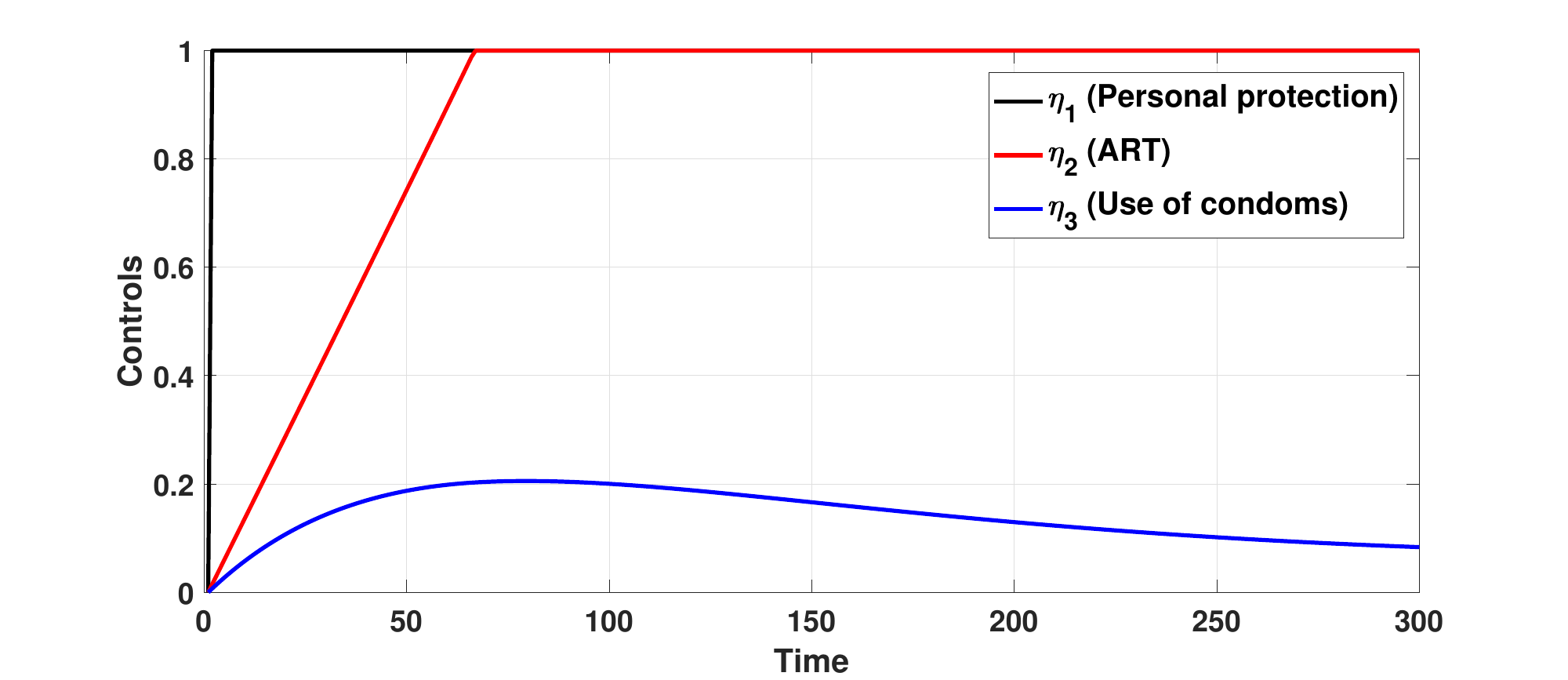}}
\subfigure[Brazil]{
\includegraphics[width=7.cm, height=4cm]{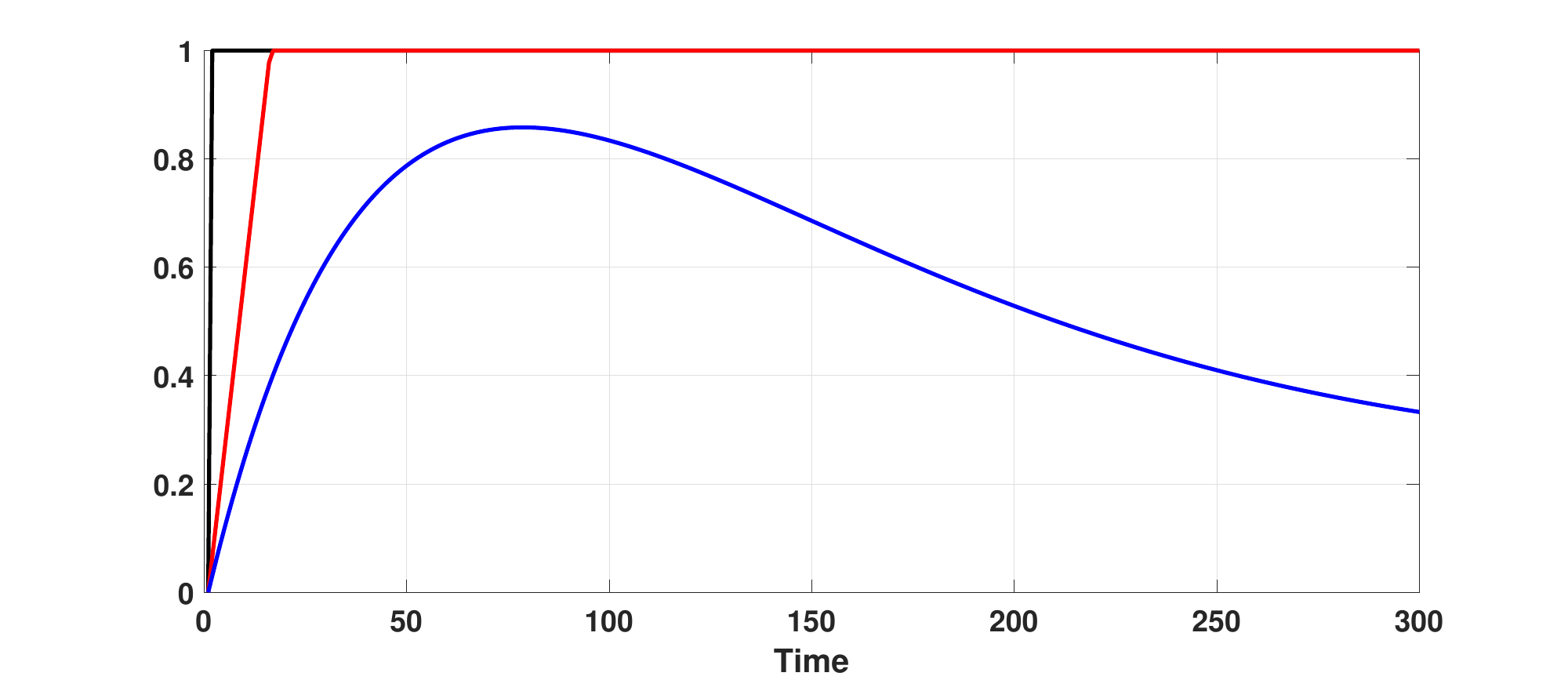}}
\caption{Simulations of the behaviour of the controls in Colombia and Brazil.}
\label{controls-both}
\end{figure}

\begin{figure}[H]
\centering
%%%%%controlled I
\subfigure{
\includegraphics[width=8.cm, height=4cm]{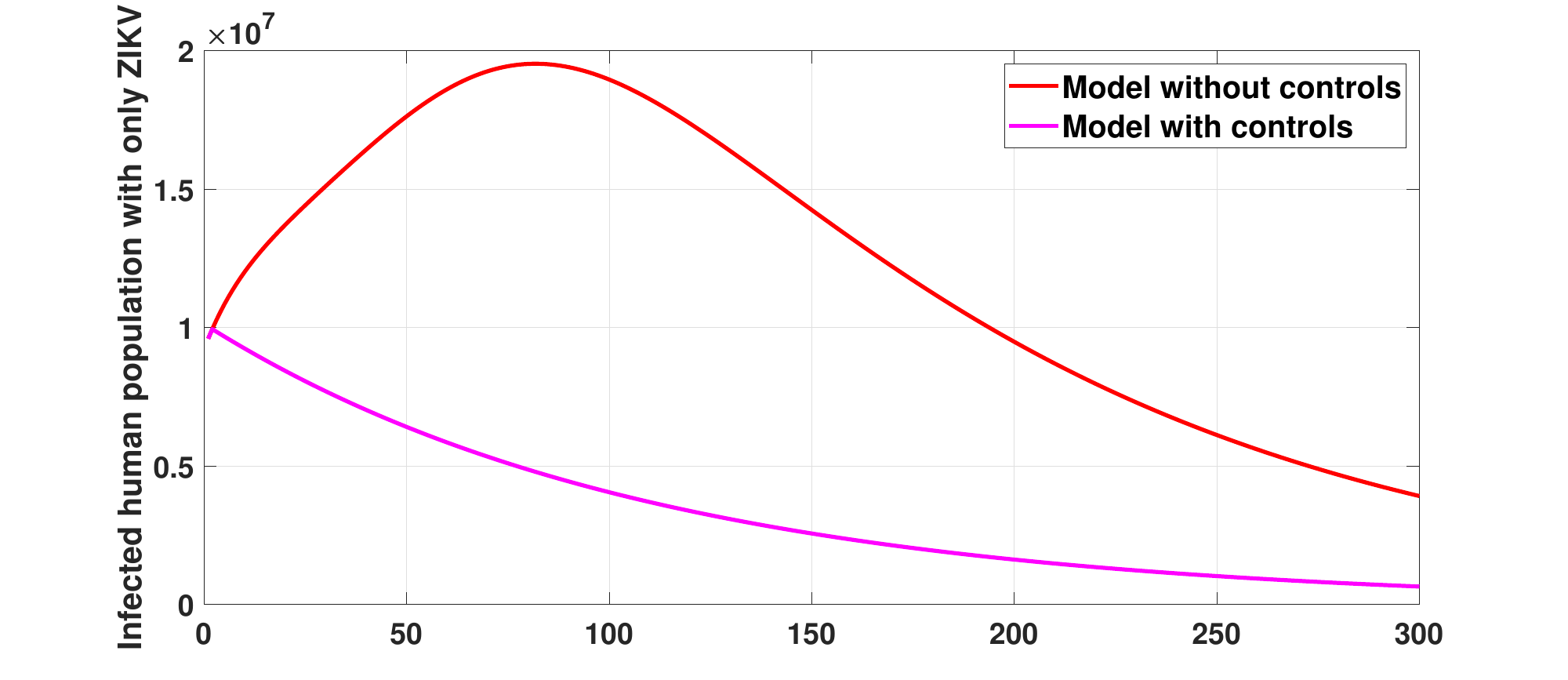}}
\subfigure{
\includegraphics[width=8.cm, height=4cm]{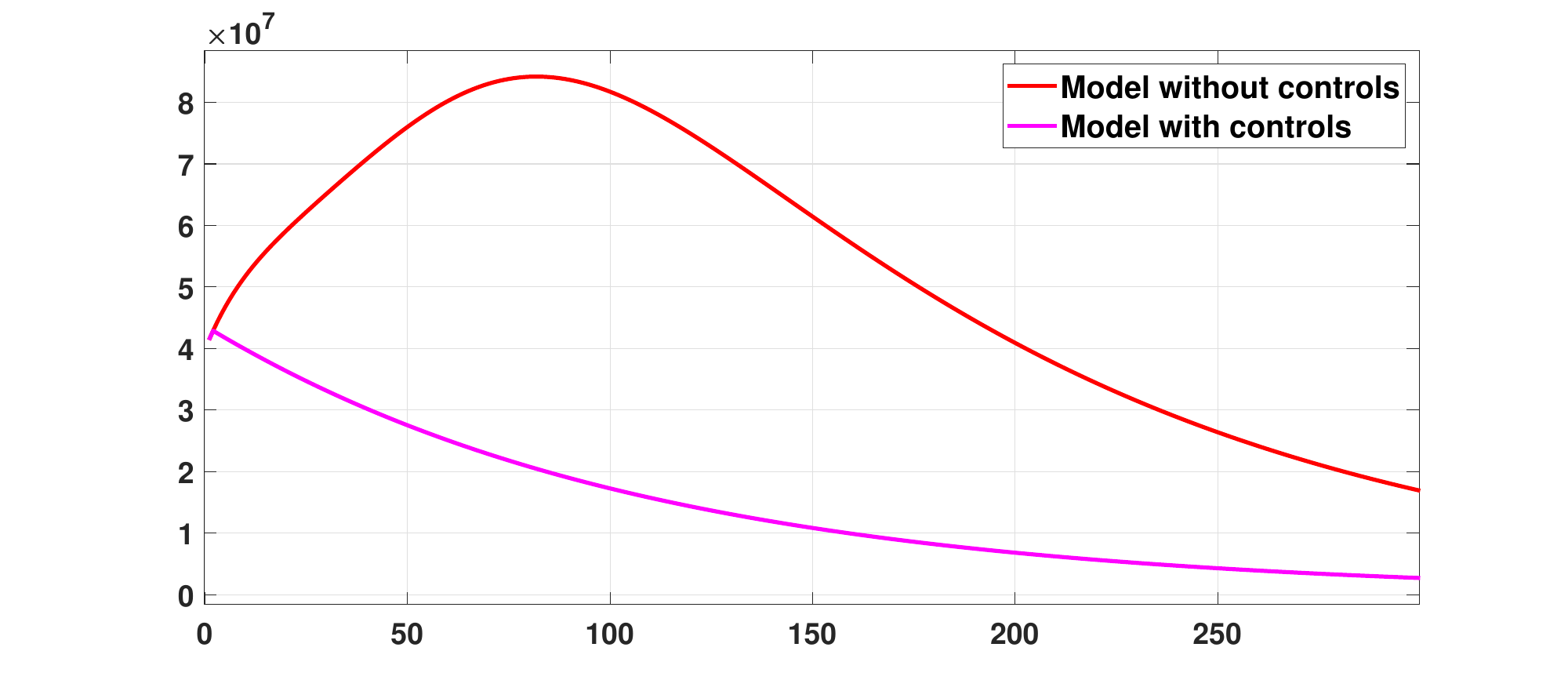}}
%%%%%%%Controlled I_h
\subfigure{
\includegraphics[width=8.cm, height=4cm]{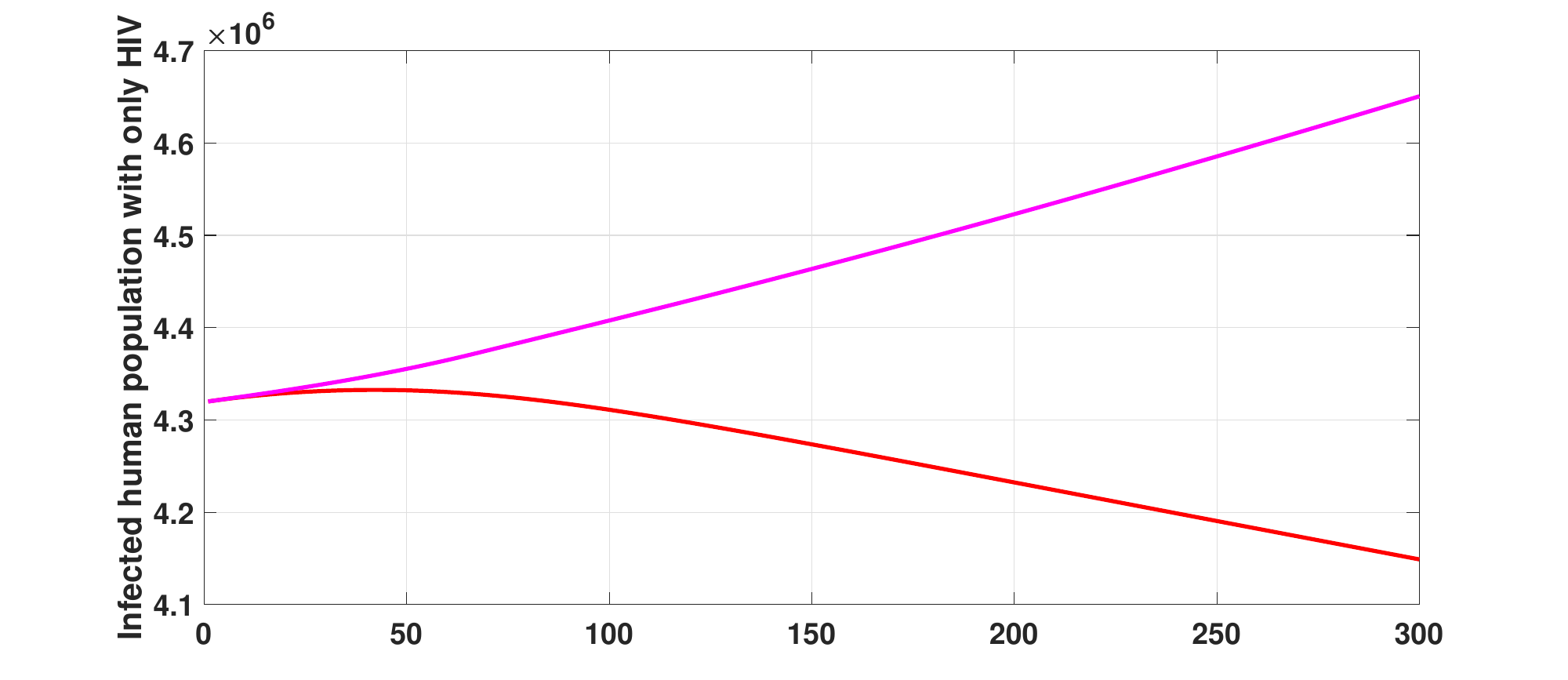}}
\subfigure{
\includegraphics[width=8.cm, height=4cm]{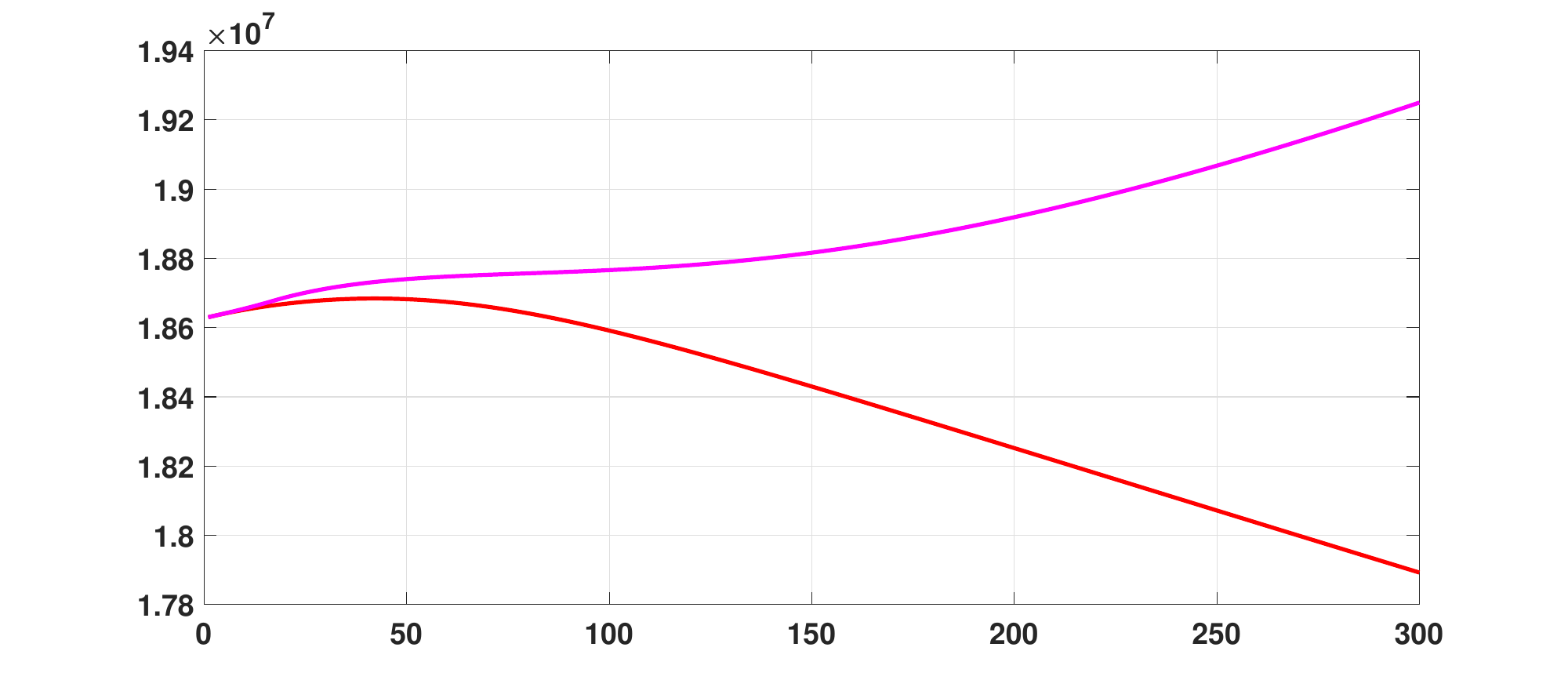}}
%%%%%%controlled A
\subfigure{
\includegraphics[width=8.cm, height=4cm]{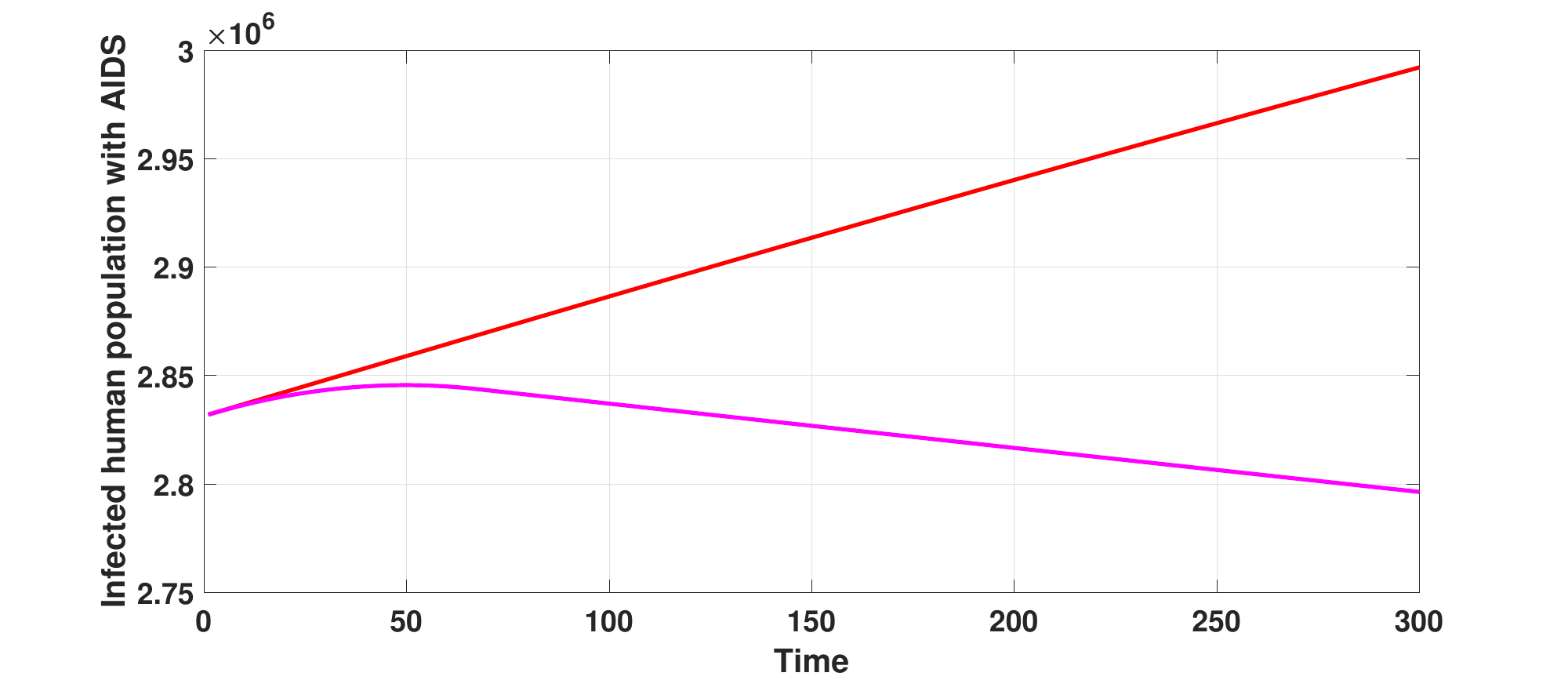}}
\subfigure{
\includegraphics[width=8.cm, height=4cm]{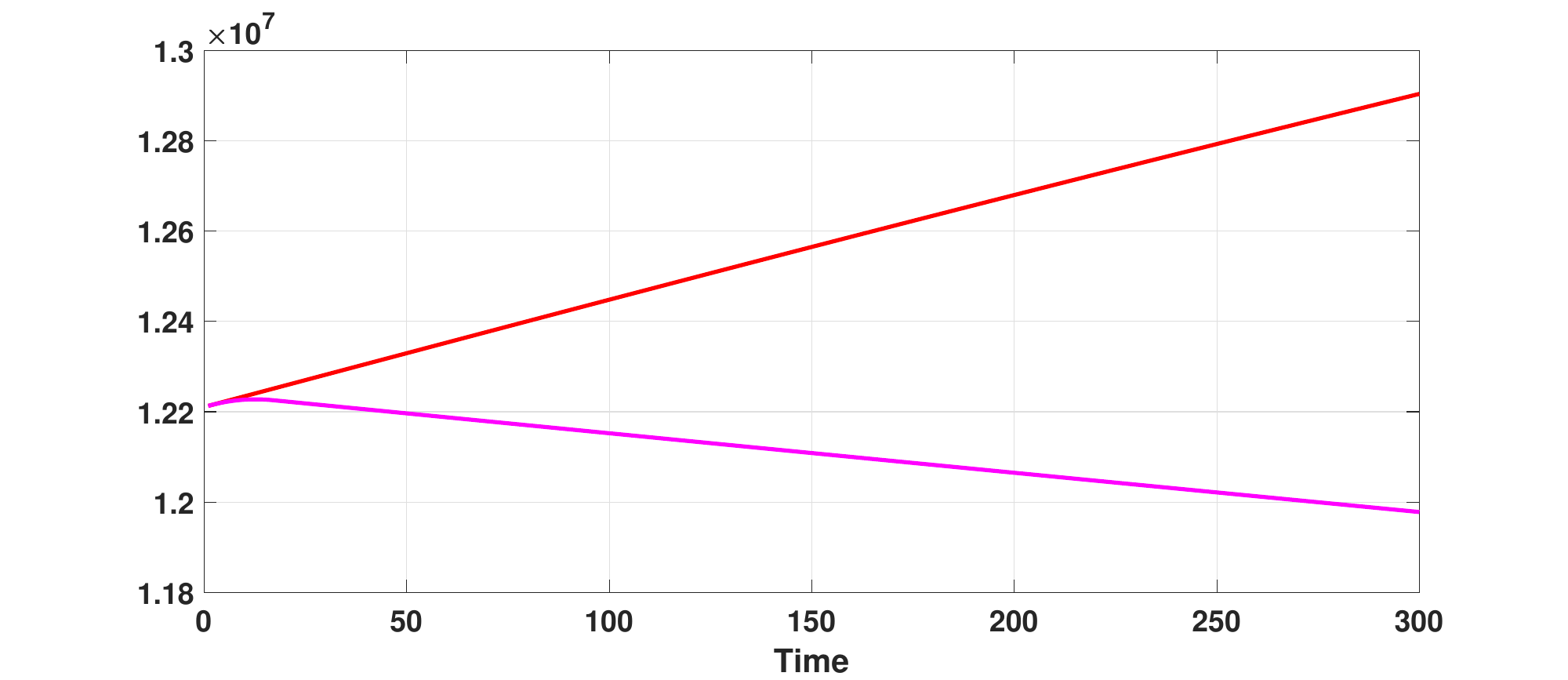}}
%%%%%%controlled I_hz
\setcounter{subfigure}{0}
\subfigure[Colombia\label{a}]{
\includegraphics[width=8.cm, height=4cm]{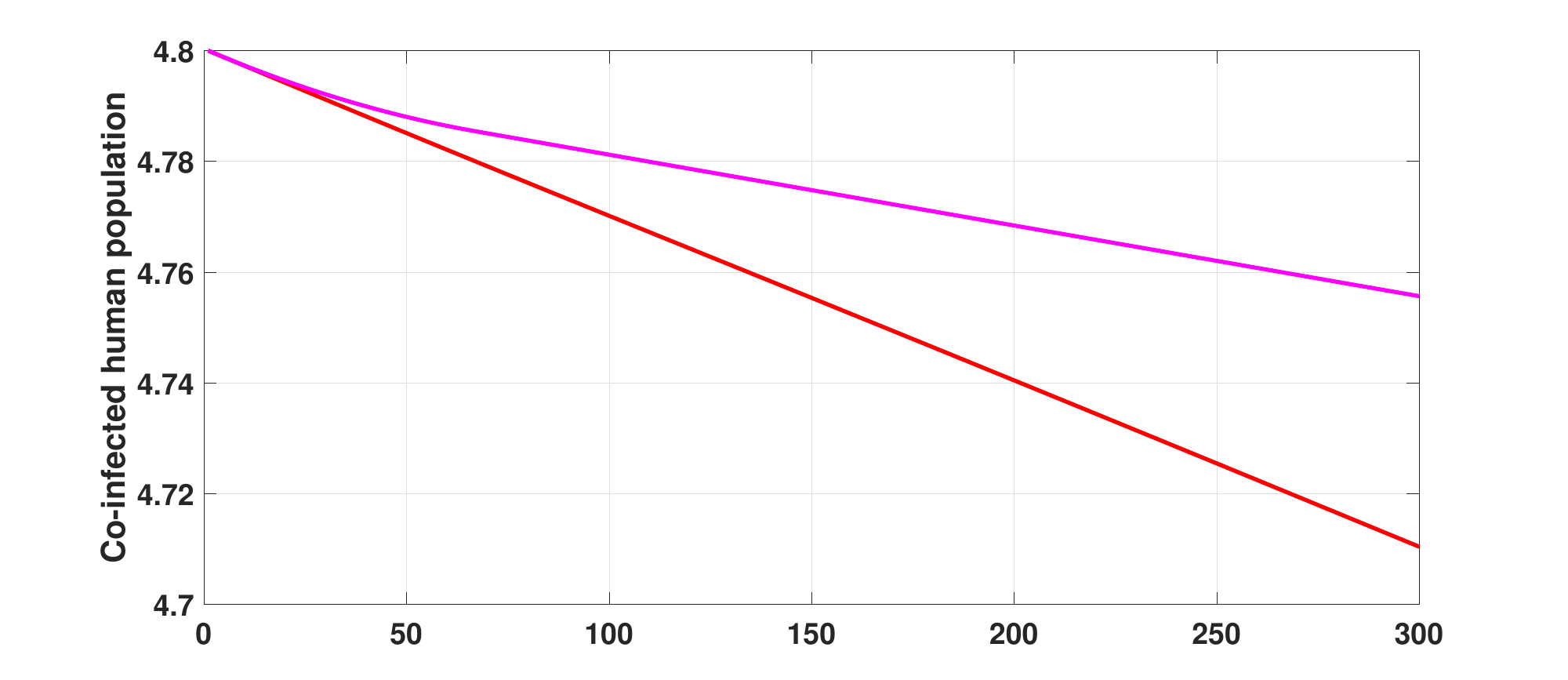}}
\subfigure[Brazil]{
\includegraphics[width=8.cm, height=4cm]{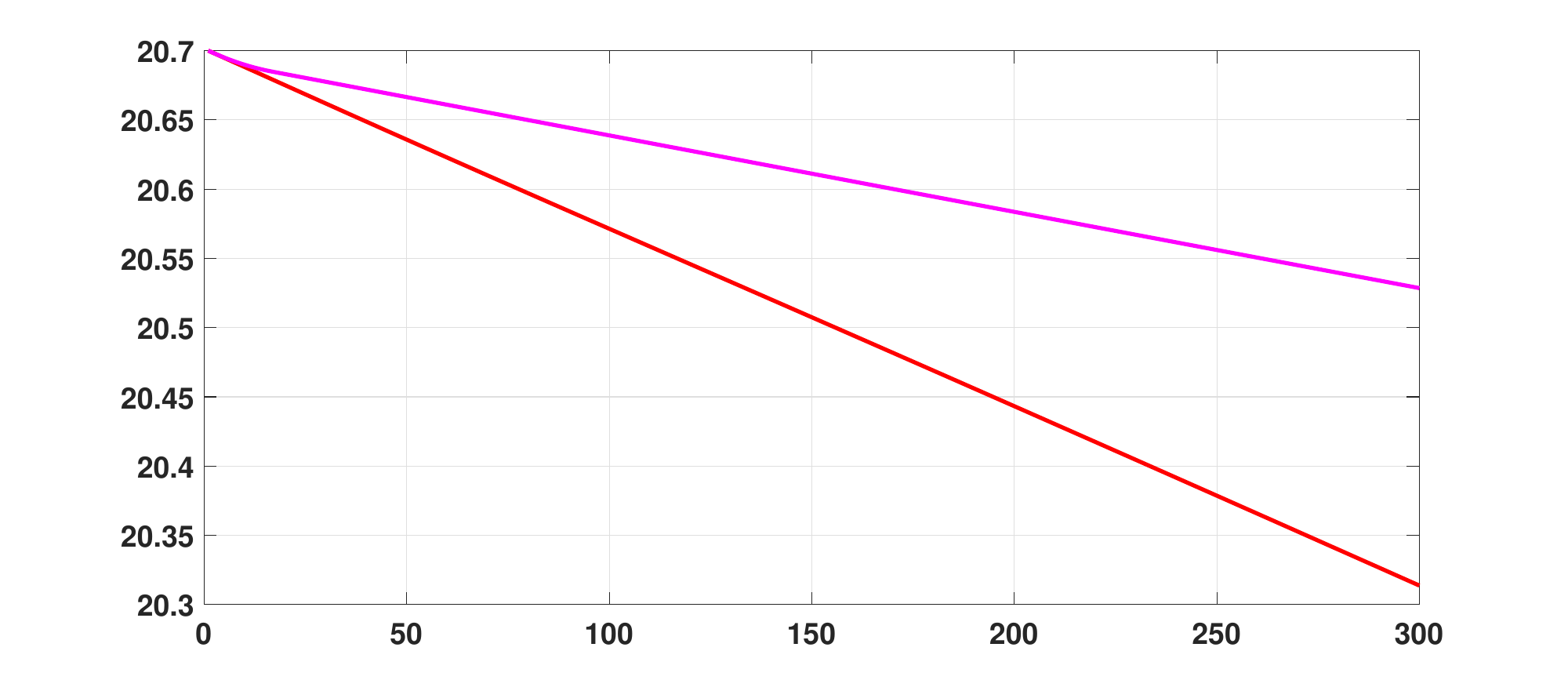}}
%%caption
\caption{Simulations of controlling infected human sizes in Colombia  and Brazil.}
\label{controledboth-humans}
\end{figure}

Hence, using the controls $\eta_1^*, \eta_2^*,$ and $\eta_3^*$ determined in Proposition \ref{Prop-OCP-2}  to optimize the controlled dynamical system \eqref{model_ocp}, we investigate the incorporation of control strategies into the simulated endemic scenarios for the uncontrolled system \eqref{model1} in Colombia and Brazil (Figure \ref{endemicboth}). Hence, we obtain in Figures  \ref{controledboth-mosquitos}--\ref{controledboth-humans} a timely controlled pattern. The sizes of humans infected with ZIKV, HIV/ZIKV, and HIV/AIDS are reduced due to mixed strategies to maintain the level of infection. However, the size of the HIV-infected population is increasing because of ART, resulting in better healthcare for the infected population, which helped to reduce the number of individuals with AIDS. In contrast, in Figure \ref{controledboth-mosquitos}, ZIKV prevention measures helped reduce the number of infected mosquitoes and their contact with susceptible individuals. 

In Figure \ref{controls-both}, the control
$\eta_1$ is required at its full maximum at the beginning by establishing measures such as mosquito repellents and nets, elimination of mosquito breeding sites, and protective clothing.
$\eta_2$ goes in three months to its maximum capacity to contain the number of individuals who attended the AIDS stage, which slows the progression of the disease, prevents further damage to the immune system, and reduces the risk of HIV transmission.
Additionally, $\eta_3$ describes a gradual campaign to promote condom use can help reduce the transmission of HIV and increase awareness about the importance of protecting oneself and one's partners from sexually transmitted infections, such as educating people about the benefits of using condoms to prevent the spread of HIV, increasing easy access to condoms and encourage their use.

%%%%
Compartments $I_z$ and $A$ are controlled to a smaller size than the $\eta_1=\eta_2=\eta_3=0$ simulation, while compartments $I_h$ and $I_{hz}$ have a larger size owing to the ART that prevents infected people from reaching the AIDS stage. However, the mixed strategies were maintained at their approximate total capacities, as illustrated in Figure \ref{controledboth-humans}. 
\par 
In addition, the size of the infected mosquito population remained below the initial condition, which prevented the spread of ZIKV due to the effect of the $\eta_1$ control.

%%%%%%%%%%%%
%%%%%%%%%%%
%%%%%%%%%%new figures

\begin{figure}[H]
\centering
%%%%%%controlled I_hz eta1, eta2
\subfigure[Colombia ($\eta_1$-$\eta_2$) ]{
\includegraphics[width=8.cm, height=4cm]{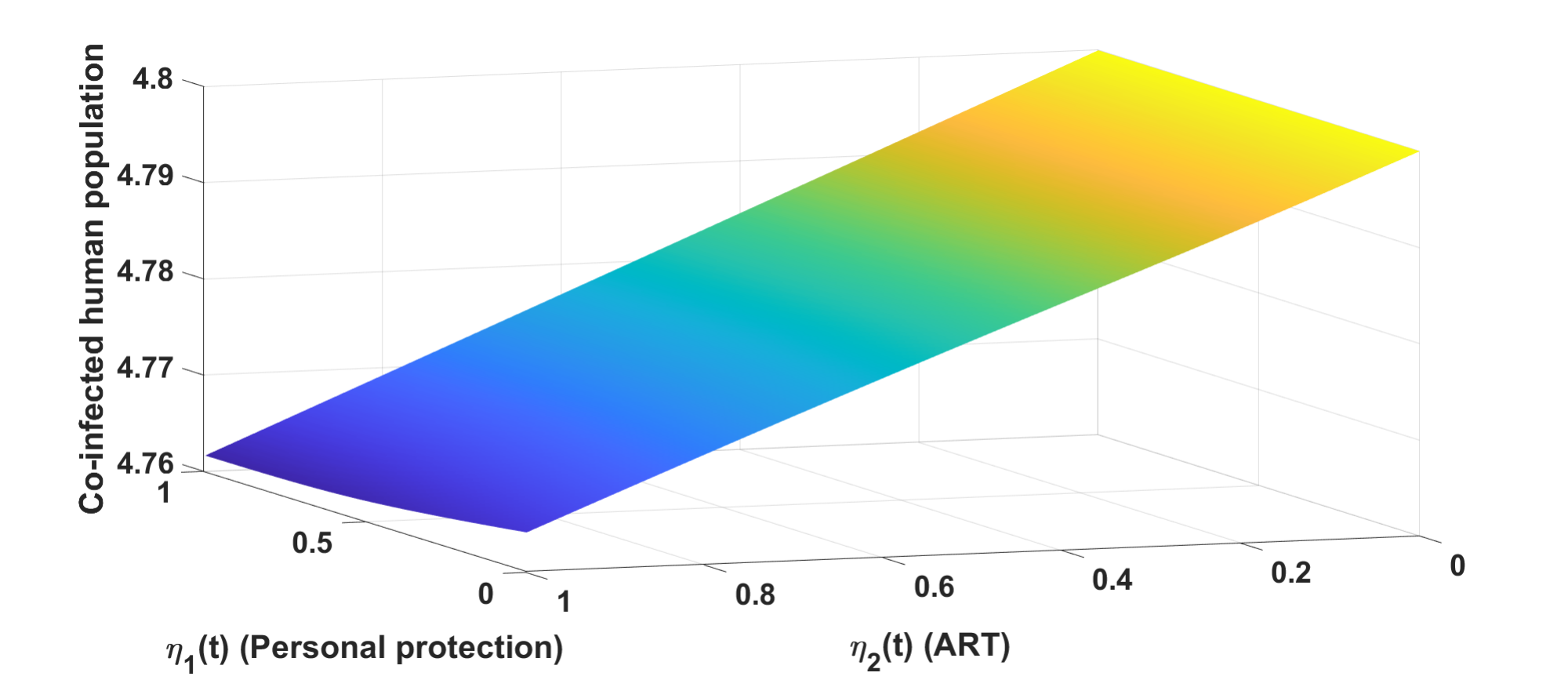}}
\subfigure[Brazil ($\eta_1$-$\eta_2$) ]{
\includegraphics[width=8.cm, height=4cm]{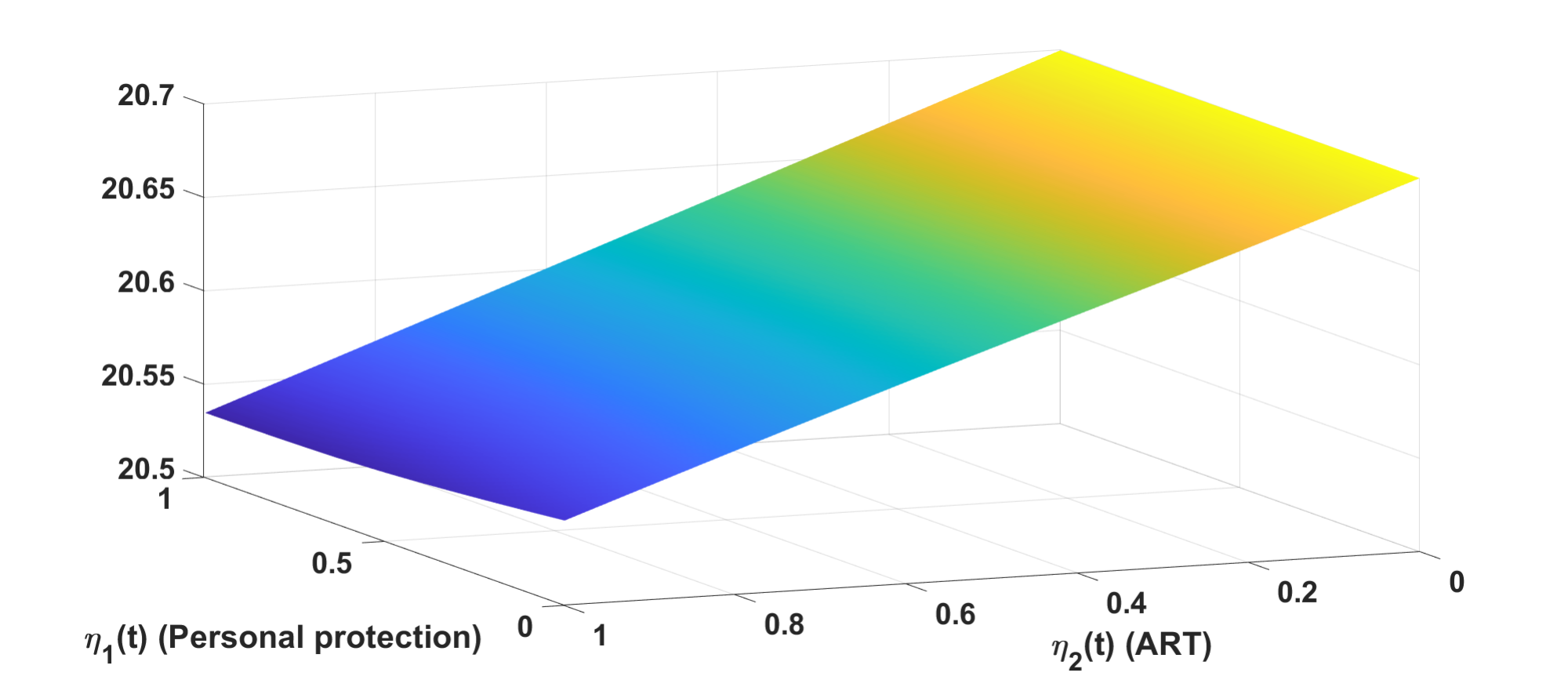}}
%%%%%%%%%%eta2 eta3
\subfigure[Colombia ($\eta_2$-$\eta_3$) ]{
\includegraphics[width=8.cm, height=4cm]{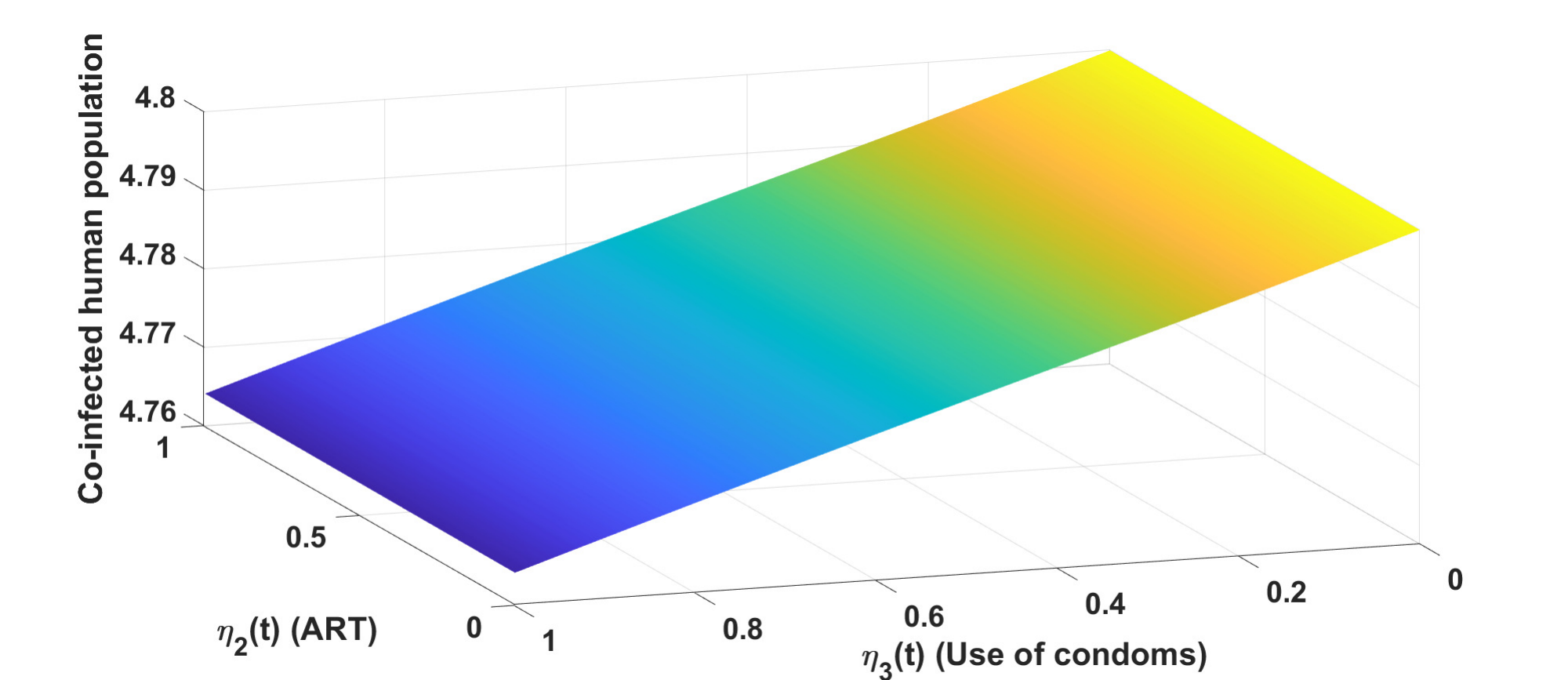}}
\subfigure[Brazil ($\eta_2$-$\eta_3$)]{
\includegraphics[width=8.cm, height=4cm]{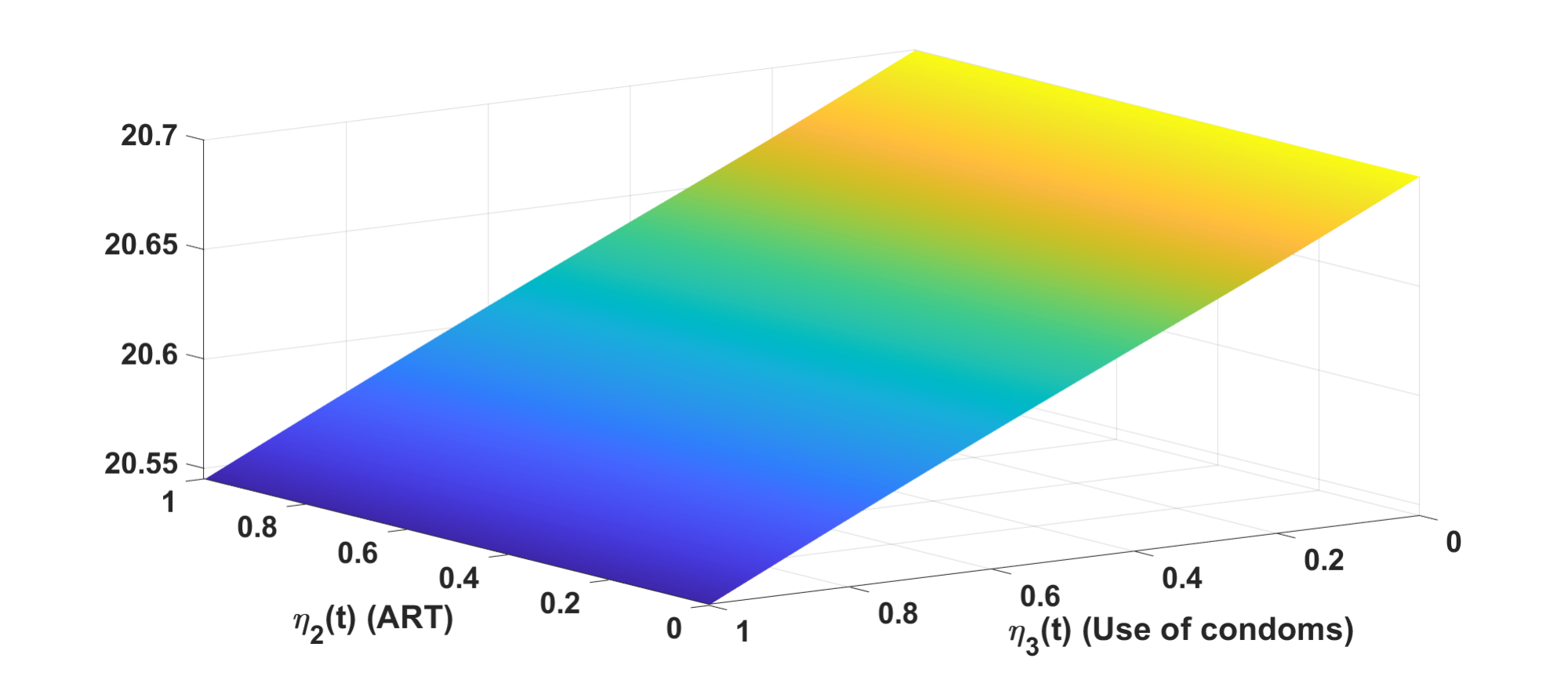}}
%%caption
\caption{Simulation of the effects of combined $\eta_1$-$\eta_2$ and $\eta_2$-$\eta_3$ on the co-infected human population in Colombia and Brazil. }
\label{3d-controledboth-humans}
\end{figure}

\begin{figure}[H]
\centering
%%%eta1 and eta2
\subfigure[Colombia ($\eta_1$-$\eta_2$)]{
\includegraphics[width=7.cm, height=4cm]{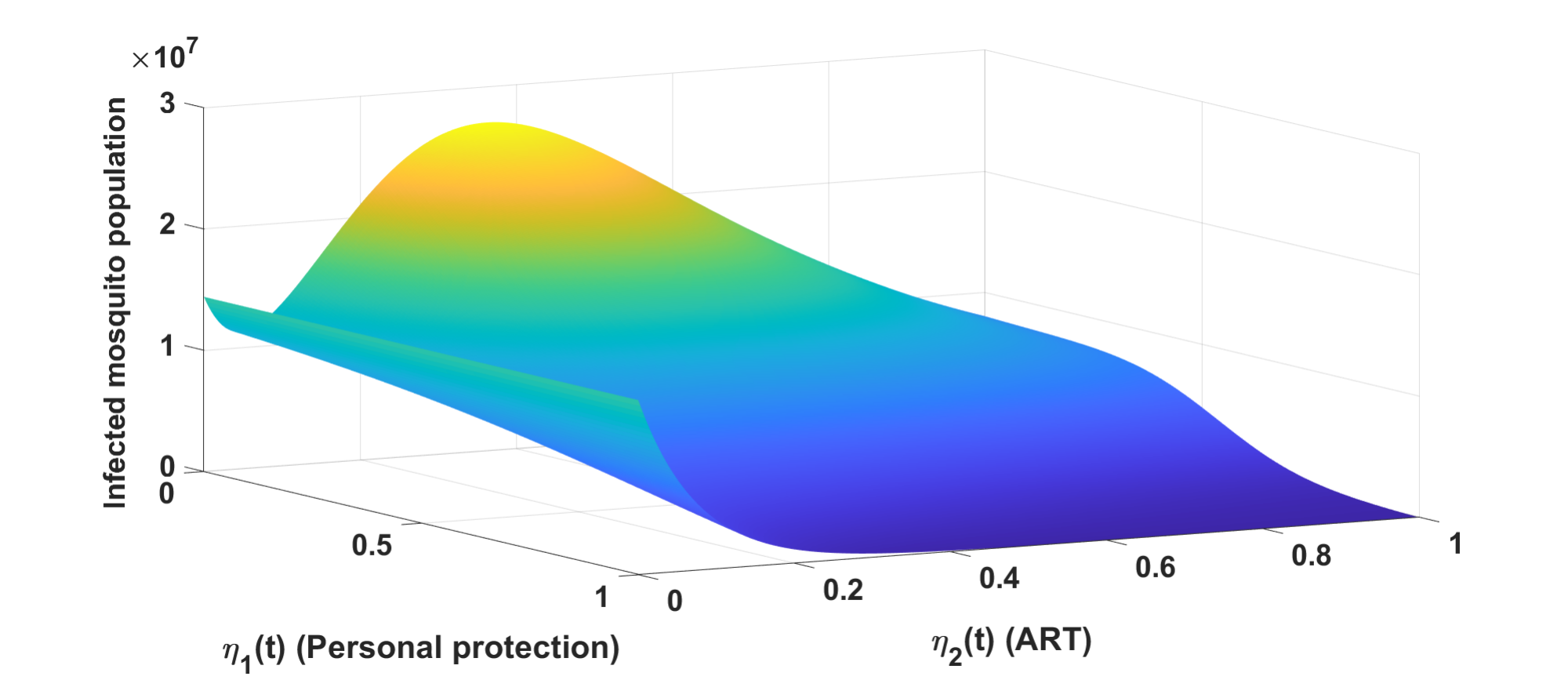}}
\subfigure[Brazil($\eta_1$-$\eta_2$)]{
\includegraphics[width=7.cm, height=4cm]{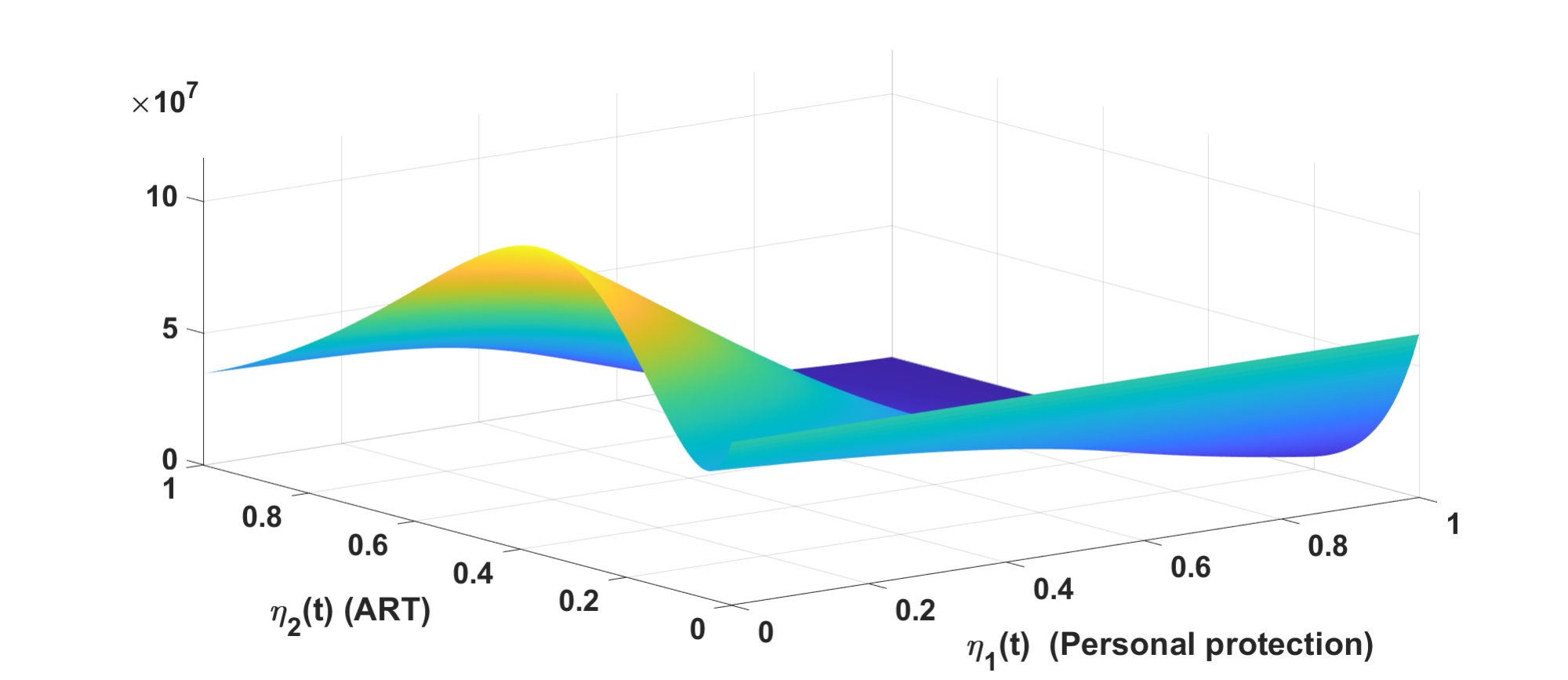}}
%%%%eta2 and eta3
\subfigure[Colombia ($\eta_2$-$\eta_3$)]{
\includegraphics[width=7.cm, height=4cm]{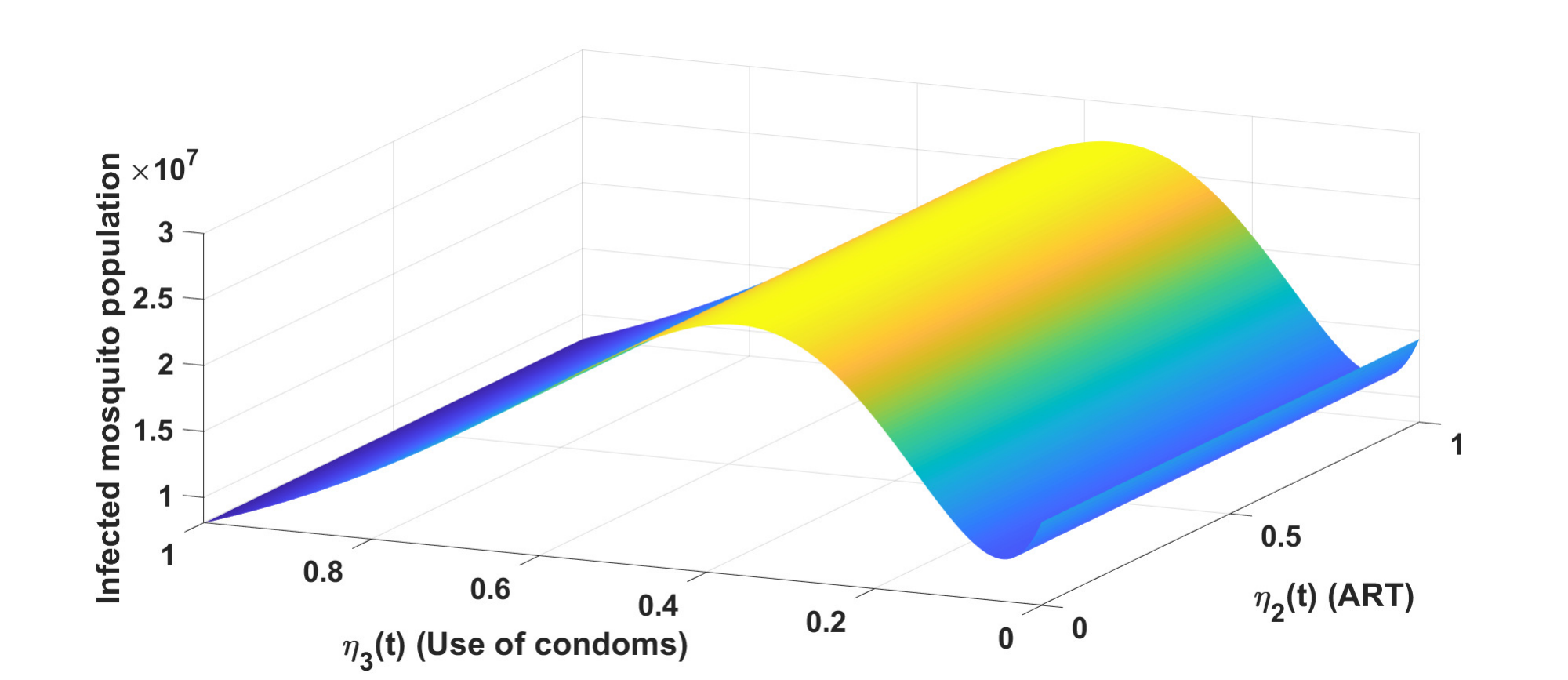}}
\subfigure[Brazil ($\eta_2$-$\eta_3$) ]{
\includegraphics[width=7.cm, height=4cm]{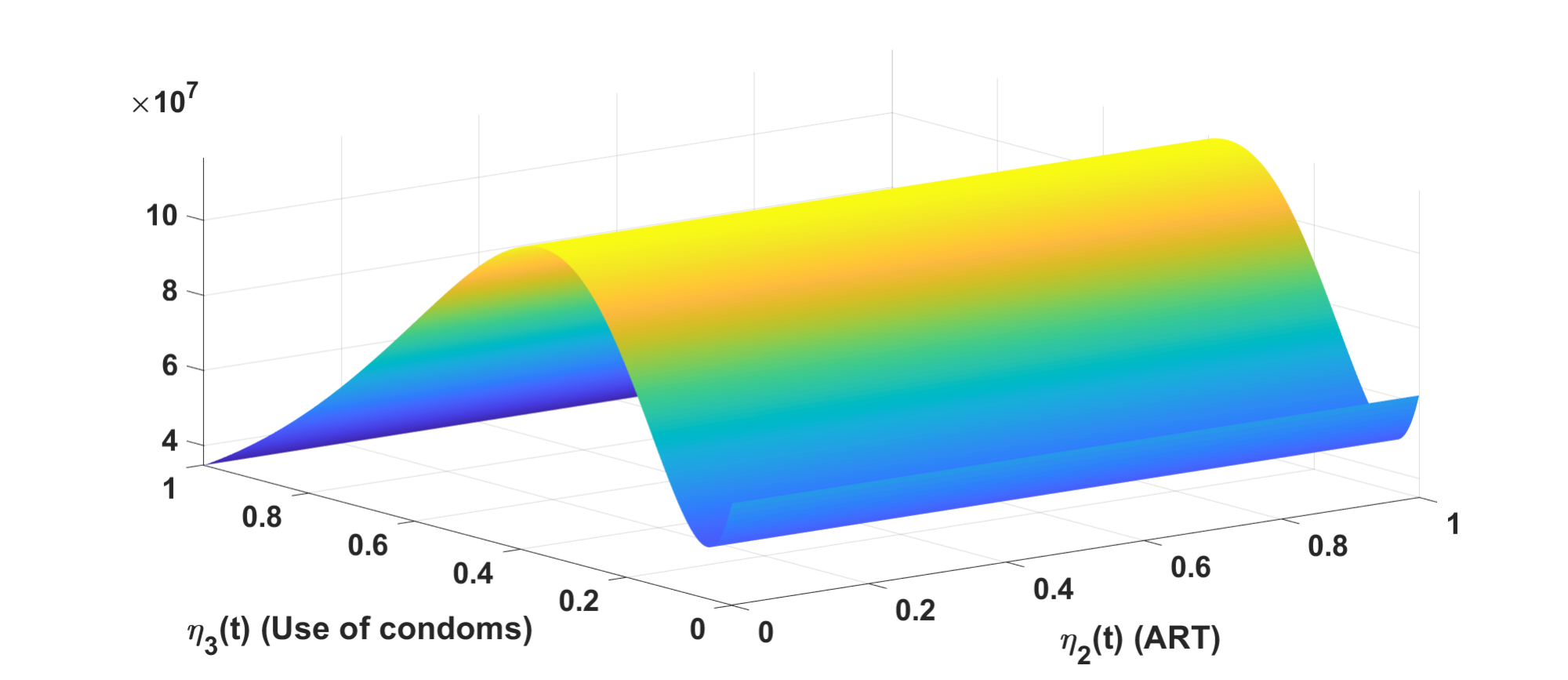}}
\caption{Simulation of the effects of combined $\eta_1$-$\eta_2$ and $\eta_2$-$\eta_3$ on the infected mosquito population in Colombia and Brazil.}
\label{3d-controledboth-mosquitos}
\end{figure}

Finally, to corroborate the behaviour of the co-infected human and the mosquitoes populations with respect to the variation of the combined controls $\eta_1$, $\eta_2$ and $\eta_3$, (see Figures \ref{3d-controledboth-humans} and \ref{3d-controledboth-mosquitos}), we have performed some additional numerical experiments. Figure \ref{3d-controledboth-humans} shows that the concurrent activation of $\eta_1$ (personal protection) and $\eta_2$ (ART) generate comparable trends in the co-infected individuals with Zika and HIV within the contexts of Colombia and Brazil. Hence, in  Figures \ref{3d-controledboth-humans} (a)-(b), we investigated the co-infection patterns by fixing the control $\eta_1$ for personal protection and increasing gradually the ART control from 0 to its maximum value 1 with an increment value of $\Delta \eta_1=\Delta \eta_2=0.003$.
In both cases for Brazil and Columbia, the increase of the personal protection did not play a significant role as the gradual increase of ART for the co-infection. However, the personal protection plays an important role to reduce the overall ZIKV infected population.
Analogously, the simultaneous implementation of ART and condoms use ($\eta_3$) in Colombian and Brazil reduce the number of co-infected individuals burden. Hence, in Figures \ref{3d-controledboth-humans} (c-d), the condom use plays a more significant role in reducing the size of co-infected size.

 This supports the outcomes derived from the simulations presented in Figure \ref{controledboth-humans}, wherein the control strategy involving the simultaneous application of all controls above was previously established.
The dynamics exhibited by the population of Zika virus-carrying mosquitoes, when subjected to concurrent interventions involving personal protection, ART, and condom use, diverge from those observed within the co-infected human population. This contrast is evident upon examining Figure \ref{3d-controledboth-mosquitos}, wherein distinct patterns manifest for Colombia and Brazil. Notably, the impact of personal protection emerges as more pronounced than that of ART, as underscored by the observation that mosquito density exhibits heightened levels when the efficacy of personal protection approaches to zero. The limited influence of sexual transmission dynamics of Zika and HIV on mosquito density substantiates this outcome.
Conversely, upon simultaneous implementation of ART and condom use in Colombia and Brazil, the peaks in mosquito densities materialize when both controls are administered at 50\%. In contrast, the lowest densities align with situations with no controls or controls implemented at 100\%.
These observations significantly explain the complex interplay between control strategies and their varying effects on human and mosquito populations in the HIV/ZIKV co-infection context. This information could guide policymakers and public health authorities in making informed decisions regarding the most effective strategies for managing and controlling these diseases.

%%%%%%%%

%%%%%%%
\section{Discussion and concluding remarks}\label{sec-discussion}
%%%%%%%%

The first documented cases of HIV/ZIKV co-infection in Colombia and Brazil highlighted the potential interactions between these two viruses. The co-circulation of both illnesses in South America presented an important challenge for the public health authorities, as both viruses share the same transmission mechanism, which is sexual contact, and overlap clinical symptoms. 
\par
The mathematical modelling conducted in this study  was used to theoretically represent  the transmission dynamics of both viruses in co-infected individuals, allowing for the evaluation of different intervention scenarios. The findings of this study were particularly relevant during the 2015-2016 Zika outbreak in South America, which coincided with the region's chronic HIV epidemic. By identifying the patterns of this co-infection and their potential impact on transmission, the results of this study could inform public health strategies to control and prevent the spread of both viruses.
\par 
The numerical simulations conducted with parameter values for Colombia and Brazil offered useful information into the potential impact of HIV/ZIKV co-infection. By comparing three distinct scenarios about co-infection probabilities (high, medium, and low), the study demonstrated that an increase in these parameters led to a higher number of co-infected individuals. Through the qualitative behaviour of our mathematical model, we found the following results: First,  both countries exhibited similar trends in spreading and stabilizing these viruses, with differences primarily liable to the population size. Second, individuals infected with ZIKV demonstrated greater variability in their behaviour than those infected with HIV/AIDS. This suggests that ZIKV infection displayed a diverse and fluctuating progression with various manifestations, potentially influenced by multiple factors. In contrast, individuals infected with HIV/AIDS exhibited more consistent and stable behaviour over time.
Third, the numerical experiment results demonstrated that increasing the values of the co-infection probabilities $\omega_1$ and $\omega_2$ could lead to a higher population size of co-infected individuals. This finding underscores the importance of implementing effective strategies to prevent and control the spread of these viruses, especially in more vulnerable populations, such as pregnant women. Fourth, the main strategies for preventing HIV and Zika infections included avoiding mosquito bites, using condoms during sexual activity, and treatment strategies such as antiretroviral therapy (ART). 
\par 
Given that Zika is mainly spread by mosquitoes, we found that timely control measures are of the personal protection type, which includes, but is not limited to, the use of repellents, clothing that covers the entire body, and staying indoors during peak mosquito biting hours. We also illustrated a control for condoms use to lower the risk of transmission of Zika, which can also spread sexually. Similarly, since sexual contact is the primary way HIV spreads, the most effective strategy to prevent HIV transmission during sex is to use condoms appropriately and regularly. Furthermore, we incorporated a control variable for ART, which included the use of a variety of HIV medications to successfully suppress the virus, maintain or enhance immune function, and reduce the risk of transmission.
\par

While this study has provided insightful information, it is crucial to recognize several limitations in its methodology. First, the mathematical modelling approach, which is essential to capture the innate complexity of viral transmission dynamics, introduces certain simplifications. These oversimplifications result from the challenge of modelling a biological phenomenon in a more accessible manner, avoiding potential barriers to general reader comprehension and unnecessary complexities in the presentation of results. The interactions among susceptible, co-infected, and infected individuals with a single pathogen can yield a multitude of scenarios, making it extremely challenging to predict outcomes owing to the complex nature of infection pathways and potential combinations of results. For instance, a susceptible individual engaging in sexual contact with a co-infected person may contract Zika, HIV, or both simultaneously, creating a scenario that is not easily representable by mathematical equations. Consequently, our model assumed that co-infected individuals could transmit the virus only through vectorial contact. Second, the study's reliance on numerical simulations with parameter values specific to Colombia and Brazil reduces the generalizability of the findings to other geographical locations. Diverse epidemiological landscapes, varying healthcare practices, and demographic gaps across regions may result in distinct patterns of Zika and HIV co-infection dynamics. Additionally, this study's dependence on limited historical data and assumptions regarding intervention effectiveness may not fully capture the dynamic nature of evolving public health strategies. Current developments in medical interventions, shifts in public health policies, and the emergence of new viral variants can significantly affect the effectiveness of the proposed intervention measures. These limitations emphasize the need for care when extending the study's findings. Additionally,  more research is necessary to improve the mathematical models' applicability under different scenarios and the challenges in terms of public health
.
\par 
In summary, this study highlighted the need for continued research on the transmission dynamics of Zika and HIV/AIDS and developing effective intervention strategies to control and prevent their spread. Future work in this field plans to incorporate compartments of women giving birth to babies with and without congenital malformations to understand the impact of co-infection in children better. Including these compartments would allow for a more detailed assessment of the long-term effects of co-infection on child health outcomes, including the potential for developmental delays, neurological deficits, and other complications. This approach could also facilitate the development of more targeted prevention and treatment strategies for the affected children and their families. Ultimately, this research is critical for improving our understanding of the complex interactions between HIV and Zika and developing effective public health interventions to mitigate their impact on affected individuals and communities.

%%%%%%%%agradecimientos
\section*{Acknowledgements}
The authors appreciate the support provided by the One Health Modelling Network for Emerging Infections (OMNI-RÉUNIS) and Mathematics for Public Health (MfPH), which are financially supported by t Natural Sciences and Engineering Research Council of Canada (NSERC) and the Public Health Agency of Canada (PHAC).

\section*{Data Availability }
All data involved during this study are included in this published article.

\section*{Conflicts of Interest }
Authors declare that  have NO conflicts of interest to disclose in relation to this work.
%%%%%%%%

%%%%%machine

%%%%%%%%%References
\bibliographystyle{ieeetr}
\bibliography{Biblio}
\newpage

%%%%%%%%%%%%%%%%%%
%%%%%%%%%%%APENDIX
%\section*{Appendix}
\appendix 

\section{Appendices}
\subsection{Proof that $\mathbf{E}_{h_0}$ is GAS when $\mathcal{R}_h<1$}\label{appendix:A1}

Let $\textbf{X}$ be	the vector field given by  the right side hand of System \eqref{planar-hiv}. It is enough to prove the existence of a Lyapunov function  for the translated system 

\begin{equation*}
\dot{\textbf{X}}=F(\textbf{X}+\textbf{E}_{h_0})-F(\textbf{E}_{h_0})=f(\textbf{X}),
\end{equation*}
where the system $\dot {\textbf{Y}}=F(\textbf{Y})$ has $\textbf{Y}=0$ as equilibrium solution. Let us consider the function 
\begin{equation*}\label{lyapunov-hiv}
V^*(S, I_h, A)=\dfrac{1}{\sigma_1+\mu}I_h, 
\end{equation*}
and define 
\begin{equation}\label{lyapunov-hiv}
V(\bar S, \bar I_h, \bar A)=V^*\left(S-\frac{\Lambda}{\mu}, I_h, A\right).
\end{equation}
We verify that the function $V$ defined on \eqref{lyapunov-hiv} is a Lyapunov function.  Indeed, $V$ is positive definite, that is, $V(\textbf{E}_{h_0})=V^*(\textbf{0})=0$ and $V>0$ for all $(\bar S, \bar I_h, \bar A)\neq \textbf{E}_{h_0} in \,\Omega_h$. Additionally, the orbital derivative of $V$ along the trajectories of system \eqref{model-hiv} is

\begin{equation*}
\begin{array}{ll}
\dot V&=\dfrac{\partial V^*}{\partial \left(S-\frac{\Lambda}{\mu}\right)}\dot S+\dfrac{\partial V^*}{\partial I_h}\dot I_h+\dfrac{\partial V^*}{\partial A}\dot A \\ \\
 &=\dfrac{1}{\sigma_1+\mu}\left(\beta_h I_hS-(\sigma_1+\mu)I_h  \right)\\ \\
&\leq \dfrac{1}{\sigma_1+\mu}(\sigma_1+\mu)\left( \dfrac{\beta_h \frac{\Lambda}{\mu}}{\sigma_1+ \mu}-1\right)I_h \\ \\ 
&= (\mathcal{R}_h-1)I_h \leq 0,\quad \forall I_h\geq 0.
\end{array}
.\end{equation*}

Let $\triangle =\{  (S, I_h, A)\in \mathbb{R}_+^3: \dot V=0\}\subset \{ (S, I_h, A)\in \mathbb{R}_+^3: I_h=0  \}$ and $\triangle ^\prime \subset \triangle$ the largest invariant set respect to \eqref{model-hiv}.  It can be easily showed that $\triangle^\prime=\{ \textbf{E}_{h_0}\}$.  Therefore by the LaSalle's invariance principle \cite{teschl2012ordinary}, $\textbf{E}_{h_0}$ is a global attractor whenever $\mathcal{R}_h<1$.

%%%%%%%%%
\subsection{Proof that $\mathbf{E}_{h}^*$ is GAS when $\mathcal{R}_h>1$}\label{appendix:A2}

Note that the third equation of System \eqref{model-hiv} is uncoupled in the variable $S$ and its  equilibrium  is $I_h=\frac{\mu_h+\mu}{\sigma_1}A$.
Replacing this value in the two first equations, we obtain the planar system

\begin{equation}\label{planar-hiv}
\left\{
\begin{array}{ll}
 \dfrac{dS}{dt}&=\Lambda-\beta_h \dfrac{\mu_h+\mu}{\sigma_1}SA-\mu S \\ \\
\dfrac{dA}{dt}&=\beta_h AS-(\sigma_1+\mu )A. 
\end{array}
\right.
\end{equation}

Therefore, solutions to system \eqref{model-hiv} tend asymptotically to those of the planar system \eqref{planar-hiv} (see e.g. \cite{castillo1994asymptotically}).  
The Dulac criterion \cite{mccluskey1998bendixson} claims that if there exists a real continuously differentiable function $\phi(S,A)$ such that $\nabla\cdot[\phi(S,A)\textbf{X}(S,A)]\neq 0$, where $\textbf{X}(S,A)=(F(S,A), G(S,A)$	is the right side hand of system \eqref{planar-hiv}, then there are no periodic orbits contained entirely inside $\Omega_h$. Let

\begin{equation*}
\phi(S,A)=\dfrac{1}{SA} \quad \text{for} \quad S>0, \; A>0,
\end{equation*}

then 

\begin{equation*}
\begin{array}{ll}
\nabla\cdot [\phi(S,A)\textbf{X}(S,A)] &=\dfrac{\partial(F\phi)}{\partial S}+\dfrac{\partial (G\phi)}{\partial A} \\ \\
&=\dfrac{\partial}{\partial S}\left( \frac{\Lambda}{SA}-\frac{\beta_h (\mu_h+\mu)}{\sigma_1 SA}SA-\frac{\mu S}{SA} \right)+\dfrac{\partial}{\partial A}\left( \frac{\beta_h AS}{SA}-\dfrac{\sigma_1+\mu}{SA}A \right) \\ \\
&=\dfrac{\partial}{\partial S}\left( \frac{\Lambda}{SA}-\frac{\beta_h (\mu_h+\mu)}{\sigma_1}-\frac{\mu}{A}\right)+\dfrac{\partial}{\partial A}\left(\beta_h-\frac{\sigma_1+\mu}{S}\right)\\ \\
&= -\dfrac{\Lambda}{AS^2}<0, \, \text{for} \, S,A>0.
\end{array}
.
\end{equation*}

Thus, there are no periodic orbits in $\Omega_h$. Given that $\Omega_h$ is
positively invariant, and the endemic equilibrium exists if $\mathcal{R}_h>1$, it
follows from the Poincaré-Bendixson Theorem \cite{mccluskey1998bendixson} that all solutions of the
system starting in $\Omega_h$ remain in $\Omega_h$ for all $t$. Thus, because of the absence of periodic
orbits in $\Omega_h$, this implies that the unique endemic equilibrium of System \eqref{model-hiv} is \textit{GAS} when  $\mathcal{R}_h>1$.

%%%%%%%%%%%%%%%%%%%%%%
\subsection{Proof of Lemma \ref{lemmaR*}}\label{appendix:A3}
%\section{Proof of Lemma \ref{lemmaR*}}\label{appendix:A3}
We prove item $i)$; all others are proved in a similar way. On the one hand, from the definition of $\mathcal{R}_z^*$,  we immediately have that if $\mathcal{R}_z^*=2\mathcal{R}_{z_1}+\bar{\mathcal{R}}_{z_2}<1$, then $2\mathcal{R}_{z_1}<1.$   On the other hand, 
\begin{equation*}
\begin{array}{ll}
\mathcal{R}_z^*<1 &\Rightarrow \mathcal{R}_z^2+2\mathcal{R}_{z_1}(1-\mathcal{R}_z)<1 \\ \\
&\Rightarrow \mathcal{R}_z(1-2\mathcal{R}_{z_1})+2\mathcal{R}_z\\ \\
&\Rightarrow \mathcal{R}_z(1-2\mathcal{R}_{z_1})<1-2\mathcal{R}_z,  \quad  (2\mathcal{R}_{z_1}<1)\\ \\
&\Rightarrow \mathcal{R}_z<1.
\end{array}
\end{equation*}

%%%%%%%%%%%%
\subsection{Proof of the condition  $R^*<R_{max}$}\label{appendix:A4}
\begin{figure}[H]
\centering
\includegraphics[width=11cm, height=6cm]{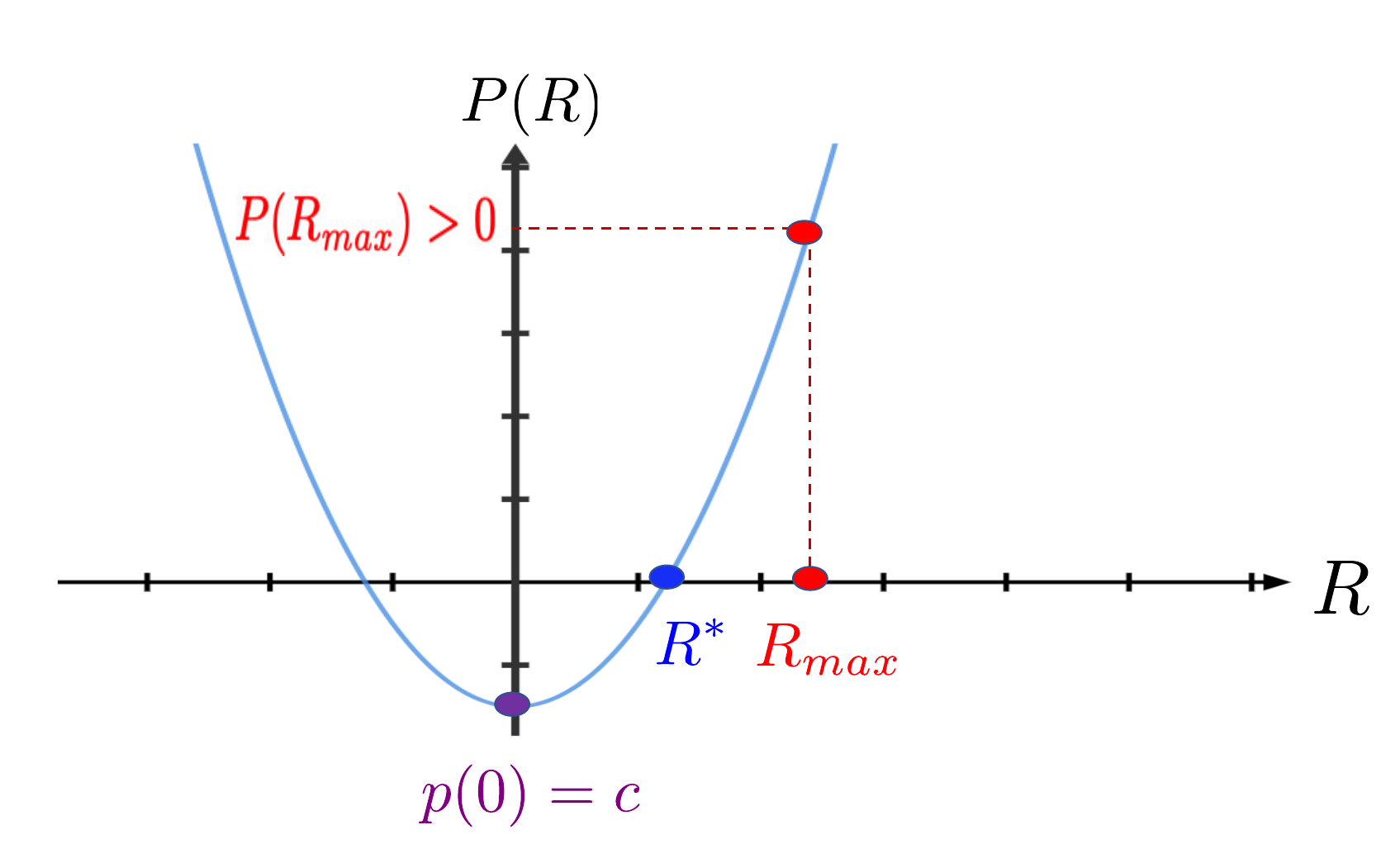}
\caption{Polynomial $p(R)$ associated to system \eqref{R-zikv}.}
\label{fig02}
\end{figure}

We have to prove that  $R^*<R_{max}$ is satisfied. To this end, we consider the polynomial $p(R)$ associated to system \eqref{R-zikv}, whose coefficients are in equation  \eqref{coeficients-R-zikv}.  The  graph of $p(R)$ is a parabola that opens upward  (see Figure \ref{fig02}). Since $R^*>0$ and $p(0)=c<0$ as long as $\mathcal{R}_{z_1}+\mathcal{R}_{z_2}>1 $, then  to ensure that  $R^*<R_{max}$, it is enough to prove that $p(R_{max})>0$. In fact

\begin{equation*}
\begin{array}{ll}
p(R_{max})&=aR_{max}^2+bR_{max}+c \\ \\
                  &=\dfrac{\beta_z\Lambda\alpha_m\mu}{\delta_z R_{max}} R_{max}^2+\left[\left(  \dfrac{\beta_m\alpha_m\Lambda_m}{\mu_m}+\beta_z\mu_m \right)\dfrac{\Lambda}{R_{max}}+\left[\alpha_m\mu(\mu_z+\delta_z+\mu)-\beta_z\Lambda\alpha_m\right]\dfrac{\mu}{\delta_z}\right]R_{max} \\ \\
   &+(\mu_z+\delta_z+\mu)\mu\mu_m\left(1- \mathcal{R}_z^* \right) \\\\           
                &=\dfrac{\beta_z\Lambda\alpha_m\mu}{\delta_z } R_{max}+\left(  \dfrac{\beta_m\alpha_m\Lambda_m}{\mu_m}+\beta_z\mu_m \right)\Lambda+\left[\alpha_m\mu(\mu_z+\delta_z+\mu)-\beta_z\Lambda\alpha_m\right]\dfrac{\mu}{\delta_z}R_{max} \\ \\
                  &+(\mu_z+\delta_z+\mu)\mu\mu_m\left(1- \mathcal{R}_z^* \right) \\ \\
                  &=  \left(  \dfrac{\beta_m\alpha_m\Lambda_m}{\mu_m}+\beta_z\mu_m \right)\Lambda+\dfrac{\alpha_m\mu^2(\mu_z+\delta_z+\mu)R_{max}}{\delta_z} +(\mu_z+\delta_z+\mu)\mu\mu_m\left(1- \mathcal{R}_z^* \right)\\ \\
                   &=(\mu_z+\delta_z+\mu)\mu\mu_m(\mathcal{R}_z^*) (\mathcal{R}_z^*-\mathcal{R}_z^*) +\alpha_m\Lambda\mu^2+(\mu_z+\delta_z+\mu)\mu\mu_m \\ \\
                   &=\alpha_m\Lambda\mu^2+(\mu_z+\delta_z+\mu)\mu\mu_m >0.
\end{array}
\end{equation*}

%%%%%%%%%%%%%%%%%%%%%%%%
\subsection{Proof that $\textbf{E}_{z_0}$ is GAS when  $\mathcal{R}_{z}^*<1$}\label{appendix:A5}
We will use $\kappa=\mu_z+\delta_z+\mu$ and the notation and reasoning analogous to that used in Appendix \ref{appendix:A1}, with functions

\begin{equation*}
V^*(S, I_z, R, S_m, I_m)=\left(S-\dfrac{\Lambda}{\mu}\log{\dfrac{S\mu}{\Lambda}}\right)+I_z+\dfrac{\beta_m\Lambda}{\mu\mu_m}\left(S_m-\dfrac{\Lambda_m}{\mu_m}\log{\dfrac{S_m\mu_m}{\Lambda_m}}\right)+\dfrac{\beta_m\Lambda}{\mu\mu_m}I_m
\end{equation*}
and 
\begin{equation}\label{lyapunov-zikv}
V(\bar S, \bar I_z, \bar R, \bar S_m, \bar I_m)=V^*\left(S-\frac{\Lambda}{\mu}, I_z, R, S_m,  I_m-\frac{\Lambda_m}{\mu_m}\right).
\end{equation}
We must prove that $V$ defined in \eqref{lyapunov-zikv} is a Lyapunov function.  $V$ is positive definite and  the orbital derivative of $V$ along the trajectories of \eqref{model-zikv} is

\begin{equation*}
\begin{array}{ll}
\dot V&=\left(1-\frac{\Lambda}{\mu S} \right)\dot S+\dot I_z+\frac{\beta_m\Lambda}{\mu\mu_m}\left(1-\frac{\Lambda_m}{\mu_mS_m}\right)\dot S_m+\frac{\beta_m\Lambda}{\mu\mu_m}\dot I_m \\ \\
%%%%l1
&=\left(1-\frac{\Lambda}{\mu S} \right)(\Lambda-\beta_mI_mS-\beta_zI_zS-\mu S)+ (\beta_mI_mS+\beta_zI_zS-\kappa I_z) \\ \\
&+\frac{\beta_m\Lambda}{\mu\mu_m}\left(1-\frac{\Lambda_m}{S_m\mu_m}\right)(\Lambda_m-\alpha_mI_zS_m-\mu_mS_m)+\frac{\beta_m\Lambda}{\mu\mu_m}(\alpha_mI_zS_m-\mu_mI_m) \\ \\
%%%%%l2
&=(\Lambda-\mu S)-\frac{\Lambda}{\mu S}(\Lambda-\mu S)+\frac{\Lambda}{\mu S}(\beta_mI_m+\beta_z I_z)S-\kappa I_z+\frac{\beta_m\Lambda}{\mu\mu_m}(\Lambda_m-\mu_m S_m) \\ \\
&-\frac{\beta_m\Lambda\Lambda_m}{\mu\mu_mS_m\mu_m}(\Lambda_m-\mu_m S_m)+\frac{\beta_m\Lambda\Lambda_m\alpha_mS_mI_z}{\mu\mu_mS_m\mu_m}-\frac{\beta_m\Lambda\mu_mI_m}{\mu\mu_m}\\ \\
%%%%l3
&=-\frac{(\Lambda-\mu S)^2}{\mu S}-\frac{\beta_m\Lambda}
{\mu\mu_m^2}\frac{(\Lambda_m-\mu_mS_m)^2}{S_m}+\left(\frac{\Lambda}{\mu}\beta_z+\frac{\beta_m\alpha_m\Lambda_m\Lambda}{\mu\mu_m^2}-\kappa\right)I_z \\ \\
%%%%l4
&=-\frac{(\Lambda-\mu S)^2}{\mu S}-\frac{\beta_m\Lambda}
{\mu\mu_m^2}\frac{(\Lambda_m-\mu_mS_m)^2}{S_m}+\kappa\left(\mathcal{R}_z^*-1\right)I_z.
\end{array}
\end{equation*}

Note that the last expression in the above inequality is  negative if  $\mathcal{R}_z^*<1$ and for $S=\frac{\Lambda}{\mu}$, $S_m=\frac{\Lambda_m}{\mu_m}$ and $I_z=I_m=0$.  Finally, using  the LaSalle's invariance principle \cite{teschl2012ordinary}, $\textbf{E}_{z_0}$ defined in \eqref{DFE-zikv} is a global attractor whenever $\mathcal{R}_z^*<1$.

%%%%%%%%%%%%%%%%%%%%
\subsection{Sensitivity index of $\mathcal{R}_z$  with respect to  $\beta_z$}\label{appendix:A6}

We have
\begin{equation*}
\begin{array}{ll}
\dfrac{\partial \mathcal{R}_z}{\partial \beta_z}&=\dfrac{\partial}{\partial \beta_z}\left(  \dfrac{\beta_z\Lambda}{2\mu\kappa_1}+ \sqrt{\left(\dfrac{\beta_z\Lambda}{2\mu\kappa_1}\right)^2+\dfrac{\alpha_m\beta_m\Lambda_m\Lambda}{\mu\mu_m^2\kappa_1}}\right)  \\ \\
&=\dfrac{\Lambda}{2\mu (\delta_z + \mu + \mu_z)}+ \dfrac{\beta_z \Lambda^2}{
 4 \mu^2 (\delta_z + \mu + \mu_z)^2 \sqrt{\dfrac{\beta_z^2 \Lambda^2}{
   4 \mu^2 (\delta_z + \mu + \mu_z)^2} + \dfrac{\alpha_m \beta_m \Lambda \Lambda_m}{
   \mu \mu_m^2 (\delta_z + \mu + \mu_z)}} }.
\end{array}
\end{equation*}
Then, it is enough to compute the expression for

\begin{equation*}
\dfrac{\partial \mathcal{R}_z}{\partial \beta_z}\dfrac{\beta_z}{\mathcal{R}_z}=\dfrac{\partial \mathcal{R}_z}{\partial \beta_z}\dfrac{\beta_z}{ \dfrac{\beta_z\Lambda}{2\mu\kappa_1}+ \sqrt{\left(\dfrac{\beta_z\Lambda}{2\mu\kappa_1}\right)^2+\dfrac{\alpha_m\beta_m\Lambda_m\Lambda}{\mu\mu_m^2\kappa_1}}}.
\end{equation*}

%%%%%%%%%%%%%%%%%%%%
\subsection{Proof of  Proposition  \ref{Prop_existence_OCP}}\label{appendix:A7}
All state variables and controls  are non-negative and, for $i=\{1,2,3\}$, the set of control variables $\eta_i \in \mathcal{A}$ is also convex and closed. We note that  the boundedness of the optimal system \eqref{model_ocp} determines the compactness for the existence of the optimal control. 
Moreover, there exists a constant $\nu>1$, $\omega_1=\min ( d_1,d_2,d_3 )$, and $\omega_2>0$ such that
\begin{equation}
       \mathcal{J}(\eta) \geq \omega_1 ||\eta ||^{\nu} - \omega_2.
\end{equation}
Therefore, according to \cite{lukes1982differential}, the controlled system \eqref{model_ocp} admits an optimal control solution $\eta^*$.

%%%%%%%%%%%%%%%%%%%%
\subsection{Proof of Proposition  \ref{Prop-OCP-2}}\label{appendix:A8}
We have
\begin{eqnarray*}
     \dfrac{p_1}{dt}&=-\dfrac{\partial H}{\partial S}=& p_1 \left[(1-\eta_1) \beta_m I_m+(1-\eta_3)(\beta_z I_z+\beta_h I_h) +\mu \right] - p_2 \left[(1-\eta_1)\beta_mI_{m}+(1-\eta_3)\beta_z I_z \right] \nonumber \\
     & & - p_3 \left[ (1-\eta_3)\beta_hI_h  \right], \\
    \dfrac{p_2}{dt}&=-\dfrac{\partial H}{\partial I_z}=& -c_1 + p_1  (1- \eta_3) \beta_z S - p_2 \left[ (1- \eta_3) \beta_z S - \omega_2 (1- \eta_3) \beta_h I_h -( \mu_z + \delta_z + \mu )\right] - p_3 ( 1 - \eta_3) \beta_z I_h  \\
    & & -p_6 \delta_z + p_7 (1- \eta_1) \alpha_m S_m- p_8 (1- \eta_1 ) \alpha_m S_m, \\
    \dfrac{p_3}{dt}&=-\dfrac{\partial H}{\partial I_h}=& -c_2 + p_1 (1- \eta_3) \beta_h S + p_2 \omega_2 (1- \eta_3) \beta_h I_z - p_3 \left[ \omega_1[(1-\eta_1)\beta_mI_m+(1-\eta_3)\beta_zI_z] \right. \\
    & & \left.- (1- \eta_2) \sigma_1 - \mu \right] -p_4 \left[ \omega_1[(1-\eta_1)\beta_mI_m+(1-\eta_3)\beta_zI_z]+\omega_2(1-\eta_3)\beta_h I_z\right] \\
    & & -p_5 (1- \eta_2) \sigma_1, \\
     \dfrac{p_4}{dt}&=-\dfrac{\partial H}{\partial I_{hz}}=& -c_3+p_4 \left( \epsilon\delta_z+(1-\eta_2)\sigma_2+\mu_{hz}+\mu\right) - p_5(1-\eta_2) \sigma_2 - p_6 \epsilon \delta_z + p_7 (1- \eta_1) \alpha_m S_m \\
     & & - p_8 (1- \eta_1) \alpha_m S_m,\\
    \dfrac{p_5}{dt}&=-\dfrac{\partial H}{\partial A}=& - c_5 + p_5 (\mu_h + \mu ), \\
     \dfrac{p_6}{dt}&=-\dfrac{\partial H}{\partial R}=& p_6 \mu, \\
     \dfrac{p_7}{dt}&=-\dfrac{\partial H}{\partial S_m}=& p_7 \left( (1-\eta_1)\alpha_m (I_z+I_{hz})  + \mu_m \right) - p_8   (1-\eta_1)\alpha_m (I_z+I_{hz}),  \\
      \dfrac{p_8}{dt}&=-\dfrac{\partial H}{\partial I_m}=&  -c_4 + p_1 (1- \eta_1) \beta_m S - p_2 ( 1- \eta_1 ) \beta_m S + p_3 (1- \eta_1) \beta_m I_h - p_4 \omega_1 (1- \eta_1) \beta_m I_h + p_8 \mu_m,
\end{eqnarray*}
with transversality conditions $p_i(T)=0$, for $i=\{ 1,2,3,4,5,6,7,8\}$. According to PMP, the optimal conditions are
\begin{eqnarray*}
     \dfrac{\partial H}{\partial \eta_1} & = &  d_1 \eta_1 - \left( (p_2 - p_1) S +  (p_4 - p_3) I_h \right)  \beta_m I_m - (p_8 - p_7) \alpha_m (I_z + I_{hz}) S_m= 0,  \\  
     \dfrac{\partial H}{\partial \eta_2} & = &   d_2 \eta_2 - \left( p_5 - p_3  \right)  \sigma_1 I_h - \left( p_5 - p_4  \right)  \sigma_2 I_{hz}= 0, \\
     \dfrac{\partial H}{\partial \eta_3} & = &  d_3 \eta_3 - (p_2 - p_1) \beta_z I_z S   - (p_3 - p_1) \beta_h I_h S - (p_4 - p_3) \omega_1 \beta_z I_z I_h  - (p_4 - p_2) \omega_2 \beta_h I_z I_h = 0.  
\end{eqnarray*}
Hence, we get assertions \eqref{asser1-3}. This completes the proof.

\end{document}